\documentclass[english,aip,article]{revtex4}
\usepackage[T1]{fontenc}
\usepackage[latin9]{inputenc}
\usepackage{verbatim}
\usepackage{textcomp}
\usepackage{relsize}
\usepackage{amsmath}
\usepackage{graphicx}
\usepackage{amssymb}
\usepackage{esint}

\makeatletter

\DeclareRobustCommand{\lyxmathsym}[1]{\ifmmode\begingroup\def\b@ld{bold}
  \def\rmorbf##1{\ifx\math@version\b@ld\textbf{##1}\else\textrm{##1}\fi}
  \mathchoice{\hbox{\rmorbf{#1}}}{\hbox{\rmorbf{#1}}}
  {\hbox{\smaller[2]\rmorbf{#1}}}{\hbox{\smaller[3]\rmorbf{#1}}}
  \endgroup\else#1\fi}

\providecommand{\tabularnewline}{\\}

\usepackage{rotate}
\usepackage{epsfig} 
\usepackage{graphicx} 
\usepackage{graphics} 

\makeatother

\usepackage{babel}

\begin{document}

\title{Two hard spheres in a pore:\\ Exact Statistical Mechanics for different
shaped cavities}

\author{Ignacio Urrutia}

\affiliation{Consejo Nacional de Investigaciones Científicas y Técnicas, Av. Rivadavia
1917, RA-1033 Buenos Aires, Argentina. Departamento de Física, Comisión
Nacional de Energía Atómica,\\ Av. Gral. Paz 1499, RA-1650 San Martín,
Buenos Aires, Argentina.\\ Departamento de Física, Facultad de Ciencias
Exactas y Naturales, Universidad de Buenos Aires, Ciudad Universitaria,
RA-1428 Buenos Aires, Argentina. iurrutia@cnea.gov.ar}
\begin{abstract}
The Partition function of two Hard Spheres in a Hard Wall Pore is
studied appealing to a graph representation. The exact evaluation
of the canonical partition function, and the one-body distribution
function, in three different shaped pores are achieved. The analyzed
simple geometries are the cuboidal, cylindrical and ellipsoidal cavities.
Results have been compared with two previously studied geometries,
the spherical pore and the spherical pore with a hard core. The search
of common features in the analytic structure of the partition functions
in terms of their length parameters and their volumes, surface area,
edges length and curvatures is addressed too. A general framework
for the exact thermodynamic analysis of systems with few and many
particles in terms of a set of thermodynamic measures is discussed.
We found that an exact thermodynamic description is feasible based
in the adoption of an adequate set of measures and the search of the
free energy dependence on the adopted measure set. A relation similar
to the Laplace equation for the fluid-vapor interface is obtained
which express the equilibrium between magnitudes that in extended
systems are intensive variables. This exact description is applied
to study the thermodynamic behavior of the two Hard Spheres in a Hard
Wall Pore for the analyzed different geometries. We obtain analytically
the external work, the pressure on the wall, the pressure in the homogeneous
zone, the wall-fluid surface tension, the line tension and other similar
properties.
\end{abstract}
\maketitle

\section{Introduction}

The exact analytical evaluation of the partition function and thermodynamic
properties in systems of confined particles is a new trend in statistical
mechanics. Due to the inherent difficulties in searching the exact
solution of three dimensional systems, the interest is focused in
few confined particles, and is restricted to Hard Spherical particles.
Systems composed by many Hard Spheres (HS) have attracted the interest
of many people because of they constitute a prototypical three dimensional
simple fluid \cite{Lowen_2000}. Even, though its apparent simplicity
only a few exact analytical results are known. In the limit of large
homogeneous systems, only the first four virial coefficients in the
pressure virial series for the monodisperse system are known (see
\cite{Nairn_1972} and references therein). Similarly, the fourth
order coefficient for the polydisperse systems were also obtained
\cite{Blaak_1998}. It is interesting to note that the exact equation
of state (EOS) for the HS is unknown although an approximate, simple,
analytical, and accurate EOS was found by Carnahan and Starling \cite{Carnahan_1969}.
The earlier published works on HS were specially devoted to the analysis
of uniform fluid properties, as it was the classical Molecular Dynamic
experiment on fluid particles of Alder and Wainwright \cite{Alder1957}.
Gradually, the focus of succeeding works turns to inhomogeneous systems.
In the last decades a great effort were devoted to the understanding
of HS inhomogeneous fluid systems, in part because such system are
the starting point of several density functional theories \cite{Lowen_2002,Roth_2010}.
These general theories deal with a large class of simple and complex
fluid systems with successfully results in the study of the substrate-fluid
behavior including wetting, capillary condensation, and adsorption
phenomena. Recent advances in the analysis of fluid adsorption in
porous matrix were supported by developments in this field \cite{Neimark_2003,Neimark_2006,Neimark_2009}.
In last years much attention was focused to small systems of HS confined
in vessels. The study of simple fluids constrained to small cavities
of various shapes has enlightening fundamental questions of statistical
mechanics and thermodynamics (for example about phase transitions
\cite{Kegel_1999,Neimark_2006}), but only recently the relevance
of few body systems was recognized.

Few bodies confined systems is a topic of statistical mechanics which
belong at the opposite of the thermodynamic limit. The study of such
systems is becoming technologically interesting because the manipulation
of matter in the microscopic and nanoscopic scales shows that they
can be built. Besides, the design of new nano-devices could take advantage
of its properties. From that point of view, the use of simple hard-core
potentials enables a schematic description of the interactions between
particles and with the container. As we will see below, this simplified
picture makes analytically tractable the two-body system. Interestingly,
colloidal particles with HS-like interaction has been produced and
studied experimentally \cite{Kruglov_2005,Pusey_1994,Pusey_1986}.

In few bodies systems different ensembles are not equivalents. The
correct ensemble to describe the properties of a given system is such
that better simulates its real properties. Thus, the canonical partition
function of the confined few-HS systems attempts to describe the statistical
mechanics properties of this system kept at constant temperature.
Besides, exact canonical ensemble studies of few bodies confined systems
provides the building blocks for an exact grand canonical study of
them. The grand ensemble is important because the statistical mechanics
theory of macroscopic liquids is largely developed in such framework.
We recognize that the absence of exact results for inhomogeneous fluids
in this framework is an obstacle which difficult the theoretical improvement
of the theory of liquids. Thus, we expect that in the near future
the connection between exactly solved few bodies systems and the theory
of macroscopic fluids can provide new theoretical insight.

From now on we will focus on the analytical exact solution of few
HS system in a pore making emphasis on canonical ensemble results.
Until present only the two HS (2-HS) system was tackled. Recently,
the canonical ensemble 2-HS confined in a spherical cavity was solved
\cite{Urrutia_2008}, and also, the system confined in a spherical
cavity with a hard internal core was evaluated \cite{Urrutia_2010}.
In both works the principal result is the analytic expression of the
configuration integral (CI), but the one body density distribution
and pressure tensor were analyzed too. Studies of the same system
in the framework of the microcanonical ensemble has also been done
\cite{Uranagase_2006}. The present work (PW) is devoted to the exact
solution of the statistical mechanic properties of 2-HS into hard
wall simple pores in the framework of the canonical ensemble. We expose
new results for the cuboidal, the cylindrical and the ellipsoidal
cavities. We should mention that the microcanonical ensemble CI of
2-HS in a cuboidal cavity found in \cite{Uranagase_2006} is formally
identical to that analyzed in PW for the same vessel. However, we
present a different approach to the integral evaluation and a simpler
and more explicit expression of the CI. We have checked that both
solutions are equivalent.

In section \ref{sec:Two-bodies} we show how a hard wall cavity that
contains 2-HS can be treated as another particle. There, we show explicit
expressions of the canonical configuration integral for 2-HS into
three pores of simple shape. We study the confinement in a cuboidal,
cylindrical, and spheroidal, cavities. The obtained exact CI are functions
of a set of parameters $\mathbf{X}$ which characterize the different
shapes of the cavity. In Sec. \ref{sec:Distribution-function} we
analyze both, the one body distribution function and the pressure
tensor, for some of the studied cavities. In this Section we also
obtain an analytic expression for the intersecting volume between
a cuboid and a sphere which appears to be a novel geometrical result.
Sec. \ref{sec:Analysis} is devoted to the search of some universal
features in the CI of the 2-HS system constrained to simple geometric
cavities including the cuboidal, cylindrical, spherical, ellipsoidal,
and also the spherical cavity with a concentric hard core. A discussion
of how to obtain a thermodynamic description of the system by transforming
the CI from $Z_{2}(\mathbf{X})$ to a more interesting description
$Z_{2}(\mathbf{M})$ where $\mathbf{M}$ is a set of thermodynamic
measures is done in Sec. \ref{sec:Thermodynamic-study}. There, we
find the equations of state of the 2-HS system in the studied cavities
and obtain some exact results for the many HS system in contact with
curved walls. Final remarks are shown in Sec. \ref{sec:Conclusions}.

\section{Two bodies in a pore\label{sec:Two-bodies}}

The canonical partition function of two distinguishable particles
in a pore is $Q_{2}=\Lambda^{-6}Z_{2}$ being $\Lambda$ the thermal
de Broglie wavelength and $Z_{2}$ the CI, which may be expressed
as a three nodes graph\begin{equation}
{\psfig{figure=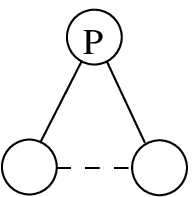,height=1.2cm}}=Z_{2}=\int\int\, e(\mathbf{r}_{1})e(\mathbf{r}_{2})\, e(\mathbf{r}_{12})\, d\mathbf{r}_{1}d\mathbf{r}_{2}\;.\label{eq:Q0}\end{equation}
Here $e(\mathbf{r}_{i})=e_{i}=Exp(-\beta U(\mathbf{r}_{i}))$ with
$i=1,2$, $e(\mathbf{r}_{12})=e_{12}=Exp(-\beta\varphi(\mathbf{r}_{12}))$,
$\mathbf{r}_{12}=\mathbf{r}_{1}-\mathbf{r}_{2}$, $U$ is the external
potential acting on each particle, $\varphi$ is the interparticle
potential, and the integration must be performed over the infinite
space. The accessible region of space for the $i$th-particle, $\Omega$,
is the region where $e_{i}>0$, and its boundary is $\partial\Omega$.
In PW we assume that $\Omega$ and $\partial\Omega$ are the same
for all (two) particles. The labeled $P$ node in Eq. (\ref{eq:Q0})
that represents the pore is linked to the particles by the $e_{i}$
bonds drawn with continuous lines in Eq. (\ref{eq:Q0}). Particles
are linked each to other by the $e_{ij}$ bond drawn with dashed line.
Pores with hard walls have $e_{i}=\{1\textrm{ if }\mathbf{r}_{i}\in\Omega,\textrm{ and }0\textrm{ otherwise}\}$
and then the $e_{i}$ bonds fulfils the in-pore condition. For hard
spherical particles $e_{ij}=\Theta(r_{ij}-\sigma)$, where $\Theta$
is the Heaviside function, $r_{ij}=\left|\mathbf{r}_{ij}\right|$
and $\sigma$ is the hard repulsion distance (it is also the diameter
of one HS). Therefore, the $e_{ij}$ bond fulfils the non-overlap
between particles condition being null if particles overlap each other.
Both conditions are mandatory for the non-null value of the integrand
in Eq. (\ref{eq:Q0}). It is clear that $Z_{2}$ for a 2-HS system
confined in a hard wall cavity is by its nature a geometrical magnitude.
This means that $Z_{2}$ depends on $\sigma$ and a set of parameters
which characterize the shape and size of the cavity. Therefore, $Z_{2}$
is a piece of the bridge that links geometry and thermodynamics. We
will return to this point in Sec. \ref{sec:Thermodynamic-study}.
Before the evaluation of the integral (\ref{eq:Q0}) we may perform
some simple Mayer type transformations on it. Using the general identity
$e=1+f$ we may replace the $e_{12}$ bond and/or the $e_{i}$ bonds.
The introduced $f_{12}$ function is non-null only if both particles
are overlapping while $f_{i}$ is null if $i$-particle is in the
pore. We will draw the functions $e_{i}$ and $f_{12}$ with continuous
line while we will draw the functions $f_{i}$ and $e_{12}$ with
dashed line. Following this procedure we obtain
\begin{equation}
{\psfig{figure=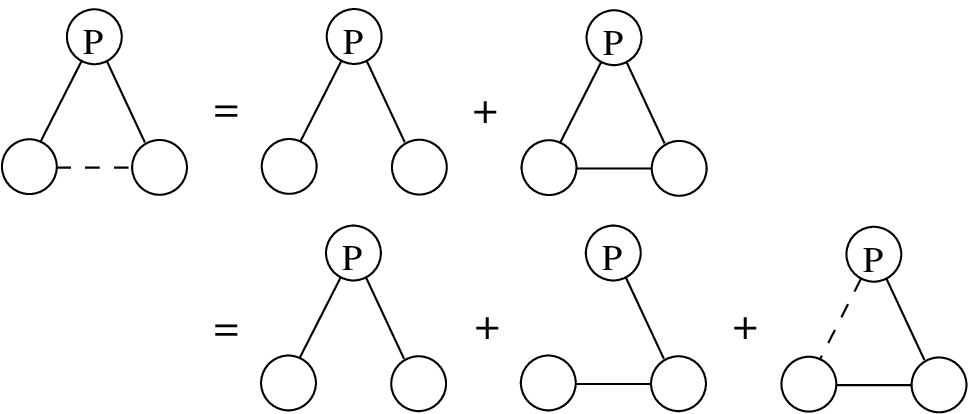,width=8.0cm}}\;,\label{eq:tri1}\end{equation}
where each graph with an articulation node can be factorized and easily
evaluated \cite{Hill56} taking into account some trivial identities
\begin{equation}
{\psfig{figure=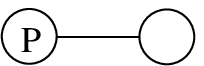,width=1.5cm}}=Z_{1}\:,\label{eq:2cont}\end{equation}
\begin{equation}
{\psfig{figure=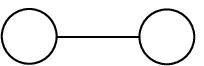,width=1.5cm}}=-2b_{2}=-\frac{4\pi}{3}\sigma^{3}\:.\label{eq:2cont2}\end{equation}
Here $Z_{1}$ is the CI of the one particle system, $-b_{2}$ is
the usual second virial coefficient, and $\sigma$ plays the role
of exclusion radius. Note that $Z_{1}$ depends both, on the shape
of the empty cavity and the HS size $\sigma$.In the first row of
Eq. (\ref{eq:tri1}) it was explicitly separated the independent
particle term $Z_{1}^{2}$ from the second term that concentrates
all corrections to this simple picture. This term is $2Z_{1}b_{2}(pore)$
being $b_{2}(pore)$ the first cluster integral with the complete
dependence on the size and shape of the pore \cite{Hill56}. Therefore
the first row of Eq. (\ref{eq:tri1}) is
\begin{equation}
Z_{2}=Z_{1}^{2}-2Z_{1}b_{2}(pore)\;.\label{eq:b00}\end{equation}
From an opposite point of view we may regard the p-node as it was
a particle. This allow us to recognize that the right hand side term
in the first row of Eq. (\ref{eq:tri1}) is part of the third virial
coefficient of a fluid mixture \cite{Urrutia_2008,Urrutia_2010}.
In the second row of Eq. (\ref{eq:tri1}) was also extracted the first
non-ideal gas term, $-2Z_{1}b_{2}$, which contains the usual second
virial coefficient for homogeneous systems. Therefore, the third term
contains the nontrivial core of the problem involving a complex dependence
on the pore's shape parameters. It hides the inhomogeneous system
dependencies and takes the control over the entire density regime,
from low density (or large pore size) to the close packing condition.
Moreover, this term produces ergodic-non-ergodic transitions and dimensional
crossovers. To make a contribution to the last graph in the second
row Eq. (\ref{eq:tri1}), one particle must be outside of $\Omega$
(or inside $\bar{\Omega}$, the complement of $\Omega$) while the
other particle must be inside of $\Omega$, and also both particles
must be near each other. This explains that for large pores the term
scales with the surface area of the container which is a measure of
the size of $\partial\Omega$. Even more interesting, this graph remains
unmodified if we turn to the conjugate system of 2-HS \emph{confined}
in $\bar{\Omega}$, i.e. the graph is symmetric with respect to the
in-out inversion. More explicitly, we introduce a partition of the
euclidean space $\mathbb{E}^{3}=\Omega\cup\bar{\Omega}$ being $V_{\infty}=Z_{1}(\Omega)+Z_{1}(\bar{\Omega})$
the volume of the space. The Eq. (\ref{eq:b00}) is valid for $Z_{2}=Z_{2}(\Omega)$
and $Z_{1}=Z_{1}(\Omega)$ as was already stated, but also for $Z_{2}=Z_{2}(\bar{\Omega})$
and $Z_{1}=Z_{1}(\bar{\Omega})$. This is the in-out symmetry of the
2-HS system confined in a hard wall cavity \cite{Urrutia_2008}.

\begin{figure}
\begin{centering}
\includegraphics[scale=0.4]{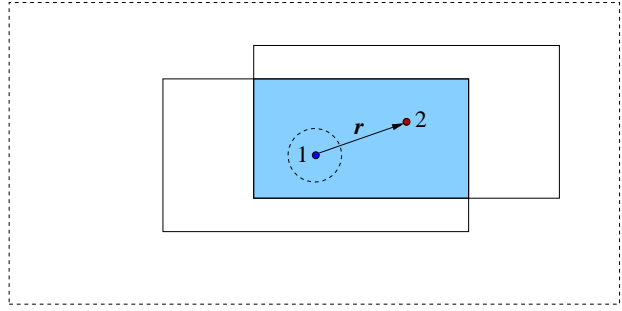}
\par\end{centering}

\caption{(color online) Representation of the integration procedure utilized
in the evaluation of $Z_{2}$. The drawn points indicate particles
1 and 2. The solid line draw cavity-1 and -2, while the overlap region
with volume $W$ is shaded. Dashed line plots the ES and EB. \label{fig:Integration-technike-picture.}}

\end{figure}

Now we concentrate in the evaluation of Eq. (\ref{eq:Q0}). In principle,
the integration is over the positions of both particles (with a fixed
pore position), however, it can be rewritten as an integration over
the coordinates of the pore and one particle (by fixating the second
particle). Hence, we first fix both particles coordinates and integrate
over the pore center position which allow that both particles be inside
the cavity. The result of the integration is the volume $W$. To build
the region with volume $W$ we can follow a simple geometrical recipe.
Choose one point of the cavity, e.g. the center, and draw two cavities
centered at particle-1 and particle-2 positions. The cavity-2 must
be the translation in $\mathbf{r}$ of the cavity-1, i.e., they must
be equally oriented. The overlap between cavities -1 and -2 is the
available region for the pore center. In Fig. \ref{fig:Integration-technike-picture.}
we show a schematic picture for a cuboidal pore. The result of the
first integration is the overlap volume $W$, the grey region (color
online) defined by the overlap of cavities -1 and -2. At a second
stage, we should integrate over the position of particle-2 with coordinate
$\mathbf{r}$. The integration domain is the region outside the exclusion
sphere (ES) with radius $\sigma$ and inside the zone of vanishing
$W$ which defines the external boundary (EB). The EB is determined
by the region enclosed by all the positions of particle-2 when we
support cavity-2 on cavity-1 and translate it in all possible directions
by keeping in touch the boundary of both cavities. Through this padding
procedure the obtained EB is the region enclosed by the dashed line
in Fig. \ref{fig:Integration-technike-picture.}. The CI of the system
reads \begin{equation}
Z_{2}=\intop W(\mathbf{r})\, e(r)\, d^{3}r\:,\label{eq:Q}\end{equation}
where $e(r)=\Theta(r-\sigma)$. By integrating only the pore center
position we find an unnormalized two body density distribution, $g(\mathbf{r})=Z_{2}^{-1}W(\mathbf{r})\, e(r)$.
Interestingly, $Z_{2}$ and $W(\mathbf{r})$ of the 2-HS confined
system are related to the CI of other systems as it is the confined
stick-particle (or dumbbell) \cite{Urrutia_2010} which may be obtained
by a sticky-bond transformation. The one body density distribution
$\rho(\mathbf{r})$ will be analyzed in Sec. \ref{sec:Distribution-function}.
A simple consequence of Eqs. (\ref{eq:Q0}, \ref{eq:Q}) is that $Z_{2}$
depends on the $\mathbf{X}$ parameters (introduced by the $e_{i}$
bond and $W$ volume), which characterize the geometry of each pore.
Now we are ready to solve Eq. (\ref{eq:Q}) for some simple cavities.
As we mentioned above PW is mainly devoted to study the 2-HS confined
system of distinguishable particles. Even so, at the end of Sec. \ref{sec:Two-bodies}
we make a brief comment about the 2-HS system of indistinguishable
particles.

\subsection{CI of 2-HS in a cuboid pore\label{sub:CI-2HS-CUB}}

The empty cavity is characterized by the length parameters $L'_{x}$,
$L'_{y}$, and $L'_{z}$. We introduce the effective cavity length
parameters $L_{i}=L'_{i}-\sigma$, which characterize the available
space for the center of one particle and the dimensionless lengths
$l_{i}=L_{i}/\sigma$ with $i=x,\, y,\, z$. Then we obtain for the
cuboid shaped pore $Z_{1}=L_{x}L_{y}L_{z}$, and\begin{equation}
W(\mathbf{r})=\left|L_{x}-x\right|\left|L_{y}-y\right|\left|L_{z}-z\right|\;,\label{eq:Wcub}\end{equation}
where $\mathbf{r}=(x,y,z)$ (see Fig. \ref{fig:Integration-technike-picture.}).
The EB is a cuboid with doubled length sides and the $W(\mathbf{r})$
dependence turns convenient to integrate Eq. (\ref{eq:Q}) over $0\leq x\leq L_{x}$,
$0\leq y\leq L_{y}$, $0\leq z\leq L_{z}$ and multiply by $8$. Although,
$W(\mathbf{r})$ is positive defined for any $\mathbf{r}$ and it
is a non analytic function. For practical purposes we will extend
analytically $W$ to enlarge the integration domain outside the EB
box to $x\geq0$, $y\geq0$, and $z\geq0$.  Assuming $L_{x}\leq L_{y}\leq L_{z}$
it is necessary to analyze the integral (\ref{eq:Q}) for different
pore size domains in the parameter space $\mathbf{X}=(L_{x},L_{y},L_{z})$
or the similar $\mathbf{X}=(l_{x},l_{y},l_{z})$. Introducing the
directions: $\hat{xy}\propto L_{x}\hat{x}+L_{y}\hat{y}$, $\hat{xz}\propto L_{x}\hat{x}+L_{z}\hat{z}$,
$\hat{yz}\propto L_{y}\hat{y}+L_{z}\hat{z}$, and $\hat{xyz}\propto L_{x}\hat{x}+L_{y}\hat{y}+L_{z}\hat{z}$
we may distinguish eight different regions of $\mathbf{X}$.

\subsubsection*{Region 1}

The \textit{large pore domain} is defined by the condition that ES
is completely enclosed into the EB, i.e., that $l_{i}\geq1$ with
$i=x,\, y,\, z$. The integral (\ref{eq:Q}) splits into the simpler
ones,

\begin{equation}
\mathcal{I}_{1}=8\int_{0}^{L_{x}}dx\,\int_{0}^{L_{y}}dy\,\int_{0}^{L_{z}}\, W\, dz=Z_{1}^{2}\;,\label{eq:I1}\end{equation}

\begin{eqnarray}
\mathcal{I}_{2} & = & 8\int_{0}^{\sigma}dx\,\int_{0}^{\sqrt{\sigma^{2}-x^{2}}}dy\,\int_{0}^{\sqrt{\sigma^{2}-x^{2}-y^{2}}}\, W\, dz\;,\nonumber \\
 & = & \frac{4\pi}{3}\sigma^{3}L_{x}L_{y}L_{z}-\frac{\pi}{2}\sigma^{4}(L_{x}L_{y}+L_{x}L_{z}+L_{y}L_{z})+\nonumber \\
 &  & \frac{8}{15}\sigma^{5}(L_{x}+L_{y}+L_{z})-\frac{1}{6}\sigma^{6}\;,\label{eq:I2}\end{eqnarray}
which in terms of dimensionless length variables is\begin{eqnarray}
\mathcal{I}_{2} & = & \frac{4\pi}{3}\sigma^{6}l_{x}l_{y}l_{z}-\frac{\pi}{2}\sigma^{6}(l_{x}l_{y}+l_{x}l_{z}+l_{y}l_{z})+\nonumber \\
 &  & \frac{8}{15}\sigma^{6}(l_{x}+l_{y}+l_{z})-\frac{1}{6}\sigma^{6}\;.\label{eq:I2b}\end{eqnarray}
Then, we find 

\begin{equation}
Z_{2}=\mathcal{I}_{1}-\mathcal{I}_{2}=Z_{1}^{2}-2\, Z_{1}b_{2}(cub)\;.\label{eq:Z01}\end{equation}
Last expression is similar to Eq. (\ref{eq:b00}).

\subsubsection*{Region 2}

The \textit{ES exceeds only two faces of the EB domain}. Here, the
exclusion sphere showed in Fig. \ref{fig:Integration-technike-picture.}
should extend beyond the EB at most in one direction normal to the
faces of the box. As far as this direction was labeled as $\hat{x}$,
then we have $l_{x}\leq1$, $l_{y}\geq1$, and $l_{z}\geq1$. We define
the auxiliary integral $\mathcal{I}_{3x}$, its integration domain
is the spherical cup outside the EB box in the $\hat{x}$ direction

\begin{eqnarray}
\mathcal{I}_{3x} & = & 8\int_{L_{x}}^{\sigma}dx\,\int_{0}^{\sqrt{\sigma^{2}-x^{2}}}dy\,\int_{0}^{\sqrt{\sigma^{2}-x^{2}-y^{2}}}\, W\, dz\;,\nonumber \\
 & = & \sigma^{6}\left\{ -\frac{1}{30}\left(1-l_{x}\right)^{4}\left(5+4l_{x}+l_{x}^{2}\right)-\frac{\pi}{6}\, l_{y}l_{z}\left(1-l_{x}\right)^{3}\left(l_{x}+3\right)+\right.\;,\nonumber \\
 &  & \left.\left(l_{y}+l_{z}\right)\left[\frac{1}{15}\,\sqrt{1-l_{x}^{2}}\,\left(8+9\, l_{x}^{2}-2\, l_{x}^{4}\right)-l_{x}\left(\frac{\pi}{2}-\arcsin(l_{x})\right)\right]\right\} \label{eq:I3x}\end{eqnarray}

\begin{equation}
Z_{2}=\mathcal{I}_{1}-\mathcal{I}_{2}+\mathcal{I}_{3x}\;,\label{eq:Z02}\end{equation}
In the same sense, we define $\mathcal{I}_{3i}$ with $i=x,y,z$.
The integration domain of $\mathcal{I}_{3i}$ corresponds to the spherical
cup outside the EB in the $\hat{i}$ direction, which completes the
description of the set of functions $\{\mathcal{I}_{3x},\,\mathcal{I}_{3y},\,\mathcal{I}_{3z}\}$.

\subsubsection*{Region 3}

We consider the situation when \textit{ES exceeds only four faces
of the EB}, where the exclusion sphere must extend beyond the cuboidal
EB in $\hat{x}$ and $\hat{y}$ directions but not in directions $\hat{z}$
and $\hat{xy}$. In this case $l_{x}\leq1$, $l_{y}\leq1$, $l_{z}\geq1$
and $l_{x}^{2}+l_{y}^{2}\geq1$. The CI is\begin{equation}
Z_{2}=\mathcal{I}_{1}-\mathcal{I}_{2}+\mathcal{I}_{3x}+\mathcal{I}_{3y}\;.\label{eq:Z03}\end{equation}

\subsubsection*{Region 4}

The next domain to consider is when \textit{ES exceeds all six faces
of EB but any more}. It goes beyond the EB in $\{\hat{x},\hat{y},\hat{z}\}$
directions but not in $\{\hat{xy},\hat{xz},\hat{yz}\}$. Then, we
need $l_{i}\leq1$ and $l_{i}^{2}+l_{j}^{2}\geq1$, for $i,j=x,\, y,\, z$
with $i\neq j$, therefore\begin{equation}
Z_{2}=\mathcal{I}_{1}-\mathcal{I}_{2}+\mathcal{I}_{3x}+\mathcal{I}_{3y}+\mathcal{I}_{3z}\;.\label{eq:Z04}\end{equation}

\subsubsection*{Region 5}

In this region \textit{ES exceeds at four faces and four edges of
EB}. The sphere fall off the EB in $\{\hat{x},\hat{y},\hat{xy}\}$
directions but not in $\hat{z}$. Then we have obtain $l_{x}^{2}+l_{y}^{2}\leq1$
and $l_{z}\geq1$. We define the auxiliary integral $\mathcal{I}_{3xy}$,
its integration domain is the right angle spherical wedge outside
the EB in both $\hat{x}$ and $\hat{y}$ directions. Note that the
edge of the spherical wedge does not cross the sphere center. In addition,
we define $\mathcal{I}_{2xy}$ , its integration domain is the space
outer to ES and inner to EB, \begin{equation}
\mathcal{I}_{3xy}=8\int_{L_{x}}^{\sqrt{\sigma^{2}-L_{y}^{2}}}dx\,\int_{L_{y}}^{\sqrt{\sigma^{2}-x^{2}}}dy\,\int_{0}^{\sqrt{\sigma^{2}-x^{2}-y^{2}}}\, W\, dz\;,\label{eq:I3xy0}\end{equation}
\begin{eqnarray}
\mathcal{I}_{2xy} & = & 8\int_{0}^{L_{x}}dx\,\int_{0}^{L_{y}}dy\,\int_{0}^{\sqrt{\sigma^{2}-x^{2}-y^{2}}}\, W\, dz\;,\label{eq:I2xy0}\end{eqnarray}
both integrals are related by\begin{equation}
\mathcal{I}_{2xy}=\mathcal{I}_{2}-\mathcal{I}_{3x}-\mathcal{I}_{3y}+\mathcal{I}_{3xy}\;.\label{eq:I2xyI3xy}\end{equation}
For $\mathcal{I}_{2xy}$ we found \begin{eqnarray}
\mathcal{I}_{2xy} & = & \sigma^{6}\left\{ -\frac{1}{6}l_{x}^{2}l_{y}^{2}\left(6-l_{x}^{2}-l_{y}^{2}\right)+\frac{8}{15}l_{z}+\right.\;,\nonumber \\
 &  & \left.\frac{1}{15}l_{z}\,\sqrt{1-l_{x}^{2}-l_{y}^{2}}\,\left[8+9\,(l_{x}^{2}+l_{y}^{2})-2\,(l_{x}^{4}+l_{y}^{4})+6l_{x}^{2}l_{y}^{2}\right]-\right.\nonumber \\
 &  & \left.\frac{8}{3}l_{x}l_{y}l_{z}\arctan\left(\frac{l_{x}l_{y}}{\sqrt{1-l_{x}^{2}-l_{y}^{2}}}\right)+l_{z}\left[\mathcal{H}(l_{x},l_{y})+\mathcal{H}(l_{y},l_{x})\right]\right\} \:,\label{eq:I2xy}\end{eqnarray}
\begin{equation}
\mathcal{H}(u,v)=-\frac{1}{15}\,\sqrt{1-u^{2}}\,\left(8+9\, u^{2}-2\, u^{4}\right)-u\,\arcsin(u)+\frac{1}{3}v\left(3+6\, u^{2}-u^{4}\right)\arcsin\left(\frac{v}{\sqrt{1-u^{2}}}\right)\:.\label{eq:F}\end{equation}
The straight forward generalization of $\mathcal{I}_{2xy}$ and $\mathcal{I}_{3xy}$
in Eqs. (\ref{eq:I3xy0}, \ref{eq:I2xy0}) defines the set of functions
$\{\mathcal{I}_{2xy},\,\mathcal{I}_{2xz},\,\mathcal{I}_{2yz},$ $\,\mathcal{I}_{3xy},\,\mathcal{I}_{3xz},\,\mathcal{I}_{3yz}\}$.
The zone of the phase space with non-null integrand in Eq. (\ref{eq:Q0}),
i.e. the available phase space of the system (APS) breaks or fragments
in two equal unlinked zones because the pair of particles can not
interchange its positions anymore. In this sense we refer to an ergodicity
breaking in the canonical ensemble, which introduce an overall factor
$\xi=1/2$ in the CI, therefore 

\begin{equation}
Z_{2}=\xi*\left(\mathcal{I}_{1}-\mathcal{I}_{2xy}\right)\;.\label{eq:Z05}\end{equation}
Note that $\xi$ was not explicitly written in Eq. (\ref{eq:b00})
and then a $\xi=1$ value was there assumed.

\subsubsection*{Region 6}

In this region \textit{ES exceeds at six faces and only four edges
of EB}. Here, the sphere should exceeds the EB in $\{\hat{x},\hat{y},\hat{z},\hat{xy}\}$
directions but not in $\{\hat{xz},\hat{yz}\}$. Then, the region in
the parameter space is $l_{x}^{2}+l_{y}^{2}\leq1$, $l_{z}\leq1$,
$l_{x}^{2}+l_{z}^{2}\geq1$, and $l_{y}^{2}+l_{z}^{2}\geq1$. For
this region the particles can not interchange its positions, thus,
the APS breaks into two equal and unlinked zones. The ergodicity breaking
introduce the overall factor $\xi=1/2$,\begin{equation}
Z_{2}=\xi*\left(\mathcal{I}_{1}-\mathcal{I}_{2xy}+\mathcal{I}_{3z}\right)\;.\label{eq:Z06}\end{equation}

\subsubsection*{Region 7}

When \textit{ES exceeds at six faces and only eight edges but any
vertex of EB} we have the seventh region. Here, the sphere exceeds
the EB in $\{\hat{x},\hat{y},\hat{z},\hat{xy},\hat{xz}\}$ directions
but not in $\hat{yz}$. The parameter domain is $l_{z}\leq1$, $l_{x}^{2}+l_{y}^{2}\leq1$,
$l_{x}^{2}+l_{z}^{2}\leq1$, and $l_{y}^{2}+l_{z}^{2}\geq1$. Again,
APS breaks but now into four equal and unlinked zones each one characterizing
a set of microestates, which is non-symmetric under some of the symmetries
of the cuboid cavity. This is a spontaneous symmetry breaking phenomena.
The ergodicity breaking produces a factor $\xi=1/4$, and the CI reads\begin{eqnarray}
Z_{2} & = & \xi*\left(\mathcal{I}_{1}-\mathcal{I}_{2}+\mathcal{I}_{3x}+\mathcal{I}_{3y}+\mathcal{I}_{3z}-\mathcal{I}_{3xy}-\mathcal{I}_{3xz}\right)\;,\nonumber \\
 & = & \xi*\left(\mathcal{I}_{1}-\mathcal{I}_{2xy}+\mathcal{I}_{3z}-\mathcal{I}_{3xz}\right)\;.\label{eq:Z07}\end{eqnarray}

\subsubsection*{Region 8}

The last region considered is when \textit{ES exceeds at six faces,
twelve edges but any vertex of EB}. Then, the sphere exceeds the EB
box in $\{\hat{x},\hat{y},\hat{z},\hat{xy},\hat{xz},\hat{yz}\}$ direction
but not in $\hat{xyz}$. Then, $l_{i}^{2}+l_{j}^{2}\leq1$ for $i,j=x,\, y,\, z$
($i\neq j$), and $l_{x}^{2}+l_{y}^{2}+l_{z}^{2}\geq1$. With this
conditions the APS breaks into eight equal and unlinked zones which
also involves a spontaneous symmetry breaking. The factor introduced
by the ergodicity breaking is $\xi=1/8$, while CI is\begin{equation}
Z_{2}=\xi*\left(\mathcal{I}_{1}-\mathcal{I}_{2}+\mathcal{I}_{3x}+\mathcal{I}_{3y}+\mathcal{I}_{3z}-\mathcal{I}_{3xy}-\mathcal{I}_{3xz}-\mathcal{I}_{3yz}\right)\;.\label{eq:Z08}\end{equation}
Finally, in the case that ES exceeds the EB also in $\hat{xyz}$ direction,
the partition function becomes null because both particles do not
fit into the cavity.

\subsection{CI of 2-HS in a cylindrical pore\label{sub:CI-2HS-CYL}}

Let us define the usual length parameters, height and radius, that
characterize an empty cylindrical cavity $L_{h}'$, $R'$. The effective
cavity length parameters are then $L_{h}=L_{h}'-\sigma$, $R=R'-\sigma/2$
and the dimensionless ones are given by $h=L_{h}/\sigma$, $\mathsf{R}=\mathsf{s}^{-1}=2R/\sigma$.
For the cylindrical shaped pore we have $Z_{1}=\pi L_{h}\, R^{2}$.
As it was above mentioned, we need to know the volume defined by the
intersection of two equal and parallel cylinders, $W(\mathbf{r},L_{h},R)$.
It is related to the intersection of two disks of equal radii $R$
and separated by a distance $r$, $W_{disk}(r,R)=2R^{2}\left[\arccos(\mathsf{r})-\mathsf{r}\left(1-\mathsf{r}^{2}\right)^{1/2}\right]$,
where $\mathsf{r}=r/(2R)$ by \begin{equation}
W(\mathbf{r},L_{h},R)=\left|L_{h}-z\right|W_{disk}(r,R)\;.\label{eq:Wcyl}\end{equation}
Note that $W$ is a well defined function of $\mathsf{r}$ only for
the range $0<\mathsf{r}<1$. The EB is a cylinder of double lengths
and the $W$ dependence turns convenient to integrate over $0\leq z\leq L_{h}$,
$0\leq r\leq2R$ and multiply by $4\pi$. The analytic extension of
$W$ for values $z\geq0$ will be considered when it becomes necessary.
We need to analyze the integral considering the parameters $\mathbf{X}=(R,L_{h})$
which define the allowed pore size domain. Defining $\hat{rz}\propto R\,\hat{r}+(L_{h}/2)\,\hat{z}$
we distinguish four regions.

\subsubsection*{Region 1}

The \textit{large pore domain} is defined by the condition that the
ES is completely enclosed into the EB, i.e., that $h\geq1$ and $\mathsf{R}\geq1$.
The CI splits into,\begin{equation}
\mathcal{I}_{1}=4\pi\int_{0}^{2R}r\, dr\,\int_{0}^{L_{h}}\, W\, dz=Z_{1}^{2}\;,\label{eq:Ic1}\end{equation}

\begin{eqnarray}
\mathcal{I}_{2} & = & 4\pi\int_{0}^{\sigma}r\, dr\,\int_{0}^{\sqrt{\sigma^{2}-r^{2}}}\, W\, dz\;,\nonumber \\
 & = & \pi^{2}R^{2}L_{h}\frac{4}{3}\sigma^{3}-\pi^{2}R^{2}\frac{1}{2}\sigma^{4}-\pi^{2}R\, L_{h}\frac{1}{2}\sigma^{4}\left[1-\,_{2}F_{1}(-\frac{1}{2},\frac{1}{2};3;\mathsf{s}^{2})\right]\nonumber \\
 &  & +\frac{\pi}{12}\sigma\,\sqrt{(2R)^{2}-\sigma^{2}}(-6R^{4}+5R^{2}\sigma^{2}+\sigma^{4})\nonumber \\
 &  & +\pi R^{2}(2R^{4}-2R^{2}\sigma^{2}+\sigma^{4})\arcsin(\mathsf{s})\;,\label{eq:Ic2}\end{eqnarray}
where $_{2}F_{1}(-\frac{1}{2},\frac{1}{2};3;a)$ is the Gauss hypergeometric
function which can also be written in terms of complete elliptic integrals
\cite{Abramowitz72,functionswolfram}. The CI is then\begin{equation}
Z_{2}=\mathcal{I}_{1}-\mathcal{I}_{2}\;.\label{eq:Zc1}\end{equation}

\subsubsection*{Region 2}

The \textit{ES exceeds only the bases of EB, domain}. Here, the exclusion
sphere should go beyond the EB only in the $\hat{z}$ direction and
then $h\leq1$, $\mathsf{R}\geq1$. We define the auxiliary integral
$\mathcal{I}_{3z}$, its integration domain is the spherical cup outside
the upper base of the EB\begin{eqnarray}
\mathcal{I}_{3z} & = & 4\pi\int_{0}^{\sqrt{\sigma^{2}-L_{h}^{2}}}r\, dr\,\int_{L_{h}}^{\sqrt{\sigma^{2}-r^{2}}}\, W\, dz\;,\nonumber \\
 & = & R^{6}\pi/45\left\{ -256h\mathsf{s}^{2}(-2+7\mathsf{s}^{2}+3\mathsf{s}^{4})\, E\left(\arcsin\left(\mathsf{s}\,\sqrt{1-h^{2}}\right),\mathsf{s}^{-2}\right)\right.\nonumber \\
 &  & \left.+15\pi\,(3-12(1+h^{2})\mathsf{s}^{2}+64h\mathsf{s}^{4})-256h\mathsf{s}^{3}(1+2s^{2}-3\mathsf{s}^{4})\, F\left(\arccos\left(h\right),\mathsf{s}^{2}\right)\right.\nonumber \\
 &  & \left.+2\mathsf{s}\,\sqrt{(1-h^{2})(1-(1-h^{2})\mathsf{s}^{2})}\left[-45+2(75+41h^{2})\mathsf{s}^{2}-24(-5-4h^{2}+h^{4})\mathsf{s}^{4}\right]\right.\nonumber \\
 &  & \left.+30(-3+12(1+h^{2})\mathsf{s}^{2}+8(-3-6h^{2}+h^{4})\mathsf{s}^{4})\,\arccos\left(\mathsf{s}\,\sqrt{1-h^{2}}\right)\right\} \;,\label{eq:Ic3z}\end{eqnarray}
where $F(a,b)$ and $E(a,b)$ are the incomplete elliptic integrals
of the first and second kind respectively \cite{Abramowitz72}. The
CI is

\begin{equation}
Z_{2}=\mathcal{I}_{1}-\mathcal{I}_{2}+\mathcal{I}_{3z}\;.\label{eq:Zc2}\end{equation}

\subsubsection*{Region 3}

In this case \textit{ES exceeds only the curved lateral face of EB}.
The exclusion sphere should go beyond the cylindrical EB only in the
$\hat{r}$ direction being $h\geq1$ and $\mathsf{R}\leq1$. With
these conditions particles can not interchange its positions producing
that APS breaks into two equal and unlinked zones. We introduce the
auxiliary integral $\mathcal{I}_{2r}$ is given by \begin{eqnarray}
\mathcal{I}_{2r} & = & 4\pi\int_{0}^{2R}r\, dr\,\int_{0}^{\sqrt{\sigma^{2}-r^{2}}}\, W\, dz\;,\nonumber \\
 & = & \pi^{2}\,(R^{6}-R^{4}\sigma^{2}+\frac{4}{3}R^{2}L_{h}\sigma^{3})+\frac{4\pi}{45}L_{h}\sigma\,\times\nonumber \\
 &  & \left[(32R^{4}-28R^{2}\sigma^{2}-3\sigma^{4})\, E(\mathsf{s}^{-2})+(-16R^{4}-8R^{2}\sigma^{2}+3\sigma^{4})\, K(\mathsf{s}^{-2}))\right]\;,\label{eq:Ic2r}\end{eqnarray}
where $K(a)$ and $E(a)$ are the complete elliptic integrals of the
first and second kind, respectively. We may also formally define $\mathcal{I}_{3r}=\mathcal{I}_{2}-\mathcal{I}_{2r}$.
The ergodicity breaking produces a $\xi=1/2$ factor, being $Z_{2}$\begin{equation}
Z_{2}=\xi*\left(\mathcal{I}_{1}-\mathcal{I}_{2r}\right)=\xi*\left(\mathcal{I}_{1}-\mathcal{I}_{2}+\mathcal{I}_{3r}\right)\;.\label{eq:Zc3}\end{equation}

\subsubsection*{Region 4}

This region appears when \textit{ES exceeds both the bases and the
curved lateral face but not the edges of EB}. Therefore, the exclusion
sphere exceeds EB in the $\{\hat{r},\hat{z}\}$ directions, but not
in $\hat{rz}$. In consequence $h\leq1$ and $\mathsf{R}\leq1$, but
$h^{2}+\mathsf{R}^{2}\geq1$. As happens in Region 3, here the APS
breaks into two equal and unlinked zones due to the ergodicity breaking,
being $\xi=1/2$ and\begin{equation}
Z_{2}=\xi*\left(\mathcal{I}_{1}-\mathcal{I}_{2}+\mathcal{I}_{3z}+\mathcal{I}_{3r}\right)\;.\label{eq:Zc4}\end{equation}
Finally, if ES exceeds also in $\hat{rz}$ direction, the partition
function becomes null because both particles can not fit into the
pore.

\subsection{CI of 2-HS in a spheroidal pore\label{sub:CI-2HS-ELL}}

The last CI that we evaluate in PW corresponds to the ellipsoidal
pore. We restrict the study to cavities where only two principal radii
are independent, i.e. to the revolution ellipsoids also called spheroids.
Therefore, two distinct shapes the prolate and the oblate ones will
be analyzed. Let us consider an effective cavity with spheroidal shape.
The effective length parameters are the principal radii $R$ and $R_{c}$,
where $R_{c}$ is the different radius. Dimensionless parameters are
$\mathsf{R}=\mathsf{s}^{-1}=2R/\sigma$, $\mathsf{C}=2R_{c}/\sigma$,
and $\lambda=R_{c}/R$. For $\lambda<1$ we deal with the oblate,
while for $\lambda>1$ we deal with the prolate, spheroids. The configuration
integral for one particle is $Z_{1}=\frac{4\pi}{3}\, R^{2}R_{c}$.
The volume of intersection of two equally oriented spheroids, $W(r,z)$,
is related with the volume of the intersection of two spheres. In
terms of $W_{sphere}(\varrho,R)=\frac{4\pi}{3}\, R^{3}(1-\frac{3}{2}\mathsf{r}+\frac{1}{2}\mathsf{r}^{3})$
with $\varrho$ the spherical radial coordinate, and $\mathsf{r}=\varrho/(2R)$
we obtain\begin{equation}
W(r,z)=\lambda W_{sphere}(\sqrt{r^{2}+(z/\lambda)^{2}},R)\;.\label{eq:Wsphrd}\end{equation}
Function $W$ is well defined in the domain $0\leq r^{2}+(z/\lambda)^{2}\leq4R^{2}$.
The EB is a spheroid with double length radii. For this pore geometry
none analytic extension is suitable and the $W$ dependence turns
convenient to integrate over $0\leq z\leq2R_{c}$, $0\leq r\leq2R$,
and multiply by $4\pi$. We need to analyze the integral considering
the allowed values of parameters $\mathbf{X}=(R,R_{c})$ that define
the pore size domain. We may distinguish three regions.

\subsubsection*{Region 1}

In the \textit{large pore domain} ES is completely enclosed into EB,
i.e. $\mathsf{C}\geq1$ and $\mathsf{R}\geq1$. As it was above described,
the integral splits into,\begin{equation}
\mathcal{I}_{1}=4\pi\int_{0}^{2R_{c}}dz\,\int_{0}^{\sqrt{(2R)^{2}-z^{2}\lambda^{-2}}}\, W\, r\, dr=Z_{1}^{2}\;,\label{eq:Ie1}\end{equation}
\begin{eqnarray}
\mathcal{I}_{2} & = & 4\pi\int_{0}^{\sigma}dz\,\int_{0}^{\sqrt{\sigma^{2}-r^{2}}}\, W\, r\, dr\;,\nonumber \\
 & = & \frac{16}{9}\pi^{2}R^{3}\lambda\sigma^{3}-\frac{\pi^{2}\sigma^{4}}{2}\left(R^{2}-\frac{\sigma^{2}}{24}\right)\left(1+\frac{\lambda^{2}}{\sqrt{-1+\lambda^{2}}}\mathrm{arcsec}(\lambda)\right)+\frac{\pi^{2}}{72}\frac{\sigma^{6}}{\lambda^{2}}\;,\label{eq:Ie2}\end{eqnarray}
which concerns to both, oblated and prolated spheroidal cavities.
For $\lambda<1$, $(-1+\lambda^{2})^{1/2}\mathrm{arcsec}(\lambda)$
transforms to $(1-\lambda^{2})^{1/2}\mathrm{arcsech}(\lambda)$. Although,
for $\lambda\rightarrow1$ we obtain $(-1+\lambda^{2})^{1/2}\mathrm{arcsec}(\lambda)\rightarrow1$
which is consistent with the spherical pore result \cite{Urrutia_2008}.
In terms of $\mathcal{I}_{1}$ and $\mathcal{I}_{2}$ we find

\begin{equation}
Z_{2}=\mathcal{I}_{1}-\mathcal{I}_{2}\;.\label{eq:Ze1}\end{equation}
Next regions concern the situation where \textit{ES exceeds EB}. Under
such condition it becomes necessary to make a separate analysis of
prolate and oblate, ellipsoids.

\subsubsection*{Region 2 (oblate)}

Here we consider an oblate ellipsoid, $\lambda<1$. In this region
\textit{ES exceeds on top and down directions of EB}, i.e. in the
direction of the principal axis $\hat{z}$, but not in $\hat{r}$.
Therefore, we consider $\mathsf{C}\leq1$, $\mathsf{R}>1$. We find
that in this region it is simpler to deal directly with $Z_{2}$,
we obtain\begin{eqnarray}
Z_{2} & = & 4\pi\int_{0}^{z_{max}}dz\,\int_{\sqrt{\sigma^{2}-z^{2}}}^{\sqrt{(2R)^{2}-z^{2}\lambda^{-2}}}\, W\, r\, dr\;,\nonumber \\
 & = & \frac{\sigma^{6}\lambda^{2}\pi^{2}}{144\,\sqrt{1-\lambda^{2}}}\left[-\mathsf{R}\,\sqrt{\mathsf{R}^{2}-1}\left(3+16\mathsf{R}^{2}-4\mathsf{R}^{4}\right)-3\left(6\mathsf{R}^{2}-1\right)\mathrm{arcsech}(\mathsf{R}^{2})\right]\;,\label{eq:Ze2a}\end{eqnarray}
with $z_{max}=\lambda\sqrt{\frac{(2R)^{2}-\sigma^{2}}{1-\lambda^{2}}}$.
We may also, formally define $\mathcal{I}_{2z}=\mathcal{I}_{1}-Z_{2}$
and $\mathcal{I}_{3z}=\mathcal{I}_{2}-\mathcal{I}_{2z}$. In the case
that ES also exceeds EB in the $\hat{r}$ direction the partition
function becomes null because both particles can not fit into the
pore.

\subsubsection*{Region 3 (prolate)}

Here we restrict to a prolate ellipsoid, $\lambda>1$. In this region
\textit{ES exceeds in the lateral direction the surface of EB}. Then,
ES goes beyond EB only in $\hat{r}$ direction, but not in $\hat{z}$,
i.e. $\mathsf{C}>1$, $\mathsf{R}\leq1$. Under these conditions APS
breaks into two equal and unlinked zones and the ergodicity breaking
produce the factor $\xi=1/2$. Again, in this region is preferable
to deal directly with $Z_{2}$\begin{equation}
Z_{2}=\xi*4\pi\int_{0}^{r_{max}}r\, dr\,\int_{\sqrt{\sigma^{2}-r^{2}}}^{\lambda\,\sqrt{(2R)^{2}-r^{2}}}\, W\, dz\;,\label{eq:Ze2b}\end{equation}
where $r_{max}=\sqrt{\sigma^{2}-(\lambda2R)^{2}}/\sqrt{\lambda^{2}-1}$.
This integral was solved by splitting it in several parts, after some
work we obtain\begin{eqnarray}
Z_{2}\xi^{-1} & = & \mathcal{I}_{1}-\frac{16}{9}\pi^{2}R^{3}\lambda\sigma^{3}+\frac{\pi^{2}\sigma^{4}}{2}R^{2}-\frac{\pi^{2}\sigma^{6}}{24}\left(\frac{1}{2}+\frac{1}{3\lambda^{2}}\right)\nonumber \\
 &  & +\frac{\pi^{2}\sigma^{6}\lambda^{2}}{144\sqrt{-1+\lambda^{2}}}\,\mathsf{R}\,\sqrt{1-\mathsf{R}^{2}}\left(3+16\mathsf{R}^{2}-4\mathsf{R}^{4}\right)\nonumber \\
 &  & +\frac{3\sigma^{6}\lambda^{2}\pi^{2}}{144\,\sqrt{-1+\lambda^{2}}}\left(6\mathsf{R}^{2}-1\right)\left[\mathrm{arcsec}(\lambda)-\arccos(\mathsf{R})\right]\:,\label{eq:Ze2c}\end{eqnarray}
Formally, we can define $\mathcal{I}_{2r}=\mathcal{I}_{1}-\xi^{-1}Z_{2}$
and $\mathcal{I}_{3r}$=$\mathcal{I}_{2}$-$\mathcal{I}_{2r}$. In
addition, we note that if the exclusion sphere exceeds EB all around,
particles do not fit into the cavity and then the CI becomes null.
The Eq. (\ref{eq:Ze2c}) is the last result about analytic expressions
for the CI of 2-HS system confined in the studied cuboid, cylindrical
and spheroidal, cavities.

In PW we deal with a pair of distinguishable particles. Even sow,
we make a brief discussion about the CI of a system of two indistinguishable
HS (2i-HS). The canonical partition function of 2i-HS confined in
a cuboidal, cylindrical, spheroidal and other shaped cavities are
easily obtained from the CI of a 2-distinguishable-HS with the introduction
of minor modifications. The first obvious change comes in the partition
function definition because we must introduce the correct Boltzmann
factor then $Q_{2,ind}=\frac{1}{2}\Lambda^{-6}Z_{2,ind}$. Secondly,
we must analize the difference between $Z_{2}$ and $Z_{2,ind}$.
In principle, expression (\ref{eq:Q0}) gives the starting point to
define both, $Z_{2}$ and $Z_{2,ind}$. However, the evaluation of
$Z_{2}$ for the studied cavities involves the factor $\xi$ that
modifies Eq. (\ref{eq:Q0}) in some regions. We recognize that $Z_{2,ind}=Z_{2}$
in regions where no extra factor appears. A detailed inspection of
the origin of $\xi$ also shows other different situations. In some
regions the ergodicity breaking appears because particles can not
interchange their positions, but this makes non sense for indistinguishable
particles. Therefore, regions where $\xi=1/2$ corresponds to $\xi_{ind}=1$.
In other regions the ergodicity breaking also involves the spontaneous
symmetry breaking, in this regions we find $\xi_{ind}=2\xi$. In summary
\[
Z_{2,ind}=\xi_{ind}\, Z_{2}(\xi=1)\:,\textrm{ with }\xi_{ind}=2\xi\textrm{ if }\xi\neq1\:,\]
and $\xi_{ind}=2$ if $\xi=1$. Remarkably, partition function relates
simply by $Q_{2,ind}=Q_{2}$ if $\xi\neq1$, and $Q_{2,ind}=\frac{1}{2}Q_{2}$
if $\xi=1$.

\section{Local properties: Density distribution and Pressure\label{sec:Distribution-function}}

In principle, the partition function of the system holds in its global
Statistical Mechanic properties. Such properties are presumably obtainable
from some derivatives of $Q_{2}$. This makes interesting the study
of the analytical properties of $Z_{2}$ which is done in Sec. \ref{sec:Analysis}.
Now, we are also interested in the local properties of the 2-HS confined
system. Therefore, we study two functions, the one body density distribution
$\rho(\mathbf{r})$ and the pressure tensor $\mathbf{P}(\mathbf{r})$.
We begin with a general brief description of the properties of $\rho(\mathbf{r})$.
For any pore shape, $\rho(\mathbf{r})$ is \cite[p.180]{Hill56}\begin{eqnarray}
\rho(\mathbf{r}) & = & 2Z_{2}^{-1}e_{1}(\mathbf{r})\,\int\, e_{2}(\mathbf{r}_{2})\, e_{12}(\left|\mathbf{r}-\mathbf{r}_{2}\right|)\, d\mathbf{r}_{2}\:,\nonumber \\
 & = & 2Z_{2}^{-1}\, e_{1}(\mathbf{r})\,\left(Z_{1}-\mathcal{J}_{2}(\mathbf{r})\right)\:,\label{eq:rho1}\end{eqnarray}
\begin{equation}
\mathcal{J}_{2}(\mathbf{r})=-\int\, e_{2}(\mathbf{r}_{2})\, f_{12}(\left|\mathbf{r}-\mathbf{r}_{2}\right|)\, d\mathbf{r}_{2}\:,\label{eq:J2def}\end{equation}
where $\mathcal{J}_{2}(\mathbf{r})$ is the overlap volume between
the cavity and the ES (with $\sigma$ radius) at position $\mathbf{r}$.
This ES is produced by one HS-particle located there. The complete
integral is $\int\rho(\mathbf{r})\, d\mathbf{r}=2$. For an arbitrary
$\mathbf{r}$, $\mathcal{J}_{2}(\mathbf{r})$ is positive and continuous
but non-analytical and may be piecewise defined. When the particle
is placed sufficiently deep inside the cavity all the ES is inner
to the boundary. Therefore, for $\mathbf{r}$ such that the shortest
distance to the boundary is greater than $\sigma$, $\mathcal{J}_{2}(\mathbf{r})$
reaches its maximum value $\mathcal{J}_{2}(\mathbf{r})=2b_{2}$. This
means that for big enough cavities of any shape a plateau of constant
density \begin{equation}
\rho_{0}=2Z_{2}^{-1}\left(Z_{1}-2b_{2}\right)\:,\label{eq:rho0}\end{equation}
develops at a distance to the boundary grater than $\sigma$. When
$\mathbf{r}$ becomes nearer to the boundary the function $\mathcal{J}_{2}(\mathbf{r})$
decreases and $\rho(\mathbf{r})$ increases. For $\mathbf{r}$ outside
of the cavity $\rho(\mathbf{r})=0$, but we can define its continuous
extension $y(\mathbf{r})$ by dropping out the $e_{1}(\mathbf{r})$
term in Eqs. (\ref{eq:rho1}). Outside of the cavity, for distances
to the boundary greater than $\sigma$, $y(\mathbf{r})$ becomes constant
because $\mathcal{J}_{2}(\mathbf{r})=0$. The $\mathcal{J}_{2}(\mathbf{r})$
for the cuboid and cylindrical pores may be expressed by combining
the $2b_{2}$ constant and the geometrical functions $\{\mathcal{J}_{2a}(\mathbf{r}\cdot\hat{\mathbf{a}}),\,\mathcal{J}_{2ab}(\mathbf{r}\cdot\hat{\mathbf{a}},\mathbf{r}\cdot\hat{\mathbf{b}}),\,\mathcal{J}_{2abc}(\mathbf{r}\cdot\hat{\mathbf{a}},\mathbf{r}\cdot\hat{\mathbf{b}},\mathbf{r}\cdot\hat{\mathbf{c}})\}$,
where $\{a,b,c\}$ represent characteristic directions normal to the
cavity boundary with inward normal versors $\{\hat{\mathbf{a}},\hat{\mathbf{b}},\hat{\mathbf{c}}\}$.
The function $\mathcal{J}_{2a}$ is the inner overlap volume defined
by the ES and one boundary surface that intersects it, $\mathcal{J}_{2ab}$
is the inner overlap volume defined by the sphere and two intersecting
boundary surfaces, and $\mathcal{J}_{2abc}$ involves three mutually
intersecting boundary surfaces. The short cut $\mathbf{r}\cdot\hat{\mathbf{a}}$
and similar are the (minimum) distance between the ES center and a
face of the boundary with normal inward versor $\hat{\mathbf{a}}$.
Although, $\mathbf{r}\cdot\hat{\mathbf{a}}$ extends to negative values
when $\mathbf{r}$ is outside of the cavity. We may mention that inner
overlap volume clearly identify a unique volume and then this description
is non-ambiguous. When position $\mathbf{r}$ is on a cavity surface
with simple curvature and away from other surfaces (a distance greater
than $\sigma$) $\mathcal{J}_{2}(\mathbf{r})=\mathcal{J}_{2a}(\mathbf{r}\cdot\hat{\mathbf{a}}=0)$
which reduces to simple expressions. In such conditions, we have $\mathcal{J}_{2}(0)=b_{2}$
for the planar surface, $\mathcal{J}_{2,sph}(0)=b_{2}(1-\frac{3}{4}\mathsf{s})$
for a concave spherical surface and $\mathcal{J}_{2,sph}(0)=b_{2}(1+\frac{3}{4}\mathsf{s})$
for the convex one \cite{Urrutia_2008,Urrutia_2010}. For $\mathbf{r}$
on the lateral curved surface of a cylinder the analytic expression
involving elliptic integrals is known \cite{Lamarche_1990}. Its power
series is, $\mathcal{J}_{2,cyl}(0)=b_{2}(1-\frac{3}{8}\mathsf{s}-\frac{1}{32}\mathsf{s}^{3})+O(\mathsf{s}^{5})$
and $\mathcal{J}_{2,cyl}(0)=b_{2}(1+\frac{3}{8}\mathsf{s}+\frac{1}{32}\mathsf{s}^{3})+O(\mathsf{s}^{5})$
for the concave and convex cases, respectively. The question becomes
even worse for the spheroidal pore surface, where we found analytic
expressions of $\mathcal{J}_{2,sphd}(0)$ only for points on the poles
and on the equatorial line.

\subsection{Density distribution in the cuboidal pore\label{sub:Density-cube}}

For the cuboid cavity the boundary surfaces are orthogonally intersecting
planes. Therefore, in cuboidal cavities, $\mathcal{J}_{2a}$ is the
inner overlap volume defined by the ES and a plane that intersects
it, $\mathcal{J}_{2ab}$ is the volume defined by the sphere and a
right angle dihedron that intersects it, and $\mathcal{J}_{2abc}$
is the volume defined by the sphere and a right angle vertex. We must
include a brief digression about the volume of intersection of a unit
sphere and a set of mutually intersecting planes. As we are primarily
interested in the cuboid we restrict ourselves to sets of mutually
orthogonal planes with at most three planes. We introduce the function
$\mathcal{K}_{a}(\mathbf{r}\cdot\hat{\mathbf{a}})$ which measures
the volume of the \textit{spherical segment or spherical cap}, defined
by the intersection of the unit sphere at position $\mathbf{r}$ and
a half-space with inward normal $\hat{\mathbf{a}}$. The vector $\mathbf{r}$
goes from a point in the plane to the sphere center. For $\hat{\mathbf{a}}=\hat{\mathbf{x}}$
we have $\mathbf{r}\cdot\hat{\mathbf{a}}=x$ with $x>0$ if the center
of the sphere is in the positive half-space. For $-1\leq x\leq0$,\begin{eqnarray}
\mathcal{K}_{x}(x) & = & 4\int_{-x}^{1}dx'\,\int_{0}^{\sqrt{1-x'^{2}}}dy'\,\int_{0}^{\sqrt{1-y'^{2}-x'^{2}}}\, dz'\;,\label{eq:K1}\\
 & = & \frac{\pi}{3}(1+x)^{2}(2-x)\;,\label{eq:K1-1}\end{eqnarray}
but Eq. (\ref{eq:K1-1}) is also valid in the extended domain $-1\leq x\leq1$.
Naturally, \begin{equation}
4\pi/3=\mathcal{K}_{a}(\mathbf{r}\cdot\hat{\mathbf{a}})+\mathcal{K}_{a}(-\mathbf{r}\cdot\hat{\mathbf{a}})\:,\label{eq:K1-prop1}\end{equation}
\begin{equation}
\mathcal{K}_{a}(-\mathbf{r}\cdot\hat{\mathbf{a}})=\mathcal{K}_{\bar{a}}(-\mathbf{r}\cdot\hat{\mathbf{a}})\:,\label{eq:K1-prop2}\end{equation}
where the label $\bar{a}=-a$ corresponds to the inward direction
$-\hat{\mathbf{a}}$. The function $\mathcal{K}_{ab}(\mathbf{r}\cdot\hat{\mathbf{a}},\mathbf{r}\cdot\hat{\mathbf{b}})$
is the volume between the sphere and a right angle wedge when the
edge cross the sphere. The wedge is defined by the quadrant determined
by the intersection of half-spaces with inward directions $\hat{\mathbf{a}}$
and $\hat{\mathbf{b}}$. As far as the center of the sphere does not
lie on the edge this spherical wedge is different to the usual one.
For $\hat{\mathbf{a}}=\hat{\mathbf{x}}$ and $\hat{\mathbf{b}}=\hat{\mathbf{y}}$
we obtain \begin{eqnarray}
\mathcal{K}_{xy}(x,y) & = & 2\int_{x}^{\sqrt{1-y^{2}}}dx'\,\int_{y}^{\sqrt{1-x'^{2}}}dy'\,\int_{0}^{\sqrt{1-y'^{2}-x'^{2}}}\, dz'\;,\label{eq:K2}\\
 & = & \frac{1}{3}\left[\pi+2\, x\, y\,\sqrt{1-x^{2}-y^{2}}-2\,\arctan(\frac{x\, y}{\sqrt{1-x^{2}-y^{2}}})\right.\nonumber \\
 &  & \left.+x\,(3-x^{2})\,\arccos(\frac{-y}{\sqrt{1-x^{2}}})+y\,(3-y^{2})\,\arccos(\frac{-x}{\sqrt{1-y^{2}}})\right]\;,\label{eq:K2-1}\end{eqnarray}
where Eq. (\ref{eq:K2}) applies for $-1\leq x,y\leq0$, $x^{2}+y^{2}<1$
but Eq. (\ref{eq:K2-1}) is valid in the extended domain $-1\leq x,y\leq1$,
$x^{2}+y^{2}<1$. The half-length of the portion of the wedges edge
inside the sphere is $1-x^{2}-y^{2}$. The function $\mathcal{K}_{ab}(\mathbf{r}\cdot\hat{\mathbf{a}},\mathbf{r}\cdot\hat{\mathbf{b}})$
has the following properties

\begin{eqnarray}
\mathcal{K}_{a}(\mathbf{r}\cdot\hat{\mathbf{a}}) & = & \mathcal{K}_{ab}(\mathbf{r}\cdot\hat{\mathbf{a}},\mathbf{r}\cdot\hat{\mathbf{b}})+\mathcal{K}_{ab}(\mathbf{r}\cdot\hat{\mathbf{a}},-\mathbf{r}\cdot\hat{\mathbf{b}})\:,\nonumber \\
\mathcal{K}_{b}(\mathbf{r}\cdot\hat{\mathbf{b}}) & = & \mathcal{K}_{ab}(\mathbf{r}\cdot\hat{\mathbf{a}},\mathbf{r}\cdot\hat{\mathbf{b}})+\mathcal{K}_{ab}(-\mathbf{r}\cdot\hat{\mathbf{a}},\mathbf{r}\cdot\hat{\mathbf{b}})\:,\label{eq:K2-prop1}\end{eqnarray}
\begin{equation}
4\pi/3=\mathcal{K}_{ab}(\mathbf{r}\cdot\hat{\mathbf{a}},\mathbf{r}\cdot\hat{\mathbf{b}})+\mathcal{K}_{ab}(-\mathbf{r}\cdot\hat{\mathbf{a}},\mathbf{r}\cdot\hat{\mathbf{b}})+\mathcal{K}_{ab}(\mathbf{r}\cdot\hat{\mathbf{a}},-\mathbf{r}\cdot\hat{\mathbf{b}})+\mathcal{K}_{ab}(-\mathbf{r}\cdot\hat{\mathbf{a}},-\mathbf{r}\cdot\hat{\mathbf{b}})\:,\label{eq:K2-prop2}\end{equation}
\begin{eqnarray}
\mathcal{K}_{ab}(\mathbf{r}\cdot\hat{\mathbf{a}},-\mathbf{r}\cdot\hat{\mathbf{b}}) & = & \mathcal{K}_{a\bar{b}}(\mathbf{r}\cdot\hat{\mathbf{a}},-\mathbf{r}\cdot\hat{\mathbf{b}})\:,\nonumber \\
\mathcal{K}_{ab}(-\mathbf{r}\cdot\hat{\mathbf{a}},\mathbf{r}\cdot\hat{\mathbf{b}}) & = & \mathcal{K}_{\bar{a}b}(-\mathbf{r}\cdot\hat{\mathbf{a}},\mathbf{r}\cdot\hat{\mathbf{b}})\:,\nonumber \\
\mathcal{K}_{ab}(-\mathbf{r}\cdot\hat{\mathbf{a}},-\mathbf{r}\cdot\hat{\mathbf{b}}) & = & \mathcal{K}_{\bar{a}\bar{b}}(-\mathbf{r}\cdot\hat{\mathbf{a}},-\mathbf{r}\cdot\hat{\mathbf{b}})\:,\label{eq:K2-prop3}\end{eqnarray}
being the Eq. (\ref{eq:K2-prop2}) a consequence of Eqs. (\ref{eq:K1-prop1},
\ref{eq:K2-prop1}). The last $\mathcal{K}$ function in which we
are interested is $\mathcal{K}_{abc}(\mathbf{r}\cdot\hat{\mathbf{a}},\mathbf{r}\cdot\hat{\mathbf{b}},\mathbf{r}\cdot\hat{\mathbf{c}})$,
the volume defined by the sphere and a right angle vertex inner to
the sphere,\begin{eqnarray}
\mathcal{K}_{xyz}(x,y,z) & = & \int_{x}^{\sqrt{1-y'^{2}-z'^{2}}}dx'\,\int_{y}^{\sqrt{1-x'^{2}-z^{2}}}dy'\,\int_{z}^{\sqrt{1-x'^{2}-y'^{2}}}\, dz'\;,\label{eq:K3}\\
 & = & \frac{\pi}{6}+x\, y\, z-\frac{1}{4}\left[\mathcal{K}_{x}(x)+\mathcal{K}_{y}(y)+\mathcal{K}_{z}(z)\right]\;,\nonumber \\
 &  & +\frac{1}{2}\left[\mathcal{K}_{xy}(x,y)+\mathcal{K}_{xz}(x,z)+\mathcal{K}_{yz}(y,z)\right]\;.\label{eq:K3-1}\end{eqnarray}
As happened before, Eq. (\ref{eq:K3}) apply for $-1\leq x,y,z\leq0$,
$x^{2}+y^{2}+z^{2}<1$ but Eq. (\ref{eq:K3-1}) is valid in the extended
domain $-1\leq x,y,z\leq1$, $x^{2}+y^{2}+z^{2}<1$. Interestingly,
we were unable to perform the direct integration expressed in Eq.
(\ref{eq:K3}), although it was evaluated making a geometrical decomposition
into simple terms. We obtain the following properties for $\mathcal{K}_{abc}(\mathbf{r}\cdot\hat{\mathbf{a}},\mathbf{r}\cdot\hat{\mathbf{b}},\mathbf{r}\cdot\hat{\mathbf{c}})$\begin{eqnarray}
\mathcal{K}_{ab}(\mathbf{r}\cdot\hat{\mathbf{a}},\mathbf{r}\cdot\hat{\mathbf{b}}) & = & \mathcal{K}_{abc}(\mathbf{r}\cdot\hat{\mathbf{a}},\mathbf{r}\cdot\hat{\mathbf{b}},\mathbf{r}\cdot\hat{\mathbf{c}})+\mathcal{K}_{abc}(\mathbf{r}\cdot\hat{\mathbf{a}},\mathbf{r}\cdot\hat{\mathbf{b}},-\mathbf{r}\cdot\hat{\mathbf{c}})\:,\nonumber \\
\mathcal{K}_{ac}(\mathbf{r}\cdot\hat{\mathbf{a}},\mathbf{r}\cdot\hat{\mathbf{c}}) & = & \mathcal{K}_{abc}(\mathbf{r}\cdot\hat{\mathbf{a}},\mathbf{r}\cdot\hat{\mathbf{b}},\mathbf{r}\cdot\hat{\mathbf{c}})+\mathcal{K}_{abc}(\mathbf{r}\cdot\hat{\mathbf{a}},-\mathbf{r}\cdot\hat{\mathbf{b}},\mathbf{r}\cdot\hat{\mathbf{c}})\:,\nonumber \\
\mathcal{K}_{bc}(\mathbf{r}\cdot\hat{\mathbf{b}},\mathbf{r}\cdot\hat{\mathbf{c}}) & = & \mathcal{K}_{abc}(\mathbf{r}\cdot\hat{\mathbf{a}},\mathbf{r}\cdot\hat{\mathbf{b}},\mathbf{r}\cdot\hat{\mathbf{c}})+\mathcal{K}_{abc}(-\mathbf{r}\cdot\hat{\mathbf{a}},\mathbf{r}\cdot\hat{\mathbf{b}},\mathbf{r}\cdot\hat{\mathbf{c}})\:,\label{eq:K3-prop1}\end{eqnarray}
\begin{eqnarray}
4\pi/3 & = & \mathcal{K}_{abc}(\mathbf{r}\cdot\hat{\mathbf{a}},\mathbf{r}\cdot\hat{\mathbf{b}},\mathbf{r}\cdot\hat{\mathbf{c}})+\mathcal{K}_{abc}(-\mathbf{r}\cdot\hat{\mathbf{a}},-\mathbf{r}\cdot\hat{\mathbf{b}},-\mathbf{r}\cdot\hat{\mathbf{c}})+\mathcal{K}_{abc}(-\mathbf{r}\cdot\hat{\mathbf{a}},-\mathbf{r}\cdot\hat{\mathbf{b}},\mathbf{r}\cdot\hat{\mathbf{c}})\:,\nonumber \\
 &  & +\mathcal{K}_{abc}(-\mathbf{r}\cdot\hat{\mathbf{a}},\mathbf{r}\cdot\hat{\mathbf{b}},-\mathbf{r}\cdot\hat{\mathbf{c}})+\mathcal{K}_{abc}(\mathbf{r}\cdot\hat{\mathbf{a}},-\mathbf{r}\cdot\hat{\mathbf{b}},-\mathbf{r}\cdot\hat{\mathbf{c}})\:,\nonumber \\
 &  & +\mathcal{K}_{abc}(-\mathbf{r}\cdot\hat{\mathbf{a}},\mathbf{r}\cdot\hat{\mathbf{b}},\mathbf{r}\cdot\hat{\mathbf{c}})+\mathcal{K}_{abc}(\mathbf{r}\cdot\hat{\mathbf{a}},-\mathbf{r}\cdot\hat{\mathbf{b}},\mathbf{r}\cdot\hat{\mathbf{c}})+\mathcal{K}_{abc}(\mathbf{r}\cdot\hat{\mathbf{a}},\mathbf{r}\cdot\hat{\mathbf{b}},-\mathbf{r}\cdot\hat{\mathbf{c}})\:.\label{eq:K3-prop2}\end{eqnarray}
\begin{eqnarray}
\mathcal{K}_{abc}(\mathbf{r}\cdot\hat{\mathbf{a}},\mathbf{r}\cdot\hat{\mathbf{b}},-\mathbf{r}\cdot\hat{\mathbf{c}}) & = & \mathcal{K}_{ab\bar{c}}(\mathbf{r}\cdot\hat{\mathbf{a}},\mathbf{r}\cdot\hat{\mathbf{b}},-\mathbf{r}\cdot\hat{\mathbf{c}})\:,\nonumber \\
\mathcal{K}_{abc}(\mathbf{r}\cdot\hat{\mathbf{a}},-\mathbf{r}\cdot\hat{\mathbf{b}},-\mathbf{r}\cdot\hat{\mathbf{c}}) & = & \mathcal{K}_{a\bar{b}\bar{c}}(\mathbf{r}\cdot\hat{\mathbf{a}},-\mathbf{r}\cdot\hat{\mathbf{b}},-\mathbf{r}\cdot\hat{\mathbf{c}})\:,\nonumber \\
\mathcal{K}_{abc}(-\mathbf{r}\cdot\hat{\mathbf{a}},-\mathbf{r}\cdot\hat{\mathbf{b}},-\mathbf{r}\cdot\hat{\mathbf{c}}) & = & \mathcal{K}_{\bar{a}\bar{b}\bar{c}}(-\mathbf{r}\cdot\hat{\mathbf{a}},-\mathbf{r}\cdot\hat{\mathbf{b}},-\mathbf{r}\cdot\hat{\mathbf{c}})\:,\label{eq:K3-prop3}\end{eqnarray}
where other identities similar to Eq. (\ref{eq:K3-prop3}) may be
obtained by symmetry considerations. To the best of our knowledge
the basic geometrical functions $\mathcal{K}_{ab}$ and $\mathcal{K}_{abc}$
are new results never published before.

Using the $\mathcal{K}$ functions we can complete the picture of
$\rho(\mathbf{r})$ for the cuboid cavity being that functions $\mathcal{J}_{2}$
and $\mathcal{K}$ are related by \begin{eqnarray}
\mathcal{J}_{2a}(\mathbf{r}\cdot\hat{\mathbf{a}}) & = & \sigma^{3}\mathcal{K}_{a}(\tilde{\mathbf{r}}\cdot\hat{\mathbf{a}})\:,\nonumber \\
\mathcal{J}_{2ab}(\mathbf{r}\cdot\hat{\mathbf{a}},\mathbf{r}\cdot\hat{\mathbf{b}}) & = & \sigma^{3}\mathcal{K}_{ab}(\tilde{\mathbf{r}}\cdot\hat{\mathbf{a}},\tilde{\mathbf{r}}\cdot\hat{\mathbf{b}})\:,\nonumber \\
\mathcal{J}_{2abc}(\mathbf{r}\cdot\hat{\mathbf{a}},\mathbf{r}\cdot\hat{\mathbf{b}},\mathbf{r}\cdot\hat{\mathbf{c}}) & = & \sigma^{3}\mathcal{K}_{abc}(\tilde{\mathbf{r}}\cdot\hat{\mathbf{a}},\tilde{\mathbf{r}}\cdot\hat{\mathbf{b}},\tilde{\mathbf{r}}\cdot\hat{\mathbf{c}})\:.\label{eq:JK}\end{eqnarray}
with $\tilde{\mathbf{r}}=\mathbf{r}/\sigma$. We take the three orthogonal
planes at $x=0$, $y=0$, and $z=0$ with inward directions $\hat{\mathbf{x}}$,
$\hat{\mathbf{y}}$, and $\hat{\mathbf{z}}$, respectively. Thus,
$\{x,\, y,\, z\}$ represent the perpendicular distances to this set
of planes. We assume a cuboidal pore such that $L_{i}>2\sigma$
(Region 1) and $0\leq x\leq y\leq z\leq1$, therefore\begin{eqnarray}
\dfrac{\mathcal{J}_{2}(\mathbf{r})}{\sigma^{3}}= & \begin{cases}
2b_{2}\,, & x,y,z\geq1\,,\\
\mathcal{K}_{x}(x)\,, & x<1,\, y\geq1,\, z\geq1\,,\\
2b_{2}-\mathcal{K}_{x}(-x)-\mathcal{K}_{y}(-y)\,, & x,\, y<1,\, x^{2}+y^{2},\, z\geq1\,,\\
2b_{2}-\mathcal{K}_{x}(-x)-\mathcal{K}_{y}(-y)-\mathcal{K}_{z}(-z)\,, & x,\, y,\, z<1,\, x^{2}+y^{2},\, x^{2}+z^{2},\, y^{2}+z^{2}\geq1\,,\\
\mathcal{K}_{xy}(x,y)\,, & x^{2}+y^{2}<1,\, z\geq1\,,\\
\mathcal{K}_{xy}(x,y)-\mathcal{K}_{z}(-z)\,, & z,\, x^{2}+y^{2}<1,\, x^{2}+z^{2},\, y^{2}+z^{2}\geq1\,,\\
\mathcal{K}_{x}(x)-\mathcal{K}_{xy}(x,-y)-\mathcal{K}_{xz}(x,-z)\,, & x^{2}+y^{2},\, x^{2}+z^{2}<1,\, y^{2}+z^{2}\geq1\,,\\
\mathcal{K}_{y}(y)-\mathcal{K}_{xy}(-x,y)-\mathcal{K}_{yz}(y,-z)\,, & x^{2}+y^{2},\, y^{2}+z^{2}<1,\, x^{2}+z^{2}\geq1\,,\\
\mathcal{K}_{x}(x)-\mathcal{K}_{xy}(x,-y)-\mathcal{K}_{xyz}(x,y,-z)\,, & x^{2}+y^{2},\, x^{2}+z^{2},\, y^{2}+z^{2}<1\,,x^{2}+y^{2}+z^{2}\geq1\\
\mathcal{K}_{xyz}(x,y,z)\,, & x^{2}+y^{2}+z^{2}<1\:.\end{cases}\label{eq:J2PW}\end{eqnarray}
Following a similar procedure we can obtain $\mathcal{J}_{2}(\mathbf{r})$
for Regions from 2 to 8.%
\begin{figure}[b]
\begin{centering}
\includegraphics[width=12.5cm]{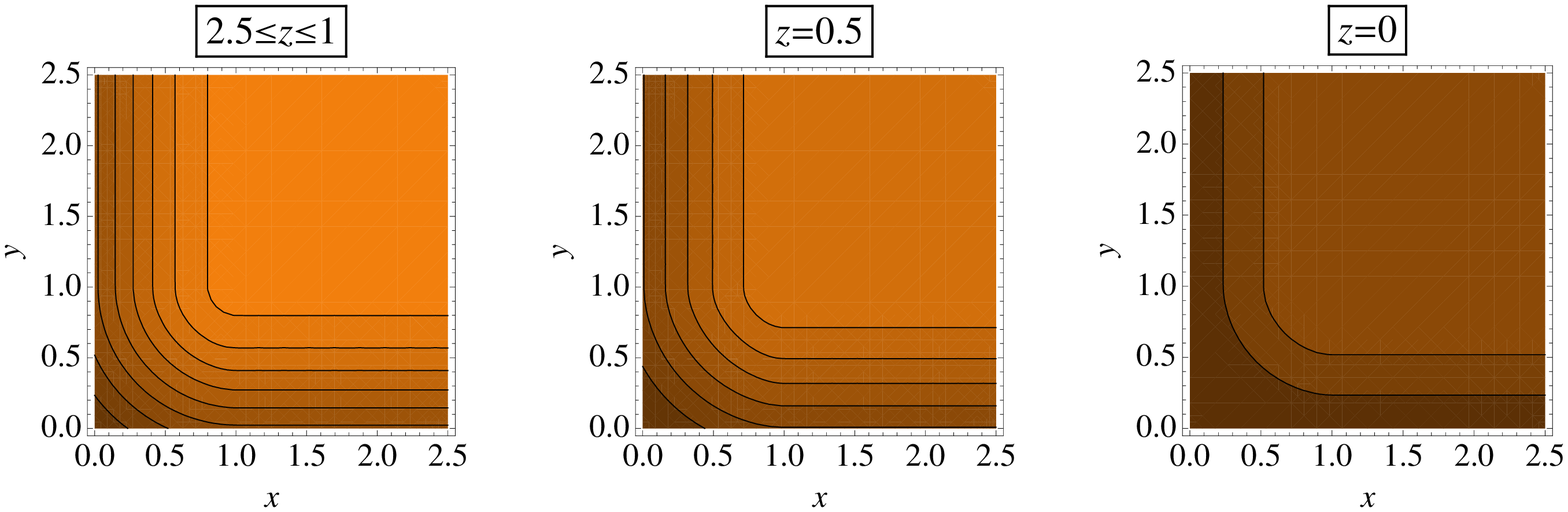}
\par\end{centering}

\caption{(Color online) Contour plot of the density distribution for a cubic
pore with $L=5\sigma$. As gray becomes darker the density becomes
higher.\label{fig:ContourPlotA}}

\end{figure}
\begin{figure}[tbh]
\begin{centering}
\includegraphics[width=5cm]{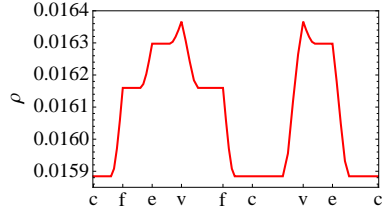}
\par\end{centering}

\caption{Density distribution for a closed path in the cubic pore with $L=5\sigma$.\label{fig:ParamPlotA}}

\end{figure}
 In Fig. \ref{fig:ContourPlotA}, we show three contour plot slices
of $\rho(\mathbf{r})$ for a cube with $L=5\sigma$. From left to
right of Fig. \ref{fig:ContourPlotA} the first slice shows the behavior
of $\rho(\mathbf{r})$ at half height of the cavity, the second one
refer to a near wall position while the third one describe the behavior
of $\rho(\mathbf{r})$ on contact with the planar wall. The nearest
line to the top-right corner of the slices corresponds to $\rho\sigma^{3}=0.0159\,,\,0.016$
and $0.0162$, respectively. The step in density between lines is
$\triangle\rho\,\sigma^{3}=0.5\,10^{-4}$. In Fig. \ref{fig:ContourPlotA}
all the relevant characteristics of the density profile $\rho(\mathbf{r})$
are apparent. We can observe the plateau of constant density at a
distance $\sigma$ from the boundary and the increasing value of $\rho(\mathbf{r})$
going from the plateau to the cuboidal cavity boundaries. Figure \ref{fig:ParamPlotA}
shows a plot of $\rho(\mathbf{r})$ for a given path in the same cubic
cavity ($L=5\sigma$). There, the path is composed by several straight
line parts. It starts at the cavity center (c), goes to the face (f)
center, next to the middle of the edge (e), and next to the vertex
(v). The rest of the path follows other highly symmetric directions
of the cube. We can observe here that even when $\rho(\mathbf{r})$
is a piecewise defined function, it is continuous and also derivable
(peaks appear because the path change its direction abruptly). The
minimum value corresponds to the plateau of constant density. For
cavities with smaller size the extent of the plateau of constant density
is more reduced. The effect of the higher confinement may be seen
at Figs. \ref{fig:ContourPlotB} and \ref{fig:ParamPlotB}, where
the density distribution for a cubic pore with $L=2\sigma$ is presented.
From the left of Fig. \ref{fig:ContourPlotB} the first slice of $\rho(\mathbf{r})$
is at half height of the cavity. Other two slices are similar to Fig.
\ref{fig:ContourPlotA}. The nearest line to the top-right corner
corresponds to $\rho\sigma^{3}=0.18\,,\,0.20$ and $0.26$, respectively.
The step in density between lines is now $\triangle\rho\,\sigma^{3}=0.02$.
As can be seen in Fig. \ref{fig:ParamPlotB} the plateau disappears,
because only for $\mathbf{r}$ at c the ES is completely inside of
the cubic cavity. It is also apparent from a comparison with Fig.
\ref{fig:ParamPlotA}. From Figs. \ref{fig:ContourPlotA}, \ref{fig:ParamPlotA},
\ref{fig:ContourPlotB} and \ref{fig:ParamPlotB} we can also smell
out the general behavior of $\rho(\mathbf{r})$ for the 2-HS in cavities
with different geometries and the effect of reducing the size of the
cavity.%
\begin{figure}[bh]
\begin{centering}
\includegraphics[width=12.5cm]{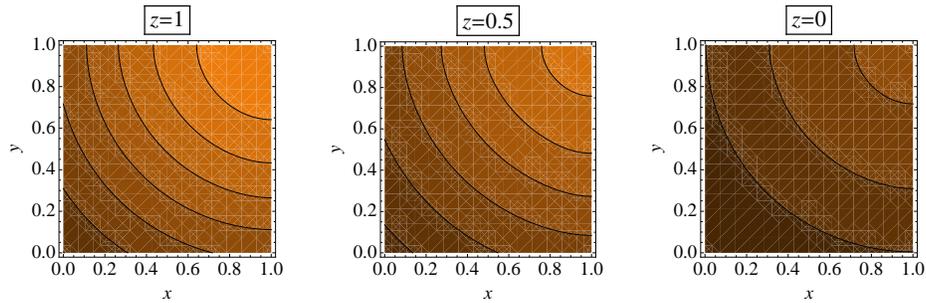}
\par\end{centering}

\caption{(Color online) Contour plot of the density distribution for a cubic
pore with $L=2\sigma$. As gray becomes darker the density becomes
higher.\label{fig:ContourPlotB}}

\end{figure}
\begin{figure}[th]
\begin{centering}
\includegraphics[width=5cm]{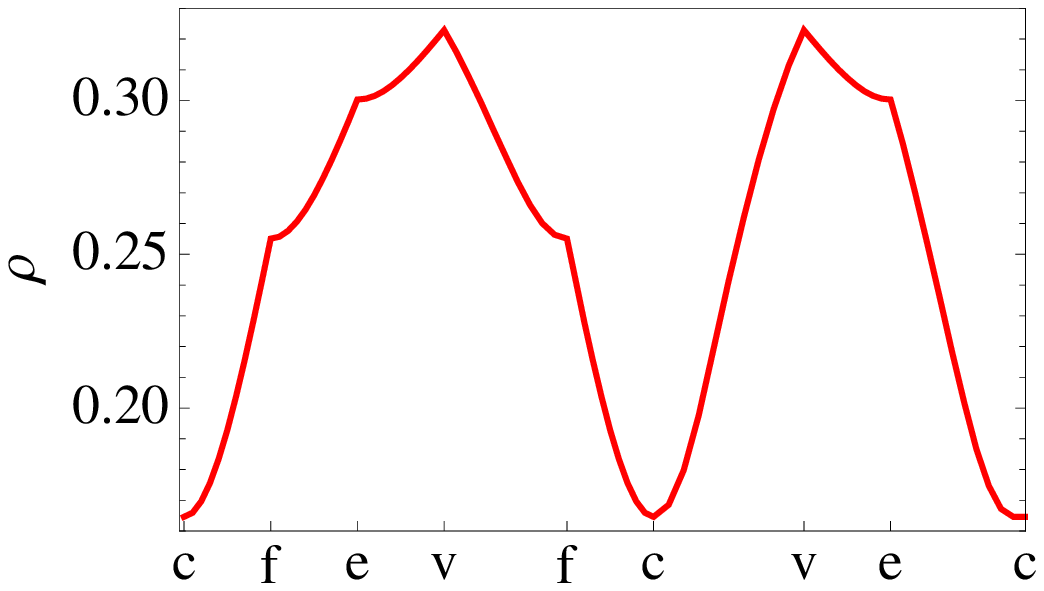}
\par\end{centering}

\caption{Density distribution for a closed path in the cubic pore with $L=2\sigma$.\label{fig:ParamPlotB}}

\end{figure}

\subsection{Density distribution in the cylindrical cavity\label{sub:Density-cyl}}

For the cylindrical pore the set of relevant functions are \begin{equation}
\{\mathcal{J}_{2z}(\mathbf{r}.\hat{\mathbf{z}}),\,\mathcal{J}_{2\bar{r}}(\mathbf{r}.\hat{\bar{\mathbf{r}}}),\,\mathcal{J}_{2z\bar{r}}(\mathbf{r}.\hat{\mathbf{z}},\mathbf{r}.\hat{\bar{\mathbf{r}}})\}=\sigma^{3}\{\mathcal{K}_{z}(\tilde{\mathbf{r}}.\hat{\mathbf{z}}),\,\mathcal{K}_{\bar{r}}(\tilde{\mathbf{r}}.\hat{\bar{\mathbf{r}}}),\,\mathcal{K}_{z\bar{r}}(\tilde{\mathbf{r}}.\hat{\mathbf{z}},\tilde{\mathbf{r}}.\hat{\bar{\mathbf{r}}})\}\;,\label{eq:JKcil}\end{equation}
where the cylinder axis is in $\hat{\mathbf{z}}$ direction and $\hat{\mathbf{r}}$
is the radial polar versor. The inward normal to the lateral face
is $\hat{\bar{\mathbf{r}}}=-\hat{\mathbf{r}}$ and $\mathbf{r}.\hat{\bar{\mathbf{r}}}$
is the shortest distance from the sphere center to the lateral surface
of the cylinder with radius $R$. Here, the functions $\{\mathcal{K}_{z}(\mathbf{r}.\hat{\mathbf{z}}),\,\mathcal{K}_{\bar{r}}(\mathbf{r}.\hat{\bar{\mathbf{r}}}),\,\mathcal{K}_{z\bar{r}}(\mathbf{r}.\hat{\mathbf{z}},\mathbf{r}.\hat{\bar{\mathbf{r}}})\}$
are defined by translating to a cylindrical cavity the description
made for the cuboidal cavity. The function $\mathcal{K}_{z}(\mathbf{r}.\hat{\mathbf{z}})$
was already analyzed in Eqs. (\ref{eq:K1}-\ref{eq:K1-prop2}). On
the basis of the analytical expression for the overlap volume between
a sphere and an infinite cylinder obtained in Ref. \cite{Lamarche_1990}
(see Eq. (3) therein) we may obtain $\mathcal{K}_{\bar{r}}(\mathbf{r}.\hat{\bar{\mathbf{r}}})$
in terms of elliptic integrals. Some properties of these functions
are\begin{equation}
4\pi/3=\mathcal{K}_{\bar{r}}(\mathbf{r}\cdot\hat{\bar{\mathbf{r}}})+\mathcal{K}_{r}(\mathbf{r}\cdot\hat{\mathbf{r}})\:,\label{eq:K1r-prop}\end{equation}
\begin{eqnarray}
\mathcal{K}_{z}(\mathbf{r}\cdot\hat{\mathbf{z}}) & = & \mathcal{K}_{z\bar{r}}(\mathbf{r}\cdot\hat{\mathbf{z}},\mathbf{r}\cdot\hat{\bar{\mathbf{r}}})+\mathcal{K}_{zr}(\mathbf{r}\cdot\hat{\mathbf{z}},\mathbf{r}\cdot\hat{\mathbf{r}})\:,\nonumber \\
\mathcal{K}_{\bar{r}}(\mathbf{r}\cdot\hat{\bar{\mathbf{r}}}) & = & \mathcal{K}_{z\bar{r}}(\mathbf{r}\cdot\hat{\mathbf{z}},\mathbf{r}\cdot\hat{\bar{\mathbf{r}}})+\mathcal{K}_{z\bar{r}}(-\mathbf{r}\cdot\hat{\mathbf{z}},\mathbf{r}\cdot\hat{\bar{\mathbf{r}}})\:,\label{eq:K2r-prop}\end{eqnarray}
\begin{equation}
4\pi/3=\mathcal{K}_{z\bar{r}}(\mathbf{r}\cdot\hat{\mathbf{z}},\mathbf{r}\cdot\hat{\bar{\mathbf{r}}})+\mathcal{K}_{z\bar{r}}(-\mathbf{r}\cdot\hat{\mathbf{z}},\mathbf{r}\cdot\hat{\bar{\mathbf{r}}})+\mathcal{K}_{zr}(\mathbf{r}\cdot\hat{\mathbf{z}},\mathbf{r}\cdot\hat{\mathbf{r}})+\mathcal{K}_{zr}(-\mathbf{r}\cdot\hat{\mathbf{z}},\mathbf{r}\cdot\hat{\mathbf{r}})\:.\label{eq:K2r-prop2}\end{equation}
We do not find an analytical expression for $\mathcal{K}_{z\bar{r}}(\mathbf{r}.\hat{\mathbf{z}},\mathbf{r}.\hat{\bar{\mathbf{r}}})$,
which implies that we are not able to describe $\rho(\mathbf{r})$
near the circular edges of the cylinder when $(\mathbf{r}\cdot\hat{\mathbf{z}})^{2}+(\mathbf{r}\cdot\hat{\bar{\mathbf{r}}})^{2}<1$.
However, the exact value of $\rho(\mathbf{r})$ on the edge is\begin{equation}
\mathcal{K}_{z\bar{r}}(0,0)=\frac{1}{2}\mathcal{K}_{\bar{r}}(0)\:.\label{eq:J2zredge}\end{equation}
For the spheroid cavity we only found analytic expressions of $\mathcal{J}_{2a}(\mathbf{r}.\hat{\mathbf{a}})$
for points on the polar axis and points on the equatorial plane, but
they are not presented here. Functions $\mathcal{J}_{2\bar{r}}(\mathbf{r}\cdot\hat{\bar{\mathbf{r}}})$
and $\mathcal{K}_{\bar{r}}(\mathbf{r}\cdot\hat{\bar{\mathbf{r}}})$
for the spherical cavity were obtained in \cite{Urrutia_2008}, and
for dimensions other than 3 in \cite{Urrutia_2008,Urrutia_2010}.
These expressions enable to obtain $\rho(r)$ near a concave or convex
spherical surface. In addition, $\rho(r)$ at the spherical pore with
a hard core can also be obtained analytically using the same $\mathcal{J}_{2\bar{r}}(\mathbf{r}\cdot\hat{\bar{\mathbf{r}}})$
and $\mathcal{K}_{\bar{r}}(\mathbf{r}\cdot\hat{\bar{\mathbf{r}}})$.

\subsection{Pressure\label{sub:PressureT}}

\begin{figure}
\begin{centering}
\includegraphics[width=8cm]{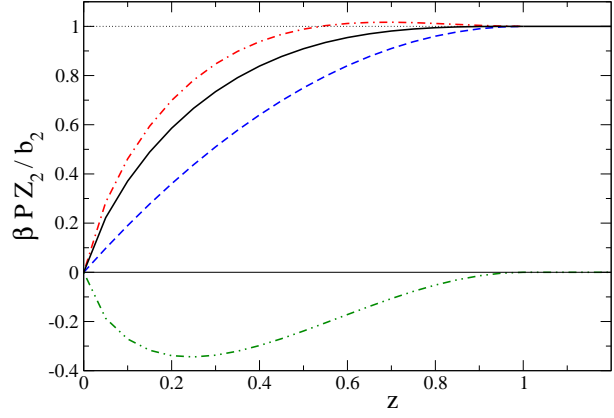}
\par\end{centering}

\caption{(color online) Position dependence of pressure tensor near a planar
wall. We have drawn magnitudes related with scalar pressure $P$ in
continuous line, $P_{N}$ in dashed line and $P_{T}$ in dot-dashed
line. The behavior of $P_{N}\left(z\right)-P_{T}\left(z\right)$ (the
integrand of Eq. (\ref{eq:SurfzofTP})) is shown in dot-dot-dashed
line. More details about the plotted functions in the text.\label{fig:Pressure-tensor.}}

\end{figure}

The analytic evaluation of the pressure tensor $\mathbf{P}(\mathbf{r})$,
a symmetric tensor of rank two, is much more difficult than the evaluation
of $\rho(\mathbf{r})$ in an inhomogeneous fluid. For that reason
we will not make a systematic search for each geometry confinement
as was done in Secs. \ref{sub:Density-cube} and \ref{sub:Density-cyl}.
Even, we only make the complete evaluation for some simple cases.
The relevant task of a detailed and systematic study of $\mathbf{P}(\mathbf{r})$
for 2-HS system near simple curved walls is planned to be presented
anywhere. We focus on the evaluation of the pressure tensor $\mathbf{P}$
of Irving and Kirkwood \cite{Irving_1950}. The components of $\mathbf{P}$
for the 2-particle system are $P_{ab}\left(\mathbf{r}\right)=\beta^{-1}\delta_{ab}\rho\left(\mathbf{r}\right)+P_{ab}^{U}\left(\mathbf{r}\right)$,
with \begin{equation}
P_{ab}^{U}\left(\mathbf{r}\right)=\left\langle r_{12}^{a}F_{12}^{b}\,\int_{0}^{1}dt\,\delta\left(\mathbf{r}-\mathbf{r}_{1}+t\mathbf{r}_{12}\right)\right\rangle _{c}\:,\label{eq:Pu_01}\end{equation}
where $\mathbf{r}_{i}$ is the coordinate of the $i$-particle, $\mathbf{r}_{12}=\mathbf{r}_{1}-\mathbf{r}_{2}$,
$r_{12}^{a}=\mathbf{r}_{12}\cdot\hat{\mathbf{a}}$, and $F_{12}^{b}=\mathbf{F}_{12}\cdot\hat{\mathbf{b}}=-\dfrac{\partial\varphi}{\partial r_{12}}\,\dfrac{r_{12}^{b}}{r_{12}}$.
By direct integration we obtain the identity\begin{equation}
I\left(\mathbf{r},\mathbf{r}_{1},\mathbf{r}_{2}\right)=\int_{0}^{1}dt\,\delta\left(\mathbf{r}-\mathbf{r}_{1}+t\mathbf{r}_{12}\right)=r_{12}^{-1}u^{-2}\delta\left(\hat{\mathbf{r}}_{12}-\hat{\mathbf{u}}\right)\Theta\left(r_{12}-u\right)\:,\label{eq:Idt}\end{equation}
with $\mathbf{u}=\mathbf{r}_{1}-\mathbf{r}=u\hat{\mathbf{u}}$. For
a fixed $\mathbf{r}$ we introduce a set of cartesian and spherical
coordinates with the usual convention for the polar angles i.e. $r_{12}^{x}=\cos\left(\theta_{12}\right)\sin\left(\phi_{12}\right)r_{12}$,
$r_{12}^{y}=\sin\left(\theta_{12}\right)\sin\left(\phi_{12}\right)r_{12}$,
and $r_{12}^{z}=\cos\left(\phi_{12}\right)r_{12}$. We can re-write
Eq. (\ref{eq:Pu_01}), and for example, the $P_{zz}^{U}$ component\begin{equation}
\beta P_{zz}^{U}\left(\mathbf{r}\right)=Z_{2}^{-1}\iint e(\mathbf{r}_{1})e(\mathbf{r}_{2})\delta\left(r_{12}-\sigma\right)\, r_{12}\cos^{2}\left(\phi_{12}\right)\, I\left(\mathbf{r},\mathbf{r}_{1},\mathbf{r}_{2}\right)\, d^{3}r_{1}d^{3}r_{2}\:.\label{eq:Puzz_01}\end{equation}
Using Eq. (\ref{eq:Idt}), changing the integration variables to $d^{3}u\, d^{3}r_{12}$,
expressing all the distances in $\sigma$ units and both variables
in spherical coordinates i.e. $d^{3}r_{12}=r_{12}^{2}\sin\left(\phi_{12}\right)dr_{12}\, d\phi_{12}\, d\theta_{12}$
and $d^{3}u=u^{2}\sin\left(\phi\right)du\, d\phi\, d\theta$, and
finally integrating on $d^{3}r_{12}$, we obtain\begin{eqnarray}
\beta P_{zz}^{U}\left(\mathbf{r}\right) & = & Z_{2}^{-1}\sigma^{3}\int e(\mathbf{r}-\mathbf{u})e\left[\mathbf{r}-\left(1-u\right)\cdot\hat{\mathbf{u}}\right]\cos^{2}\left(\phi\right)\sin\left(\phi\right)\Theta\left(1-u\right)\, du\, d\phi\, d\theta\:.\label{eq:Puzz_02}\end{eqnarray}
Note that the range of $u$ is $1$. For $\mathbf{r}$ at a distance
from the wall greater than 1 the integral $\beta P_{zz}^{U}\left(\mathbf{r}\right)$
becomes independent of $\mathbf{r}$, because for all the available
values of $\mathbf{u}$ in the integration domain we have $e(\mathbf{r}-\mathbf{u})=1$
and $e[\mathbf{r}-(\sigma-u)\cdot\hat{\mathbf{u}}]=1$. Therefore,
for such $\mathbf{r}$ in the region of constant density (see Eq.
(\ref{eq:rho0}) and comments therein) we find \begin{eqnarray}
\beta P_{zz}^{U}\left(\mathbf{r}\right) & = & Z_{2}^{-1}\sigma^{3}\int\cos^{2}\left(\phi\right)\sin\left(\phi\right)\Theta\left(1-u\right)\, du\, d\phi\, d\theta\:,\nonumber \\
 & = & Z_{2}^{-1}2b_{2}\:.\label{eq:Pu_hom}\end{eqnarray}
The other components of the tensor are $P_{xx}^{U}=P_{yy}^{U}=P_{zz}^{U}$
and $P_{xy}^{U}=P_{yz}^{U}=P_{xz}^{U}=0$. This is expected because
the pressure tensor in a region of constant density must be isotropic.
The scalar pressure and the tensor relates by $\beta P=\beta\, tr\left(\mathbf{P}\right)/3$,
where $tr$ is the trace. Therefore, the scalar pressure in the region
of constant density is \begin{equation}
\beta P_{0}=\rho_{0}+Z_{2}^{-1}2b_{2}=Z_{2}^{-1}2\left(Z_{1}-b_{2}\right).\label{eq:Pt}\end{equation}
A similar procedure was applied in \cite{Urrutia_2010} to the study
of the 2-HS system in D dimensions. There, using a different definition
of $P_{ab}^{U}$, the authors obtained the same result for $P_{0}$.
Pressure tensor near a planar wall can also be evaluated starting
from Eq. (\ref{eq:Puzz_02}). We consider a wall with inward normal
$\hat{\mathbf{z}}$ and an inner particle at a distance $\mathbf{r}.\hat{\mathbf{z}}=z$
with $0\leq z\leq1$. Integrating on a domain defined by $\left|\mathbf{r}-\mathbf{u}\right|\leq1$,
$\left|\mathbf{r}-(1-u)\cdot\hat{\mathbf{u}}\right|\leq1$, and $0<u\leq1$
we find the normal component\begin{eqnarray}
\beta P_{N}^{U}\left(\mathbf{r}\right) & = & Z_{2}^{-1}2\pi\sigma^{3}\int\cos^{2}\left(\phi\right)\sin\left(\phi\right)du\, d\phi\:,\nonumber \\
 & = & Z_{2}^{-1}b_{2}z\,\left(3-z^{2}\right)\:.\label{eq:PN}\end{eqnarray}
Such result can be easily checked. On one side, for an inhomogeneous
fluid with planar symmetry we obtain $\beta P_{N}\left(\mathbf{r}\right)=Z_{2}^{-1}2\left(Z_{1}-3b_{2}\right)$
which is independent of the position as it would be expected. On
the other side, the fact that the contact value at the wall surface
must be $\beta P_{N}^{U}\left(z=0\right)=\rho(0)$ which implies $P_{N}^{U}\left(z=0\right)=0$.
By following an identical procedure we find for both equal tangential
components that\begin{eqnarray}
\beta P_{T}^{U}\left(\mathbf{r}\right) & = & Z_{2}^{-1}\pi\sigma^{3}\int\sin^{3}\left(\phi\right)du\, d\phi\:,\nonumber \\
 & = & Z_{2}^{-1}\frac{1}{2}b_{2}z\,\left[3+z^{2}-6\,\ln\left(z\right)\right]\:.\label{eq:PT}\end{eqnarray}
For symmetry reasons the non-diagonal components are null. The scalar
pressure near a planar wall is \begin{equation}
\beta P(\mathbf{r})=\rho(z)+Z_{2}^{-1}2b_{2}z\,\left[1-\ln\left(z\right)\right]=Z_{2}^{-1}2\left[Z_{1}-b_{2}\left(1+z(1-z^{2})/2+z\,\ln\left(z\right)\right)\right]\:.\label{eq:Pt-p}\end{equation}
Finally, the wall-fluid surface tension of the 2-HS fluid in contact
with a hard planar wall and the position of the surface of tension
are\begin{equation}
\beta\gamma=\beta\sigma\int\left[P_{N}\left(z\right)-P_{T}\left(z\right)\right]dz=-Z_{2}^{-1}2a_{2}\:,\label{eq:SurfTP}\end{equation}
\begin{equation}
z_{s}=-\gamma^{-1}\sigma\int\left[P_{N}\left(z\right)-P_{T}\left(z\right)\right]z\, dz=-0.3\!\overset{\frown}{5}\:.\label{eq:SurfzofTP}\end{equation}
 In Fig. \ref{fig:Pressure-tensor.} we plot together the position
dependence for the pressure tensor components and other related magnitudes
near a planar wall. The dependence with position is highlighted by
plotting dimensionless magnitudes independent of $Z_{2}$. We plot
$\left(\beta P_{k}(z)-\rho(z=0)\right)Z_{2}/2b$ with $P_{k}=P,\, P_{N},\, P_{T}$
and $\left(\beta P_{N}(z)-\beta P_{T}(z)\right)Z_{2}/2b$ with continuous,
dashed, dot-dashed and dot-dot-dashed lines, respectively. We see
that at contact with the wall all functions go to zero with finite
slope. For $P,\, P_{N}$ and $P_{T}$ the null value at $z=0$ is
a consequence of the contact theorem. On the opposite, functions attain
their definitive homogeneous value at distance $\sigma$ from the
wall.%
 Similar to the planar case, the spherical symmetry produce only
two independent components $P_{N}^{U}$ and $P_{T}^{U}$. We have
obtained analytical expressions for the Irving-Kirkwood pressure tensor
$\mathbf{P}$ near a spherical surface. This was done for convex and
concave, surfaces. Even, the evaluation is not straightforward and
therefore the study of the pressure tensor for the 2-HS system near
a spherical wall will be presented in a future work. Near a cylindrical
wall the components of $\mathbf{P}$ involve more complex integrals
that we do not attempt to solve.

Additionally, it is interesting to note a simple relation between
pressure and density in the region of constant density. Recognizing
that $Z_{1}$ plays the role of the system volume we can define the
mean density $\bar{\rho}=2/Z_{1}$. Therefore, from Eqs. (\ref{eq:rho0},
\ref{eq:Pt-p}) we obtain the local compressibility factor in the
region of constant density \begin{equation}
\frac{\beta P_{0}}{\rho_{0}}=1+\frac{1}{2}\,\frac{b}{\bar{\rho}^{-1}-b}\:.\label{eq:CompressZ}\end{equation}
This is a \emph{local} EOS because describes the properties in certain
location of the entire 2-HS system. In Sec. \ref{sec:Thermodynamic-study}
we will study \emph{thermodynamic} or \emph{global} EOS. Expression
(\ref{eq:CompressZ}) is very similar to the EOS of a (bulk) van der
Waals system without the term of attractive force between particles.
They differ in the $1/2$ factor present on Eq. (\ref{eq:CompressZ}),
which is related to the small number of particles of the 2-HS system.
The Eq. (\ref{eq:CompressZ}) is valid for all the studied cavities,
and it was also obtained for the equivalent system of confined 2-HS
in dimensions $D\neq3$. As it was suggested in Ref. \cite{Urrutia_2010},
it seems that Eq. (\ref{eq:CompressZ}) is a universal feature of
a 2-HS system confined in a cavity with Hard Walls of any shape and
for all dimensions $D\geq1$. We note that for a small enough cavity
that produce a vanishing size density plateau the value of $\rho_{0}$
depends on the geometry of the cavity. For a spherical cavity we have
$\rho_{0}=0$ while in other cavities $\rho_{0}$ assumes positive
values.

\section{Analytic structure of CI\label{sec:Analysis}}

\begin{figure}
\begin{centering}
\includegraphics[width=14cm]{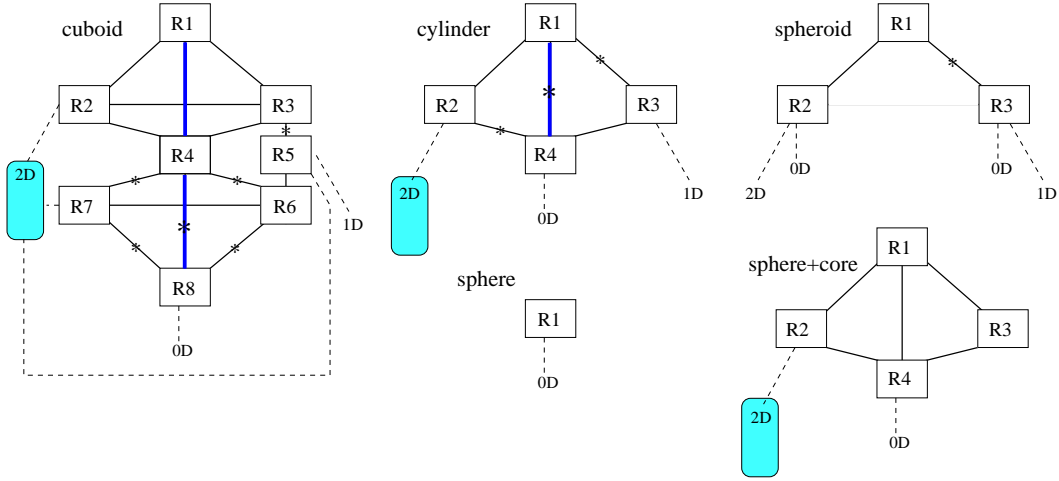}
\par\end{centering}

\caption{A simple picture representation of the $Z_{2}$ $\mathbf{X}$-space
domain for all the studied pores. From top to bottom of each graph
the volume decrease. \label{fig:Z2picture}}

\end{figure}

\begin{figure}[bh]
\begin{centering}
\includegraphics[clip,width=7cm]{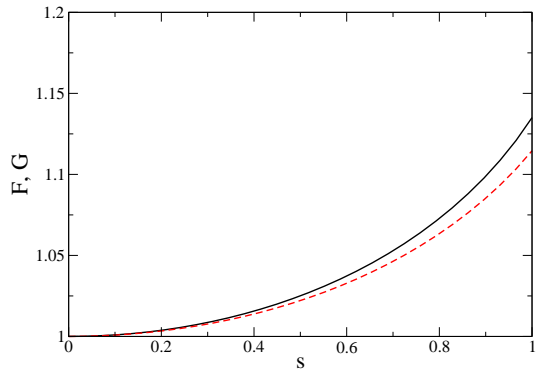}
\par\end{centering}

\caption{Shape dependent coefficients in $b_{2}(pore)$. Functions $F(\mathsf{s})$
and $G(\mathsf{s})$ of Table \ref{tab:Shape_Compar} are shown in
continuous and dashed lines respectively.\label{fig:Shape-and-size} }

\end{figure}

The usual classical statistical mechanics links some global thermodynamic
properties of any system of particles with some derivatives of $\ln(Z_{2})$,
this idea will be discussed in detail in Sec. \ref{sec:Thermodynamic-study}.
Now, we simple recognize that the analytical behavior of $Z_{2}$
is related to the physical properties of the 2-HS. Therefore, the
goal of this section is the study of the analytic structure of $Z_{2}$
as a function of pore size parameters $\mathbf{X}$, with the emphasis
in the non-analytic domain. We are interested in investigate common
features between cavities with different geometries. By including
results from \cite{Urrutia_2008,Urrutia_2010} we compare the CI for
two hard spheres constrained by five different simple geometries:
cuboid, sphere, sphere with a hard core, cylinder and spheroid shaped
pores. A picture representing the structure of the domains for those
$Z_{2}$ is shown in Fig. \ref{fig:Z2picture}. There, each box labeled
with R (R-boxes) represents a region of parameter $\mathbf{X}$ domain
studied in Sec. \ref{sec:Two-bodies} as a separate case. The analytic
domain of $Z_{2}$ is the union of the (open) domains represented
by the R-boxes. Straight line paths show the boundaries between adjacent
zones, i.e. the non analytic domain of CI, while the broaden lines
highlight paths of maximum symmetry ($L_{x}=L_{y}=L_{z}$ for cuboid
and $L_{h}=2R$ for cylinder). The stars distinguish the non-analytic
domains involving the ergodic-non-ergodic transition. Dashed lines
plot the crossover to systems with reduced dimension 0D, 1D or 2D,
the 2D effective systems are represented with dark rounded-corner-boxes.
The 2D limit for the spheroidal cavity has a different nature and
we do not draw the box for this 2D limit. %
From Fig. \ref{fig:Z2picture} we can sort the structure of the $Z_{2}$
analytic domains for the studied cavity geometries in an increasing
order of complexity: sphere, spheroid, sphere+core, cylinder, and
cuboid. The sphere is the simplest geometry, the cuboid results the
most complex while the spheroid, sphere+core and cylinder have a similar
degree of complexity. Moreover, if we restrict from the cuboid cavity
to a cube, or from the cylinder to the \emph{symmetric} cylinder,
its structure becomes much more simpler. This shows that the increment
of the symmetry result in a decrement on the number of parameters
in $\mathbf{X}$. In summary, cavities with high (poor) symmetry and
few (many) number of parameters $\mathbf{X}$ produce a simple (complex)
structure. In Fig. \ref{fig:Z2picture} we identify several interesting
common features concerning different shaped pores: (a) the large pore
domain R1, (b) its boundaries, (c) the $\textrm{Ri}\overset{*}{\rightarrow}\textrm{Rj}$,
the signature of the ergodicity breaking, (d) the $\textrm{Rj}\rightarrow\textrm{2D}$
limit that exist in cuboid, cylinder and sphere+core pores, (e) the
structure $\textrm{Ri}\overset{*}{\rightarrow}\textrm{Rj}\rightarrow\textrm{1D}$
limit, and (f) the structures $\textrm{Ri}\overset{*}{\rightarrow}\textrm{Rj}\rightarrow\textrm{0D}$
limit, $\textrm{Ri}\rightarrow\textrm{Rj}\rightarrow\textrm{0D}$
limit and particularly the last sequence $\textrm{Rj}\rightarrow\textrm{0D}$
limit. We now analyze the relevant properties for each case.
\begin{description}
\item [{(a)}] The large pore domain R1\label{des:(a)-the-large}
\end{description}
Firstly, we concentrate in large cavities. The different analyzed
geometries show that the large pore domain is the easiest to integrate
and frequently the CI has a simple functional dependence. From direct
inspection (see Eqs. (\ref{eq:I1}-\ref{eq:Z01}) and also, Refs.
\cite{Urrutia_2008,Urrutia_2010}) we note that for cuboid, spherical
and sphere+core cavities the CI is a polinomy, but a more complex
analytic dependence appears for the cylindrical and spheroidal pores.
A comparison with two dimensions shows that the CI of the system of
two hard disks into a rectangular cavity is also a polinomy, although
for a circular cavity it is not true. From all the available CI we
observe that $Z_{1}b_{2}(pore)$ of Eq. (\ref{eq:b00}) naturally
decompose in a universal way showing a simple dependence on basic
geometrical measures of the effective pore. In terms of the volume
notion $V=Z_{1}$ we obtain,\begin{eqnarray}
Vb_{2}(pore) & = & Vb_{2}-a_{2}A+\ell_{2}Le+c_{2,1}+c_{2,2}\frac{Le}{R^{2}}\;.\label{eq:b2Vgen}\end{eqnarray}
The constant coefficients $b_{2}$ (see Eq. (\ref{eq:2cont2})) and
$a_{2}=\sigma^{4}\pi/8$ are independent of the pore shape. $a_{2}$
appears in the virial expansion of the fluid-substrate surface tension
and adsorption (referred as $w_{2}$ \cite{Bellemans_1962,Bellemans_1962_b,Bellemans_1963,Sokolowski_1977})
and particularly, for a HS fluid in contact with planar and spherical
walls \cite{Bellemans_1963,Stecki_1978,Urrutia_2010}. Besides the
volume, in Eq. (\ref{eq:b2Vgen}) we introduce other geometrical characters
of the effective cavity, the area of the boundary $A$ and the total
edges length $Le$. In table \ref{tab:Shape_Compar} we present a
comparison of the set $\{\ell_{2};\, c_{2,1};\, c_{2,2}\}$ for all
the studied pore shapes, where the dependence on edges length, surface
curvature and edge curvature is traced.%
\begin{table}[h]
\begin{centering}
\begin{tabular}{|c|c|c|c|c|c|}
\hline 
 & cuboid & cylinder & spheroid & sphere & sph+core\tabularnewline
\hline
\hline 
$\ell_{2}/\sigma^{5}$ & $1/15$ & $1/15$ & $-$ & $-$ & $-$\tabularnewline
\hline 
$c_{2,1}/\sigma^{6}$ & $-1/12$ & $(L_{h}/2R)\, F(\mathsf{s})\,\pi^{2}/96$ & $H(\lambda)\,\pi^{2}/36$ & $\pi^{2}/36$ & $2\pi^{2}/36$\tabularnewline
\hline 
$c_{2,2}/\sigma^{7}$ & $-$ & $-\pi/210\, G(\mathsf{s})$ & $-$ & $-$ & $-$\tabularnewline
\hline
\end{tabular}
\par\end{centering}

\caption{Coefficients of $Vb_{2}(pore)$, dependence on the cavity shape for
the large pore region.\label{tab:Shape_Compar}}

\end{table}
 We note that $Vb_{2}(pore)$ in Eq. (\ref{eq:b2Vgen}) for cuboid,
sphere and sphere+core shaped pores involves constant coefficients
$\{\ell_{2};\, c_{2,1};\, c_{2,2}\}$. The coefficient $\ell_{2}$
that multiplies $Le$ has a unique positive value having the opposite
sign to the preceding area term. Naturally, the edges are the area
boundaries. Then, we saw the $Le$ term in Eq. (\ref{eq:b2Vgen})
as a correction to the previous one. We interpret $\ell_{2}(cub)$
and $\ell_{2}(cyl)$ coefficients as being originated in the right
dihedral edge formed by the intersection of two smooth surfaces. The
$c_{2,1}$ is in general a slowly varying function of adimensional
parameters $\mathsf{s}=\sigma/2R$ and $\lambda=Rc/R$. It is constant
for cuboid, spherical and sphere+core pores. The negative constant
$c_{2,1}(cub)$ has a sign opposite to the previous edges term. From
that we consider it as an end-of-edge correction which corresponds
to the eight right vertex of the cuboid. Then, seeking for each vertex
contribution we may write $c_{2,1}(cub)/\sigma^{6}=-8/96$ and therefore
each vertex produce $-1/96$. On the other hand $c_{2,1}(sph)$ and
$c_{2,1}(sph+core)$ are positive, i.e. they have the sign opposite
to $c_{2,1}(cub)$, and also, they are not corrections to an absent
edge term. Therefore, their nature is different to that $c_{2,1}(cub)$.
Coefficients $c_{2,1}(sph)$ and $c_{2,1}(sph+core)$ are originated
on the curvature of the surfaces and their sign is opposite to the
previous area term which corrects. Therefore, the surface curvature
should produce a negative value for $c_{2,1}$ for both, a cylindrical
and spheroidal pores. We introduce now the usual surface curvature
measures, normal curvature $j$ and Gaussian curvature $k$, which
take the values $\{j=R^{-1},\, k=0\}$ and $\{j=2R^{-1},\, k=R^{-2}\}$
for a cylinder and a sphere, respectively. We find that $c_{2,1}(sph)/\sigma^{6}=A\, R^{-2}\,\pi/144=J\!\! J(R)\,\delta^{(1)}$
and $c_{2,1}(sph+core)/\sigma^{6}=(J\!\! J(R)+J\!\! J(R-h))\,\delta^{(1)}=2c_{2,1}(sph)/\sigma^{6}$
with the extensive quadratic curvature $J\!\! J(R)=A\, j^{2}=2^{4}\pi$
and $\delta^{(1)}=3^{-2}2^{-6}\pi$ \cite{Urrutia_2010}. For cylindrical
cavities we find that $c_{2,1}(cyl)/\sigma^{6}=A_{curv}R^{-2}F(x)\,\pi/384=J\!\! J(R)\,\delta^{(1)}F(x)\,3/2$,
where $A_{curv}$ is the curved lateral surface area, $J\!\! J(R)=A_{curv}j^{2}=2\pi L_{h}/R$,
and for large radius $c_{2,1}(cyl)\sim A_{curv}R^{-2}$. An unified
description of cyl, sph and sph+core pores at large $R$ is $c_{2,1}(cyl,sph,sph+core)\,\sigma^{-6}=A_{curv}\,(\frac{3}{4}\, j^{2}+k)\,\delta^{(1)}$,
but more complex dependence exist at $c_{2,1}(sphd)$. In fact, for
large curvature radius and quasi spherical ellipsoids $\lambda\sim1$
we find $c_{2,1}(sphd)\simeq c_{2,1}(sphd)\,(1+4/5\,(1-\lambda)^{2})$.
Similarly, $c_{2,2}(cyl)$ relates with the curvature of the edges.
We may resume some characteristics of $\{F(\mathsf{s}),G(\mathsf{s}),H(\lambda)\}$,
$F(\mathsf{s})$ and $G(\mathsf{s})$ are positive and monotonically
increasing functions in the domain $\left[0,1\right]$ with asymptotic
minimum $F(0)=G(0)=1$. $H(\lambda)$ is positive in its domain $\left(0,\infty\right)$
and has a minimum at $H(1)=1$. Its asymptotic behavior is $H(\lambda\rightarrow\infty)\rightarrow\lambda\,3\pi/16$
and $H(\lambda\rightarrow0)\rightarrow\lambda^{-2}/4$. In Fig. \ref{fig:Shape-and-size}
we plot $F(\mathsf{s})$ and $G(\mathsf{s})$ adimensional functions.%

We have found a general structure of $Vb_{2}(pore)$ that is explainable
by a hierarchy of correction terms. Term $Vb_{2}$ is the homogeneous
component, it is linear in the volume and positive. The correction
to $Vb_{2}$ is the area term, the first signature of inhomogeneity.
The area term is negative and then opposite in sign to the homogeneous
term that corrects. Two types of essentially different corrections
to the area term were found they comes from the edges and the curved
area. The edge term which corrects the area term is negative and proportional
to $Le$. For right dihedral edges we found the value $-1/15$ for
the constant of proportionality, it appears for cuboidal and cylindrical
pores. The curved area term is a correction to the area term too,
and sometimes, it is independent of pore size parameters being a constant.
It is negative and approximately proportional to an extensive-like
quadratic curvature $A_{curv}j^{2}$. This term appears at cylindrical,
spherical, spherical+core and ellipsoidal pores. Noticeably, it does
not exist any extensive-like linear curvature term. Two terms which
correct the edge term were also found. They concerns an edge boundary
term and an edge curvature one. Both of them basically reproduce the
behavior of the corrections to the area term. These conclusions make
interesting the evaluation of several coefficients in other geometric
confinement, which may include, $\ell_{2}$ for the edge of an arbitrary
dihedral angle, $c_{2,1}$ for a general vertex produced by three
non-orthogonal surfaces, for the cone vertex, and the curvature correction
of the general edge.
\begin{description}
\item [{(b)}] The boundary of the large pore domain $\textrm{R1}\rightarrow\textrm{Ri}$\label{des:(b)-the-boundary}
\end{description}
In the rest of Sec. \ref{sec:Analysis} our main purpose is to study
the non-analytic behavior of CI when we go in the parameter space
from an analytic domain to a contiguous one. With this in mind, we
consider closed regions in the (Real) parameter space consisting in
a region of the analytic domain with its boundary. We introduce the
difference between the series representation of CIs, shortly $Z_{2}(\textrm{Ri})-Z_{2}(\textrm{Rj})$,
corresponding to contiguous regions and evaluated in the neighboring
of the common boundary. This may be not a well behaved magnitude.
Even that, when at least one of the CI can be analytically extended
in the contiguous region the difference $Z_{2}(\textrm{Ri})-Z_{2}(\textrm{Rj})$
is easily analyzed. More complex is the case where neither $Z_{2}(\textrm{Ri})$
nor $Z_{2}(\textrm{Rj})$ can be analytically extended in the domain
of the other. In such a case we made a careful comparison between
the coefficients in each series.

When we walk in the $\mathbf{X}$-space from R1 to its outside the
pore becomes unable to fit both particles for some fixed direction
$\hat{\mathbf{r}}_{12}$. For example, going from R1 to R3 in the
cylindrical pore becomes impossible that both particles locate in
a plane orthogonal to the central axis. The effect on the volume of
the available position phase space is not smooth enough producing
the non analytic behavior of CI. We find that the behavior of CI in
several paths of the type $\textrm{R1}\rightarrow\textrm{Ri}$ are
well described by\begin{equation}
Z_{2}(\textrm{R1})-Z_{2}(\textrm{Ri})\equiv\boldsymbol{\Delta}Z_{2}\approx-\Delta\,\varepsilon^{3}\:,\label{eq:step-b}\end{equation}
that is, for many situations we verify that CI has a discontinuous
third derivative when the large pore domain is crossed in the parameter
space. Here $\varepsilon=1-L_{i}/\sigma$, is an adimensional vanishing
parameter, $\varepsilon>0$ and $\Delta/6$ is the discontinuous step
in the third derivative in the path R1$\rightarrow$Ri. When ES exceeds
the planar regions of EB, i.e. R1$\rightarrow$R2, R1$\rightarrow$R3
and R1$\rightarrow$R4 for cuboidal pore; and R1$\rightarrow$R2 for
cylindrical pore, we obtain\begin{eqnarray}
\Delta(cub) & = & (2\pi\sigma^{4}/3)L_{y}L_{z}\,,\,\textrm{for R1}\rightarrow\textrm{R2}\:,\nonumber \\
 & = & (2\pi\sigma^{4}/3)\left(L_{y}L_{z}+L_{x}L_{z}\right)\,,\,\textrm{for R1}\rightarrow\textrm{R3}\:,\nonumber \\
 & = & (2\pi\sigma^{4}/3)\left(L_{y}L_{z}+L_{x}L_{z}+L_{x}L_{y}\right)\,,\,\textrm{for R1}\rightarrow\textrm{R4}\:,\nonumber \\
\Delta(cyl) & = & (2\pi^{2}\sigma^{4}R^{2}/3)\,,\,\textrm{for R1}\rightarrow\textrm{R2}\:,\label{eq:Dlt-cub-cyl}\end{eqnarray}
where each equation should be evaluated at $L_{i}\rightarrow\sigma$
consistent with the analyzed path. Here, the non-analyticity of $Z_{2}$
is a consequence of the limiting behavior of the functions $\{\mathcal{I}_{3x}(cub),\,\mathcal{I}_{3y}(cub),\,\mathcal{I}_{3z}(cub)\}$
and $\mathcal{I}_{3z}(cyl)$. Close to the boundary they behave \begin{eqnarray}
\mathcal{I}_{3x}(cub) & \approx & -\frac{2}{3}\pi\sigma L_{y}L_{z}\left(\sigma-L_{x}\right)^{3}=-b_{2}V\,\left(1-L_{x}/\sigma\right)^{3}\,,\,\textrm{for R1}\rightarrow\textrm{R2}\:,\nonumber \\
\mathcal{I}_{3z}(cyl) & \approx & -\frac{2}{3}\pi^{2}\sigma R^{2}\left(\sigma-L_{h}\right)^{3}=-b_{2}V\,\left(1-L_{h}/\sigma\right)^{3}\,,\,\textrm{for R1}\rightarrow\textrm{R2}\:,\label{eq:Dlt-I3}\end{eqnarray}
where $L_{x}\rightarrow\sigma$ and $L_{h}\rightarrow\sigma$ for
cuboidal and cylindrical pores, respectively. The Eqs. (\ref{eq:Dlt-cub-cyl},
\ref{eq:Dlt-I3}) may be accomplished with\begin{equation}
\Delta=\frac{\pi\sigma^{4}}{3}\, A_{+}\:,\label{eq:Dlt01}\end{equation}
\begin{equation}
\mathcal{I}_{3i}\approx-b_{2}\frac{\sigma}{2}\, A_{+}\left(1-L_{i}/\sigma\right)^{3}\:,\label{eq:Dlt-I3bis}\end{equation}
where $A_{+}$ is the total area of such cavity boundaries which can
not contain a sphere with $\sigma$ diameter. The same procedure is
feasible for non planar boundaries, R1$\rightarrow$0D in the spherical
pore, R1$\rightarrow$R2 in the sph+core pore, R1$\rightarrow$R2
and R1$\overset{*}{\rightarrow}$R3 in the spheroidal pore, and R1$\overset{*}{\rightarrow}$R3
in cylindrical pore. Taking $\varepsilon=1-2R/\sigma$ we obtain\begin{equation}
\Delta(sph)=\frac{\pi\sigma^{4}}{3}A\,,\,\textrm{for R1}\rightarrow\textrm{0D}\:,\label{eq:Dlt-sph}\end{equation}
which must be evaluated at $R=\sigma/2$. The sph+core involves two
non planar walls with different curvatures, the external spherical
wall has radius $R$ while the internal wall has radius $R_{in}$.
Both spherical walls are apart $L_{h}=R-R_{in}$. The gap in the third
derivative with $\varepsilon=1-L_{h}/\sigma$ is now\begin{equation}
\Delta(sph+core)|_{R}=\frac{\pi\sigma^{4}}{3}\,\left(A_{+}-4\pi\sigma^{2}\right)\,,\,\textrm{for R1}\rightarrow\textrm{R2}\:,\label{eq:Dlt-Sph+c}\end{equation}
where $A_{+}=4\pi(R^{2}+R_{in}^{2})$ is the total area. 

We find three situations with a different behavior, they does not
involve a finite discontinuity in the third derivative. The path R1$\rightarrow$R2
for the oblate-spheroidal pore has a discontinuous fourth derivative.
For $\varepsilon=1-\frac{2R_{c}}{\sigma}>0$ we have\begin{equation}
\boldsymbol{\Delta}Z_{2}(sphrd)\approx\frac{\pi^{2}\sigma^{6}}{12\lambda^{2}(1-\lambda^{2})}\,\varepsilon^{4}\,,\,\textrm{for R1}\rightarrow\textrm{R2}\:.\label{eq:I3z-sphrd}\end{equation}
The path $\textrm{R1}\overset{*}{\rightarrow}\textrm{R3}$ involves
an ergodicity breaking in prolate-spheroid and cylindrical, pores.
Neglecting the factor $\xi$, for $\varepsilon=1-\frac{2R}{\sigma}>0$
we obtain for the prolate-spheroid pore\begin{equation}
\boldsymbol{\Delta}Z_{2}(sphrd)\approx\frac{16\sqrt{2}\pi^{2}\lambda^{2}\sigma^{6}}{105\sqrt{-1+\lambda^{2}}}\,\varepsilon^{7/2}\,,\,\textrm{for R1}\overset{*}{\rightarrow}\textrm{R3}\:.\label{eq:Dlt-I3r-sphrd}\end{equation}
We recognize that $\Delta(cyl)$ is somewhat ill defined cause their
third lateral derivatives respect to $2R$ diverge logarithmically
to minus infinity. Even so, the difference between them becomes null.
For $\varepsilon=1-\frac{2R}{\sigma}>1$ we obtain a non-analyticity
expressible by the limiting behavior\begin{eqnarray}
\boldsymbol{\Delta}Z_{2}(cyl) & \approx & -\frac{32\,\sqrt{2}\pi\sigma^{6}}{105}\,\varepsilon^{7/2}\,,\,\textrm{for R1}\overset{*}{\rightarrow}\textrm{R3}\:.\label{eq:Dlt-I3r-cyl}\end{eqnarray}
Finally, the path $\textrm{R1}\overset{*}{\rightarrow}\textrm{R4}$
in the cylindrical pore is analyzed by a superposition of results
from Eqs. (\ref{eq:Dlt-cub-cyl}, \ref{eq:Dlt-I3r-cyl}). Its behavior
is similar to that found in path $\textrm{R1}\rightarrow\textrm{R2}$.
\begin{description}
\item [{(c)}] The path $\textrm{Ri}\overset{*}{\rightarrow}\textrm{Rj}$,
a signature of the ergodicity breaking \label{des:(c)-the-Ri*Rj}
\end{description}
The rational power in Eqs. (\ref{eq:Dlt-I3r-sphrd}, \ref{eq:Dlt-I3r-cyl})
corresponds to path with ergodicity breaking, thus, we wish to study
their characteristics. A third path with this behavior is $\textrm{R2}\overset{*}{\rightarrow}\textrm{R4}$
for cylindrical pore. Again, neglecting the $\xi=1/2$ factor we obtain
the result described in Eq. (\ref{eq:Dlt-I3r-cyl}), based on the
unanalicities of $\mathcal{I}_{3r}$. The cuboidal pore has also several
paths of this type. They are the paths $\textrm{R3}\overset{*}{\rightarrow}\textrm{R5}$,
$\textrm{R4}\overset{*}{\rightarrow}\textrm{R6}$, $\textrm{R4}\overset{*}{\rightarrow}\textrm{R7}$,$\textrm{R6}\overset{*}{\rightarrow}\textrm{R8}$,
$\textrm{R4}\overset{*}{\rightarrow}\textrm{R8}$, and $\textrm{R7}\overset{*}{\rightarrow}\textrm{R8}$.
All of them are characterized by the fact that a sphere with $\sigma$
radius fixed in the center of the cavity cross its edges. In fact,
this condition is equivalent to that described above for such a cavity
(see Sec. \ref{sub:CI-2HS-CUB}, Region 5). Here the partition functions
have an infinite discontinuous fifth derivatives as a consequence
of the analytic behavior of the family of functions $\{\mathcal{I}_{3xy},\,\mathcal{I}_{3xz},\,\mathcal{I}_{3yz}\}$\begin{eqnarray}
\mathcal{I}_{3xy}(cub) & \approx & \frac{8}{9!!}\frac{L_{z}}{L_{x}^{2}L_{y}^{2}}\left(\sigma^{2}-L_{x}^{2}-L_{y}^{2}\right)^{9/2}\:,\nonumber \\
 & \approx & \frac{2^{11/2}}{9!!}\sigma^{5}(\alpha+\alpha^{-1})^{2}\, L_{+}\,\left(1-\frac{\sqrt{L_{x}^{2}+L_{y}^{2}}}{\sigma}\right)^{9/2}\,,\,\textrm{for R3}\overset{*}{\rightarrow}\textrm{R5}\:,\label{eq:Dlt-I3xy}\end{eqnarray}
where $\alpha=L_{x}/L_{y}$ and $L_{+}$ is the total length of the
crossed right edges, i.e. in Eq. (\ref{eq:Dlt-I3xy}) $L_{+}=4L_{z}$.
Other paths are suitable analyzed by applying this result to the set
$\{\mathcal{I}_{3xy},\,\mathcal{I}_{3xz},\,\mathcal{I}_{3yz}\}$.
The path $\textrm{R4}\overset{*}{\rightarrow}\textrm{R6}$ is completely
equivalent to $\textrm{R3}\overset{*}{\rightarrow}\textrm{R5}$. Somewhat
different are the paths $\textrm{R4}\overset{*}{\rightarrow}\textrm{R7}$,
$\textrm{R6}\overset{*}{\rightarrow}\textrm{R8}$, $\textrm{R7}\overset{*}{\rightarrow}\textrm{R8}$,
and $\textrm{R4}\overset{*}{\rightarrow}\textrm{R8}$, which involves
an ergodicity breaking along with a spontaneous symmetry breaking.
Even, their analytic behavior is basically described by Eq. (\ref{eq:Dlt-I3xy}).
The path $\textrm{R7}\overset{*}{\rightarrow}\textrm{R8}$ is similar
to $\textrm{R3}\overset{*}{\rightarrow}\textrm{R5}$ with the replacement
$y\leftrightarrow z$. Paths $\textrm{R4}\overset{*}{\rightarrow}\textrm{R7}$
and $\textrm{R6}\overset{*}{\rightarrow}\textrm{R8}$ have two equal
terms with the same value of $\alpha$, the addition of both terms
makes a unique contribution identical to Eq. (\ref{eq:Dlt-I3xy})
with $L_{+}$ the total length of the four crossed edges. Last path,
$\textrm{R4}\overset{*}{\rightarrow}\textrm{R8}$ involves three terms
with $\alpha=1$, which resumes on one term with total edges length
$L_{+}=L=12\sigma/\sqrt{2}$. It is interesting to note that a similar
situation is also possible for the cylindrical pore, where the circular
edges are crossed by the sphere. It corresponds to the path $\textrm{R4}\rightarrow\textrm{0D}$
which will be studied below.
\begin{description}
\item [{(d)}] The $\textrm{Rj}\rightarrow\textrm{2D}$ limit\label{des:(d)-the-2D}
\end{description}
The equivalent of the HS system in two dimensions is the Hard Disk
system (HD). In the 2D-limit we may expect that 2-HS systems collapse
to a 2-HD system. Then, $Z_{2}$ should collapse to $Z_{2,\textrm{HD}}$
and then the CI of 2-HS in the cuboidal pore transforms to the CI
of 2-HD in a box and so on. Expressions of $Z_{2,\textrm{HD}}$ for
particles constrained in a rectangular or a circular pores, as well
as, on the surface of a sphere are well known \cite{Munakata_2002,Urrutia_2008,Urrutia_2010};
this fact allow us check several results in PW. The expected limiting
behavior of $Z_{2}$ in terms of the vanishing length parameter $\varepsilon$
is \begin{equation}
Z_{2}=\varepsilon^{2}\left[Z_{2,\textrm{HD}}+\varepsilon^{q}\, Z_{2,\textrm{HD}\, lim}\right]+O_{3+q}(\varepsilon)\:,\label{eq:lim2D}\end{equation}
where $\varepsilon=L_{x}$ and $\varepsilon=L_{h}$ for cuboidal and
cylindrical cavities, respectively. Hence, we may study the unknown
term $\varepsilon^{q}\, Z_{2,\textrm{HD}\, lim}$. For the planar
surface 2D-limit we obtain $q=2$, being for cuboidal shape\begin{eqnarray}
Z_{2,\textrm{HD}\, lim}(cub) & = & \frac{1}{6}\left[\pi L_{y}L_{z}-2\sigma(L_{y}+L_{z})+\sigma^{2}\right]\,,\,\textrm{for R2}\rightarrow\textrm{2D}\:,\nonumber \\
 & = & \frac{1}{3}\left[L_{z}\,\sqrt{\sigma^{2}-L_{y}^{2}}+L_{y}L_{z}\arcsin(L_{y}/\sigma)-\frac{1}{2}L_{y}^{2}-\sigma L_{z}\right]\,,\,\textrm{for R5}\rightarrow\textrm{2D}\:,\nonumber \\
 & = & \frac{1}{3}\left[L_{y}\,\sqrt{\sigma^{2}-L_{z}^{2}}+L_{z}\,\sqrt{\sigma^{2}-L_{y}^{2}}-\frac{1}{2}\left(\sigma^{2}+L_{y}^{2}+L_{z}^{2}\right)\right.\nonumber \\
 &  & \left.+L_{y}L_{z}\left(\arcsin(L_{z}/\sigma)+\arcsin(L_{y}/\sigma)+\pi/2\right)\right]\,,\,\textrm{for R7}\rightarrow\textrm{2D}\:,\label{eq:lim2d-cub}\end{eqnarray}
and for a cylindrical shape\begin{equation}
Z_{2,\textrm{HD}\, lim}(cyl)=\frac{4\pi}{135}\left[\left(\mathsf{R}^{2}-1\right)^{-1/2}\left(32\mathsf{R}^{4}-157\mathsf{R}^{2}-3\right)+45\mathsf{R}^{2}\,\mathrm{arcsec}\left(\mathsf{R}\right)\right]\,,\,\textrm{for R2}\rightarrow\textrm{2D}\:.\label{eq:lim2D-tapcyl}\end{equation}
In the case of a 2D-limit involving a curved surface confinement,
we obtain for the spherical+core pore $q=1$ and\begin{equation}
Z_{2,\textrm{HD}\, lim}(sph+core)=4\pi^{2}R\,[\sigma^{2}-8R^{2}]\,,\,\textrm{for R2}\rightarrow\textrm{2D}\:,\label{eq:lim2D-curv}\end{equation}
where $\varepsilon=L_{h}$. In the 2D limit of the oblate spheroidal
pore $\textrm{R2}\rightarrow\textrm{2D}$ we do not find the behavior
depicted by Eq. (\ref{eq:lim2D}).
\begin{description}
\item [{(e)}] The $\textrm{Rj}\rightarrow\textrm{1D}$ limit\label{des:(e)-the-1D}
\end{description}
The path going from R1 to the 1D-limit has an ending structure $\textrm{Ri}\overset{*}{\rightarrow}\textrm{Rj}\rightarrow\textrm{1D}$.
It means that, before to reach the limiting behavior a characteristic
ergodic-non-ergodic transition appears. Once both particles are not
able to interchange their positions the path $\textrm{Rj}\rightarrow\textrm{1D}$
can happen and the final 1D-limit may be attained. In that limit the
HS behaves like Hard Rods (HR) and $Z_{2}$ collapses to $Z_{2,\textrm{HR}}$.
The limiting behavior for $Z_{2}$ written in terms of the vanishing
length parameter $\varepsilon$ ($\varepsilon^{2}=L_{x}L_{y}$ for
a cuboid and $\varepsilon^{2}=\pi R^{2}$ for a cylinder) is \begin{equation}
Z_{2}\approx\varepsilon^{4}\left[Z_{2,\textrm{HR}}+\varepsilon^{q}\, Z_{2,\textrm{HR}\, lim}\right]\:.\label{eq:lim1D}\end{equation}
For the cuboidal pore $Z_{2,\textrm{HR}}=(L_{z}-\sigma/2)^{2}$, $q=2$,
and\begin{equation}
Z_{2,\textrm{HR}\, lim}(cub)=\frac{1}{6\sigma}(L_{z}-\sigma)\,(\alpha+\alpha^{-1})\,,\,\textrm{for R5}\rightarrow\textrm{1D}\:,\label{eq:lim1D-cub}\end{equation}
 being $\alpha=L_{y}/L_{x}$. For the cylindrical cavity we obtain
$Z_{2,\textrm{HR}}(cyl)=(L_{h}-\sigma/2)^{2}$, $q=2$, and \begin{equation}
Z_{2,\textrm{HR}\, lim}(cyl)=\frac{1}{\pi\sigma}(L_{h}-\sigma)\,,\,\textrm{for R3}\rightarrow\textrm{1D}\:.\label{eq:lim1D-cyl}\end{equation}
In addition, we may compare with the 1D-limit taken from the two dimensional
2-HD system confined into a rectangle, and from the 2-HD system confined
between two concentric circles, from Refs. \cite{Munakata_2002,Urrutia_2008}.
The 1D-limit for the 2D rectangular confinement produces $q=2$, while
the circular pore with a hard core shows $q=1$. We conclude that
the power $q=2$ is characteristic of straight line 1D-limit while
$q=1$ corresponds to curved-closed-line 1D-limit. The prolate spheroidal
pore does not behave in accordance with Eq. (\ref{eq:lim1D}).
\begin{description}
\item [{(f)}] The $\textrm{Rj}\rightarrow\textrm{0D}$ limit\label{des:(f)-the-0D}
\end{description}
The final state obtained in this limit consists of particles that
cages in a final solid or densest configuration. This densest state
of 2-HS characterizes by the complete spatial correlation of particles.
Two different paths coming from R1 and ending at the 0D-limit may
be identified, they have the structures $\textrm{Ri}\overset{*}{\rightarrow}\textrm{Rj}\rightarrow\textrm{0D}$
and $\textrm{Ri}\rightarrow\textrm{Rj}\rightarrow\textrm{0D}$. The
first case includes an ergodic-non-ergodic transition and sometimes
also includes a symmetry breaking transition, it happens for the cuboid
pore. We find that, in the 0D-limit the phase space of positions (PSP)
may collapse to three topologically different manifolds. For a cuboidal
cavity the 0D-limit shows a collapse of the PSP in a 0D-manifold,
i.e. a single point. Thus, the most compact state is a solid-like
state. For the cylindrical cavity in the 0D-limit the PSP collapse
to a 1D-manifold consisting in a simple closed line also called a
circle. Here the densest state is a rigid body which is able to rotate
with a fixed axis. For the spherical cavity the 0D-limit shows that
PSP collapse to a 2D-manifold given essentially by a spherical surface.
Therefore the densest state behaves as a freely rotating rigid body.
In the last two cases, even in the 0D-limit, particles can interchange
their positions. In general, the limiting behavior of $Z_{2}$ in
terms of some vanishing adimensional parameter $\varepsilon$ is $Z_{2}\propto\varepsilon^{q}$.
For the cuboidal cavity with $L=L_{x}=L_{y}=\alpha\, L_{z}$ and $\varepsilon=\sqrt{2+\alpha^{2}}\, L/\sigma-1$
we obtain $q=6$ and\begin{equation}
Z_{2}(cub)=\frac{(2+\alpha^{2})^{3}}{90\alpha^{2}}\sigma^{6}\varepsilon^{6}+\frac{(2+\alpha^{2})^{3}(1-\alpha^{-2}+\alpha^{-4})}{105}\sigma^{6}\varepsilon^{7}\,,\,\textrm{for R8}\rightarrow\textrm{0D}\:.\label{eq:lim0D-cub}\end{equation}
We note that $q=6$ also in the case of a general cuboid. Analyzing
the cylindrical geometry we find $q=9/2$, $L=L_{h}=\alpha\,2R$,
$\varepsilon=2R\sqrt{1+\alpha^{2}}/\sigma-1$, and\begin{eqnarray}
Z_{2}(cyl) & = & \pi\sqrt{2}\,(1+\alpha^{2})^{3/2}127575^{-1}\sigma^{6}\varepsilon^{9/2}\times\left[28350\alpha^{-6}+\right.\:,\nonumber \\
 &  & \left.84105\alpha^{-4}+98100\alpha^{-2}+40200+6688\alpha^{2}+192\alpha^{4}\right]\,,\,\textrm{for R4}\rightarrow\textrm{0D}\:.\label{eq:lim0D-cyl}\end{eqnarray}
For the spheroid, we can attain the 0D limit in two different ways,
by seeking the paths $\textrm{R2}\rightarrow\textrm{0D}$ and $\textrm{R3}\rightarrow\textrm{0D}$.
We obtain, $q=7/2$, $\varepsilon=2R/\sigma-1>0$, and\begin{equation}
Z_{2}(sphd)\approx\frac{16\sqrt{2}\pi^{2}\lambda^{2}\sigma^{6}}{105\sqrt{1-\lambda^{2}}}\,\varepsilon^{7/2}\,,\,\textrm{for R2}\rightarrow\textrm{0D}\:,\label{eq:lim0D-sphrd1}\end{equation}
and also, $q=4$, $\varepsilon=2R_{c}/\sigma-1>0$, and\begin{equation}
Z_{2}(sphd)\approx\frac{\pi^{2}\sigma^{6}}{12\lambda^{2}(-1+\lambda^{2})}\,\varepsilon^{4}\,,\,\textrm{for R3}\rightarrow\textrm{0D}\:.\label{eq:lim0D-sphrd2}\end{equation}
The 0D-limit in the spherical pore was previously studied in Ref.
\cite{Urrutia_2008}. In that work, it was found $q=3$ and $\varepsilon=(2R/\sigma-1)$.
Also, the 0D-limit of a 2D system composed by 2-HD in a circular cavity
has the same $\varepsilon$ but $q=5/2$.

We are now able to extract some minimal conclusions from this section.
Based on the analysis made in (a) we note a very general decomposition
of $Vb_{2}(pore)$ in terms of basic geometric magnitudes that characterize
the effective cavity. This decomposition could be applied in other
confinement geometries. From (b) we find a common non-analytic behavior
of $Z_{2}$ when the ES exceeds planar regions of the EB boundary.
It consists in a finite discontinuity at the third derivative with
a step proportional to the surface area of the crossed planes. We
also obtain a similar behavior for spherical surfaces and discontinuities
at higher order derivatives in other curved surfaces. In general we
observe that the paths between analytic domains involving ergodic-non-ergodic
transitions $\textrm{Ri}\overset{*}{\rightarrow}\textrm{Rj}$ are
consistent with a CI, which scales with fractional powers of the vanishing
magnitude. It is apparent in (b) where we find that a $7/2$ power
appears when ES exceeds a curved wall of the EB, and also, from (c)
and (f) (see Eqs. (\ref{eq:Dlt-I3xy}, \ref{eq:lim0D-cyl})) where
we obtain a common non-analytic behavior of $Z_{2}$ when ES exceeds
the right angle edges of the EB boundary given by a common power dependence
of $9/2$ in the vanishing length.

A general picture of the dimensional crossovers agrees with the description
given in \cite{Urrutia_2010}. Given a $N$-HS fluid system in a region
of the $D$-dimensional space, the number of total spatial (i.e. translational)
degrees of freedom is DF$=N\cdot D$. When we consider a limiting
process of dimensional crossover the dimension of the available space
reduce to $D\lyxmathsym{\textasciiacute}$ with $0\leq D\lyxmathsym{\textasciiacute}<D$.
We define the number of lost degrees of freedom (LDF) as the power
of the vanishing magnitude in the CI in the dimensional cross-over
limit. We claim that LDF$=N\lyxmathsym{\textasciiacute}\cdot(D-D\lyxmathsym{\textasciiacute})$
where $N\lyxmathsym{\textasciiacute}$ is the number of particles
constrained to the $D\lyxmathsym{\textasciiacute}$ dimensional region
being usually $N\lyxmathsym{\textasciiacute}=N$. One exception to
this rule is the 0D limit when the final densest state consists in
a rotating $N\lyxmathsym{\textasciiacute}$-particle rigid-like system.
In such a case we find LDF$=N\lyxmathsym{\textasciiacute}D-n\,3/2$,
with $n$ indicating the number of independent degrees of rotational
freedom for the caged $N\lyxmathsym{\textasciiacute}$ particles,
being $0\leq n\leq D$ \cite{Urrutia_2010}. In a unified description,
for any dimensional cross-over we obtain\begin{equation}
\textrm{LDF}=N\lyxmathsym{\textasciiacute}\cdot(D-D\lyxmathsym{\textasciiacute})-n\,3/2\:,\label{eq:LDF}\end{equation}
where $n=0$ if $D\lyxmathsym{\textasciiacute}\neq0$. Here, first
term counts the lost of translational degrees of freedom while the
second one compensates for the non-vanishing pure rotational degrees
of freedom. For PW we must fix $N=N\lyxmathsym{\textasciiacute}=2$
with a starting value of $D=3$, and analize possible values $D\lyxmathsym{\textasciiacute}=0,1,2$.
In the zero dimensional limit the 2-HS collapses to a dumbbell or
stick. Thus, $n=0$ is a non-rotating stick, $n=1$ corresponds to
a rotating stick with fixed rotation axis, and $n=2$ is a freely
rotating stick. Systems of two particles have a maximum value $n=D-1$.
Several sequences of dimensional crossovers described by Eq. (\ref{eq:LDF})
are accessible from the results exposed in PW. For example, in the
cylindrical cavity the path $\textrm{R2}\rightarrow\textrm{2D}$ involving
LDF$=2$ can be followed by a 0D-limit with LDF$=5/2$, obtained with
$D=2$, $D\lyxmathsym{\textasciiacute}=0$ and $n=1$.

\section{Thermodynamic Properties\label{sec:Thermodynamic-study}}

The aim of this section is to achieve the thermodynamic behavior of
few bodies confined systems. Along this section we use the word thermodynamic
in the sense of thermodynamic of fluids, where a fluid is a system
of particles allowed to move in a given region of the continuous space.
Our objective is to find the EOS that describe the global properties
of a few body fluid system. In order to accomplish such a goal the
discussion will be oriented towards the few and many HS system confined
in a hard wall cavity with no restriction in the number of particles.
In addition, we will keep in mind a system in a fluid-like state.
Besides these statements other systems could be included in the discussion
without much effort, such as open systems and soft interactions. Again,
we must emphasize that a few body system is far away from the thermodynamic
limit $N\rightarrow\infty$. Therefore, the thermodynamic description
developed below does not concerns to such limit. In a few body system
its different ensemble representations are not equivalent each other.
Thus, we assume that the system under interest is well described by
a certain Gibbsian ensemble and analyze the properties of this ensemble
representation. From our point of view, we obtain the EOS of the system
if we know the basic relations between the mean-ensemble values of
the thermodynamic relevant magnitudes. A rigorous discussion about
the equivalence between some mean-ensemble thermodynamic property
e.g. $U$ and the time average value $U_{\tau}$ is out of the scope
of PW. Even that, we can draw a general picture. We expect that for
cavity's size in the ergodic regime and far from an ergodic-non-ergodic
transition $U=U_{\tau}$ for times $\tau$ moderately short. For example,
in a cylindrical pore it should apply in R1 and R2, but far enough
from R3 and R4 (see Fig. \ref{fig:Shape-and-size}). In case that
the size of the cavity approaches an ergodic-non-ergodic transition
the identity $U=U_{\tau}$ only applies for increasing values of $\tau$.
For cavities with sizes in the ergodicity breaking regime $U$ and
$U_{\tau}$ may be different (e.g. R3 and R4 in the cylindrical cavity).
Next paragraphs are devoted to a general discussion about the thermodynamic
description of few body systems, while at the end of this section
we analyze the thermodynamic behavior of confined 2-HS systems in
the canonical ensemble representation making a comparison between
different shaped cavities.

The pertinence of the thermodynamic theories to small systems was
recognized by several authors, see e.g. the book of Hill \cite{Hill94}.
From this book we can extract several arguments about the relevance
of small systems to statistical mechanics and thermodynamics, and
also, we find an interesting discussion about the particularities
of the thermodynamics of small systems. Although, the central thesis
of Hill is that the \emph{macroscopic} thermodynamics must be adapted
to extend its range of validity to include small systems. His thermodynamic
approach begins with large (infinitely extended) systems and drops
to the small ones. Certainly, we adopt an opposite point of view.
We state that the first law of thermodynamics concerns to few body
systems, provided that, any assumption about the extensivity of the
energy and entropy must be avoided.

An implicit hypothesis of thermodynamics is that the equilibrium states
of a large class of fluid systems may be specified with a unique small
set of independent macroscopic quantities. A trivial example is the
class of simple homogeneous fluids usually studied by taking three
independent macroscopic magnitudes (see e.g. Callen's thermodynamics
book \cite{Callen85} pp. 13 and 283). Therefore, we say that thermodynamics
should have the \textit{Simplicity and Universality} (SU) attributes.
Usually, the studied systems involve a large number of particles,
but does not exist a minimum cutoff in this quantity. To highlight
this point, we note that in the statistical mechanics literature the
grand canonical partition function is defined by a weighted sum of
canonical partition functions over the available number of particles
in the system (see e.g. \cite{Hill56}). This sum starts from zero,
following by one, two particles, and goes usually up to infinity.
Therefore, systems with few bodies are included in the usual formulation
of the statistical mechanics. We also note that usual relations that
link statistical mechanics of partition functions and thermodynamic
magnitudes do not make any assumption about the number of particles.
This fact supports the idea that the same relations apply to systems
with few bodies. Still, any assumption of extensivity in magnitudes
like the energy, entropy, and free energies must be rejected in a
few bodies system (see e.g. \cite{Callen85} pp. 360). We understand
the thermodynamic pertinence of systems with many and few bodies as
the \textit{Size Invariance} (SI) of thermodynamics. Based on SU and
SI, we argue that a consistent thermodynamic treatment of systems
with large, many, and few number of particles should be possible using
a basic small set of independent macroscopic quantities. Naturally,
we will call to this the SUSI hypothesis.

We want to bring attention to an unsolved problem in equilibrium Statistical
Mechanics. At first sight it might be surprising that even when we
may know the exact partition function of an inhomogeneous fluid system,
their thermodynamic properties appear unrevealed. Our knowledge about
the partition function comes from the exact evaluation of an integral
(see paragraph above Eq. (\ref{eq:Q0})). As far as, the integrand
and the limits of evaluation are functions of some set of \textit{independent}
parameters $\mathbf{X}$, therefore by solving the integral we merely
obtain $Q(\mathbf{X})$. For a HS system in a hard wall cavity at
constant temperature, the discussion is mainly focused on $Z(\mathbf{X})$,
where $\mathbf{X}$ can be of geometrical nature and usually involves
proper lengths of the cavity, e.g. in a cuboidal pore $\mathbf{X}=\left\{ L_{x},\, L_{y},\, L_{z}\right\} $.
Let us suppose that, for a given $\mathbf{X}$ space with dimension
$dim(\mathbf{X})$ the canonical partition function $Q(\mathbf{X})$
for the N particles system is known within a reduced domain $\mathbb{H}$.
In such a domain we may obtain the Helmholtz free energy\begin{equation}
\beta F(\mathbf{X})=-\ln[Q(\mathbf{X})]\:,\label{eq:Th00}\end{equation}
which is related to other thermodynamic quantities by\begin{equation}
\mu=F(\mathbf{X})-F_{-}(\mathbf{X})\:,\label{eq:Th01}\end{equation}
\begin{equation}
F=U-TS\:,\label{eq:Th02}\end{equation}
\begin{equation}
dF=-S\, dT-dw\:.\label{eq:Th03}\end{equation}
In Eq. (\ref{eq:Th01}) the evaluation of the chemical potential $\mu$
assumes that the partition function for the system with N-1 particles,
$Q_{-}(\mathbf{X})$ (with Helmholtz free energy $F_{-}$) is also
known in $\mathbb{H}$. $U$, $S$, and $T$ are the energy, entropy,
and absolute temperature of the system, respectively. Lastly, $dw$
is the differential of reversible work done by the system. Eq. (\ref{eq:Th03})
shows how $F$ depends on both, $T$ and $\mathbf{X}$. The temperature
dependence gives the entropy $S$ \begin{equation}
S=-\left.\frac{\partial F}{\partial T}\right|_{\mathbf{X}}\:,\label{eq:Th04}\end{equation}
while the $\mathbf{X}$ derivative at constant $T$ is related to
the work. Let us consider two different equilibrium states $a$ and
$b$, characterized by parameters $\mathbf{X}_{a}$ and $\mathbf{X}_{b}$,
respectively. The variations $\Delta F$, $\Delta S$, $\Delta U$
in going from state $a$ to state $b$ at fixed temperature are easily
evaluated with the help of Eqs. (\ref{eq:Th01}, \ref{eq:Th02}, \ref{eq:Th04}).
We may also evaluate the reversible work $w_{ab}$ in going from $a$
to $b$\begin{equation}
w_{ab}=-\intop_{a}^{b}\nabla_{\mathbf{X}}F\cdot d\mathbf{X}=F(\mathbf{X}_{a})-F(\mathbf{X}_{b})\:,\label{eq:w_p}\end{equation}
where $\nabla_{\mathbf{X}}$ is the gradient operator with respect
to $\mathbf{X}$ parameters taken at constant $T$, and the line integral
in Eq. (\ref{eq:w_p}) does not depend on the path adopted between
$a$ and $b$. From here on, we implicitly make the same assumption
for any derivative with respect to $\mathbf{X}$. The Eq. (\ref{eq:w_p})
enable us to define the differential of reversible work \begin{equation}
dw=dw_{a\hat{\mathbf{X}}}=-\partial_{\hat{\mathbf{X}}}F\cdot dl=-\nabla_{\mathbf{X}}F\cdot\hat{\mathbf{X}}\, dl\:,\label{eq:dw_p}\end{equation}
being $\hat{\mathbf{X}}$ some unit vector in the parameter space,
$dw_{a\hat{\mathbf{X}}}$ the work to make a differential reversible
change from $\mathbf{X}_{a}$ to $\mathbf{X}_{b}=\mathbf{X}_{a}+\hat{\mathbf{X}}\, dl$,
and $\partial_{\hat{\mathbf{X}}}$ the directional derivative. Given
\textit{any volume notion} $\mathcal{V}$, which may or may not be
defined in the spirit of SUSI, we can define the overall pressure
or pressure-for-work $\bar{P}_{w,\,\hat{\mathbf{X}}}$ for an infinitesimal
transformation of the cavity \begin{equation}
\bar{P}_{w,\,\hat{\mathbf{X}}}\equiv-\frac{\nabla_{\mathbf{X}}F\cdot\hat{\mathbf{X}}}{\nabla_{\mathbf{X}}\mathcal{V}\cdot\hat{\mathbf{X}}}\:,\label{eq:Pw-0}\end{equation}
which makes sense only if $\nabla_{\mathbf{X}}\mathcal{V}\cdot\hat{\mathbf{X}}\neq0$.
For an infinitesimal transformation at constant volume we should ignore
Eq. (\ref{eq:Pw-0}). Even, we may prefer to introduce some \textit{surface
area notion} $\mathcal{A}$ and therefore we can define an external
surface tension or surface-tension-for-work by\begin{equation}
\bar{\gamma}_{w,\,\hat{\mathbf{X}}}\equiv\frac{\nabla_{\mathbf{X}}F\cdot\hat{\mathbf{X}}}{\nabla_{\mathbf{X}}\mathcal{A}\cdot\hat{\mathbf{X}}}\:.\label{eq:Gammaw-0}\end{equation}
Eqs. (\ref{eq:Pw-0}) or (\ref{eq:Gammaw-0}) are indeed physical
conventions, and therefore, we could describe the total work as it
would be produced by either an effective pressure or a surface tension.
From now on we assume that $\nabla_{\mathbf{X}}\mathcal{V}\cdot\hat{\mathbf{X}}\neq0$.
Then, the definition (\ref{eq:Pw-0}) is consistent with Eq. (\ref{eq:dw_p}),
which now reads\begin{equation}
dw_{\hat{\mathbf{X}}}=\bar{P}_{w,\,\hat{\mathbf{X}}}d\mathcal{V}_{\hat{\mathbf{X}}}\:,\label{eq:dw_PwdV-0}\end{equation}
 where $d\mathcal{V}_{\hat{\mathbf{X}}}=\nabla_{\mathbf{X}}\mathcal{V}\cdot\hat{\mathbf{X}}\, dl$.
The definition of $\bar{P}_{w,\,\hat{\mathbf{X}}}$ \textit{requires}
the introduction of a volume notion $\mathcal{V}(\mathbf{X})$. Hence,
pressure depends on both the adopted $\mathcal{V}$ and $\hat{\mathbf{X}}$.
On the opposite, even when the choice of a different $\mathcal{V}$
modifies $\bar{P}_{w,\,\hat{\mathbf{X}}}$ it does not influence $dw_{a\hat{\mathbf{X}}}$.

At this point we emphasize that, even when the above description is
exact it is not completely satisfactory. It says little about the
thermodynamic properties of the fluid inside the cavity. It depends
on $\mathbf{X}$ parameters, which do not have a universal thermodynamic
meaning. The parameters needed to describe the shape of certain cavity
are of different kind and quantity that those needed to describe other
shapes. Even worst, for a given geometry they are non unique. We may
extract some examples from the studied two particle systems. For a
cavity with spherical symmetry we may utilize $\mathbf{X}=\left\{ R+\sigma/2\right\} $
or $\mathbf{X}=\left\{ R\right\} $, but also, we may adopt $\mathbf{X}=\left\{ \pi R^{2}\right\} $
all of them with $dim(\mathbf{X})=1$. In a cuboidal cavity we may
adopt $\mathbf{X}=\left\{ L_{x},\, L_{y},\, L_{z}\right\} $ or $\mathbf{X}=\left\{ l_{x},\, l_{y},\, l_{z}\right\} $
with $dim(\mathbf{X})=3$, but also, if we are interested in $a$
and $b$ states with cubic symmetry we may choose $\mathbf{X}=\left\{ L\right\} $
with $dim(\mathbf{X})=1$. However, a somewhat more realistic cavity
model may be adopted in which the substrate atoms, HS at fixed positions,
are the building blocks of the rough confinement walls. In this case
$dim(\mathbf{X})$ could be a much larger number. In addition, the
$\mathbf{X}$-representation prevents to compare results from dissimilar
confinement conditions. Hence, the same fluid in a spherical or cuboidal
cavity produces results which inhibit any comparison between them. 

We conclude that next step forward in the thermodynamic description
of the system is out of the scope of the $\mathbf{X}$-representation.
Therefore, it is necessary to build the path between the $\mathbf{X}$-representation
of certain thermodynamic property, e.g. $Q(\mathbf{X})$, and a universal
description. Two basic questions have guided to us in the search of
such a path; i) What properties of the confined systems should depend
on the shape of the cavity? ii) What properties should depend on the
particular choice of adopted parameters $\mathbf{X}$? The rest of
this section shows some answers, which arise from our inquiries.

Being $\mathbf{X}$ an unsuitable set of parameters we must look
for a better choice. At this point we wish to extract a paragraph
from Callen's Themodynamics book, ''It should perhaps be noted that
the choice of the variables in terms of which a given problem is formulated,
while a seemingly innocuous step, is often the most crucial step in
the solution.'' (\cite{Callen85} p. 465). The interesting point
is that Callen focus on the relevance of an adequate choice of variables.
This question guide us to the concept of thermodynamic variable of
state (VOS). We are interested in such VOS that characterize the spatial
extension and other spatial features of an inhomogeneous fluid. A
long time ago, in the origins of thermodynamics, volume was recognized
as a good VOS for diluted gases as was stated in Boile's law in 1662.
A step forward was the introduction of surface area and curvature
as VOS, it is documented in the study of vapor-fluid spherical interfaces
made in 1805 and 1806 by Young and Laplace \cite{Young_1805,Laplace}.
Although, in 1875 Gibbs \cite{Gibbs1906} extended the use of curvature
measures as VOS when he analyzed non-spherical fluid-vapor and fluid-fluid
interfaces. Gibbs, also suggested the use of the length of the three
fluid interface line as VOS. This idea was further developed in 1977
by Boruvka and Newmann \cite{Boruvka_1977}, which also introduced
the curvature of such line as VOS. These VOS were extensively applied
to the thermodynamic analysis in a variety of macroscopic inhomogeneous
fluid systems including liquid-vapor and liquid-liquid interfaces,
and adsorption of fluids on solids in accordance with SU \cite{Sokolowski_1979,Henderson_1983,Henderson_2002,Henderson_2004,Blokhuis_2007},
but they were never applied to the thermodynamic analysis of few body
systems, in contradiction to SI. Besides, these thermodynamic magnitudes
are based in geometrical concepts, but even when the geometrical concepts
have a precise definition, their counterpart thermodynamic magnitudes
have usually not a precise meaning. For example, in the system of
many hard spheres in contact with a (convex) spherical wall different
choices for the locus of the so called Gibbs dividing surface is not
innocuous. A comparison between Refs. \cite{Bryk_2003} and \cite{Blokhuis_2007}
shows that the locus of this surface may modify the volume and surface
area of the inhomogeneous non-planar fluid system. Both modifications
influence the macroscopic description of the entire system, changing
the Laplace equation, the surface tension, etc. The most dramatic
change is probably in the Tolman length.

Therefore, we introduce a set $\mathbf{M}$ of thermodynamic measures,
which should be suitable VOS in accordance with SUSI requirements.
We seek for a set $\mathbf{M}$ with a precise definition which enables
an exact description of few body exactly solved systems, and also,
we expect that a good choice for $\mathbf{M}$ provides consistence
with previous well stablished known results. The homogeneous fluids
are typically described by taking $\mathbf{M}=\left\{ V\right\} $
with $dim(\mathbf{M})=1$, while for inhomogeneous systems several
authors currently add the surface area, being $\mathbf{M}=\left\{ V,A\right\} $
and $dim(\mathbf{M})=2$. The classical analysis of the ideal gas
produce an elementary EOS, $PV=NkT$. Accordingly, $\mathbf{M}$ must
include a volume measure $V$ with a pressure provided by $P=-\partial_{V}F(\mathbf{M})$
compatible with the known system pressure, yielding the expected behavior
for non interacting particles. The same thought applies for the surface
area of the substrate and the wall-fluid surface tension $\gamma$.
The discussion about the choice of $\mathbf{M}$ will be completed
later in PW. Now, assuming that we have adopted a set $\mathbf{M}$
and also that $\mathbf{M}(\mathbf{X})$ is given, we must implement
the thermodynamic description of the system using these measures.
With this purpose we need to relate the $\mathbf{X}$-representation
and the $\mathbf{M}$-representation. We state that $w_{ab}$ must
be independent of the adopted representation $\mathbf{X}$ or $\mathbf{M}$,
then we claim\begin{equation}
w_{ab}(\mathbf{M})=w_{ab}(\mathbf{X})\:,\label{eq:wab-path}\end{equation}
\begin{equation}
dw_{a\hat{\mathbf{M}}}=dw_{a\hat{\mathbf{X}}}\:,\label{eq:dw-path}\end{equation}
where we assume that $\mathbf{M}_{a}$ and $\mathbf{M}_{b}$ are well
defined quantities and also, that for all $\mathbf{X}\in\mathbb{H}$
must exist $\mathbf{M}(\mathbf{X})$. Hence, Eqs. (\ref{eq:w_p},
\ref{eq:dw_p}) transform to\begin{equation}
w_{ab}=-\intop_{a}^{b}\nabla_{\mathbf{M}}F\cdot d\mathbf{M}=-\intop_{a}^{b}\mathbf{m}\cdot d\mathbf{M}\:,\label{eq:w_m}\end{equation}
\begin{equation}
dw_{a\hat{\mathbf{M}}}=-\partial_{\hat{\mathbf{M}}}F\cdot dl=-\mathbf{m}\cdot\hat{\mathbf{M}}\, dl\:,\label{eq:dw_m}\end{equation}
where $\mathbf{m}\equiv\nabla_{\mathbf{M}}F$, and for a given direction
$\hat{\mathbf{X}}$ in the parameters space $\hat{\mathbf{M}}=\nabla_{\mathbf{X}}\mathbf{M}\cdot\hat{\mathbf{X}}$.
Comparing Eq. (\ref{eq:dw_p}) with Eq. (\ref{eq:dw_m}) we find\begin{equation}
\nabla_{\mathbf{X}}F\cdot\hat{\mathbf{X}}=\mathbf{m}\cdot\hat{\mathbf{M}}=\sum_{j}\left.\frac{\partial F}{\partial M_{j}}\right|_{\mathbf{M}-M_{j}}\left(\sum_{i}\left.\frac{\partial M_{j}}{\partial X_{i}}\right|_{\mathbf{X}-X_{i}}\hat{X}_{i}\right)\:,\label{eq:dwp-eq-dwm}\end{equation}
where $\hat{X}_{i}$ is the $i$-component of $\hat{\mathbf{X}}$,
$m_{j}=\left.\frac{\partial F}{\partial M_{j}}\right|_{\mathbf{M}-M_{j}}$,
and $\mathbf{M}-M_{j}$ means that all the measures but the $j$-component
are kept constants in the partial derivative. The Eq. (\ref{eq:dwp-eq-dwm})
is simply the chain rule for the $F$ derivatives. When we adopt the
volume notion of Eq. (\ref{eq:Gammaw-0}) as the volume measure $M_{1}=V=\mathcal{V}$
we obtain $P=-m_{1}$, and also from Eqs. (\ref{eq:dwp-eq-dwm}, \ref{eq:Gammaw-0})\begin{equation}
P-\bar{P}_{w,\,\hat{\mathbf{X}}}=\Delta P_{\hat{\mathbf{X}}}=\sum_{j=2}^{dim(\mathbf{M})}m_{j}\,\frac{\nabla_{\mathbf{X}}M_{j}\cdot\hat{\mathbf{X}}}{\nabla_{\mathbf{X}}V\cdot\hat{\mathbf{X}}}\:,\label{eq:Lapl00}\end{equation}
which is a Laplace-like equation for a fluid-substrate interface \cite{Blokhuis_2007}.
The Eqs. (\ref{eq:Gammaw-0}, \ref{eq:Lapl00}) show that $\left|\hat{\mathbf{X}}\right|$
is irrelevant and therefore the restriction to unit modulus in Eq.
(\ref{eq:dw_PwdV-0}) is superfluous. An interesting point is that
$\bar{P}_{w,\,\hat{\mathbf{X}}}$ and $P$ can be measured both experimentally
and with molecular dynamic simulations.

Now, to make a practical use of Eq. (\ref{eq:Lapl00}) the unknowns
$m_{j}$, i.e. the EOS of the system, should be revealed. Therefore,
we need $F(\mathbf{M})$ (see Eq. (\ref{eq:w_m})). In general the
set $\mathbf{M}$ may include \textit{dependent} magnitudes and then
$dim(\mathbf{M})\neq dim(\mathbf{X})$ showing that relation $\mathbf{M}\leftrightarrow\mathbf{X}$
is not a one-to-one or biyective relation. Thus, the transformation
$F(\mathbf{X})\rightarrow F(\mathbf{M})$ is not a simple change of
variables, which disable us to obtain $F(\mathbf{M})=F(\mathbf{X}(\mathbf{M}))$.
We need a procedure to identify the hidden dependence of $F(\mathbf{X})$
in $\mathbf{M}$. Accordingly, we must overcome two difficulties,
find a good set $\mathbf{M}(\mathbf{X})$ and obtain $F(\mathbf{M})$.
Now, we can show that the selection of measures $\mathbf{M}$ and
the identification of $F(\mathbf{M})$ are not independent questions.
To proceed, we analyze some results for the 2-HS confined system.

We are mainly interested in fluid-like systems where particles can
move freely and are able to interchange their positions. Then, we
look for measures $\mathbf{M}$ that enable the thermodynamic description
of systems in this regime. Certainly this $\mathbf{M}$ may or may
not be suitable to describe other situations as solid-like or dense
systems. The graph decomposition presnted in Sec. \ref{sec:Two-bodies}
(see also \cite{Urrutia_2008}) and the analysis performed in Sec.
\ref{sec:Analysis}.\ref{des:(a)-the-large} show that some thermodynamic
measures $\mathbf{M}$ appear naturally in $F$ for cavity sizes in
the \textbf{Region 1}. For higher confinement conditions, as in \textbf{Region
2 and 3}, some characteristics surface areas and lengths of the cavity
also emerge as thermodynamic measure candidates. We focus on the results
for \textbf{Region 1} where any characteristic length of the cavity
is greater than $\sigma$. The list of measures candidates starts
with the volume $V\equiv Z_{1}=\int e(r)\, d\mathbf{r}$, suggested
by the graph decomposition in Eqs. (\ref{eq:2cont}, \ref{eq:2cont2}).
This volume appears usually in the study of inhomogeneous fluids \cite{Sokolowski_1977,Sokolowski_1978,Sokolowski_1979,Sokolowski_1981,Rowlinson_1985}
of different nature. Interestingly, for fluid systems in contact with
hard walls, this $V(\mathbf{X})$ makes that $\bar{P}_{w,\,\hat{\mathbf{X}}}$
reduces to the contact pressure on the hard-wall. In fact, it reproduces
exactly the hard-wall pressure contact theorem for planar, spherical
and cylindrical hard walls, but also for much more complex geometrical
shapes of the cavity \cite{Urrutia_2010c}. Other magnitudes are also
suggested by the Eqs. (\ref{eq:b00}, \ref{eq:b2Vgen}), e.g. the
surface area measure defined as $A\equiv\int\nabla e(r)\cdot\hat{n}\, d\mathbf{r}$.
We also consider $Le$, the measure of total edges length with right
internal angle. More measures could be added, the number of right
vertex, $N_{vert}$, some measure of the surface curvature e.g. $\mathsf{M}\equiv\int(\frac{3}{4}\, j^{2}+k)\, d\mathbf{S}$,
and a measure of the edge's curvature.

Finally, even for \textbf{Region 1}, to ensure the exactness of Eq.
(\ref{eq:Lapl00}) in principle we should include $\mathbf{X}$ in
the set of measures. With all these measures we may conform a complete
measure set $\mathbf{M}_{c}=\left\{ V,A,L_{e},N_{vert},\mathsf{M},\mathbf{X}\right\} $,
which is certainly not a small set of measures. We note that a hierarchy
exist in $\mathbf{M}_{c}$, the most important therm is $V$, the
second in relevance is $A$. Both of them have been defined in detail,
and its definition can be applied to a large class of systems. Next
terms, $Le$, $N_{vert}$ and $\mathsf{M}$ behave less important
and their definition concern particular characteristics of confinement
cavity. Finally, the last added terms to $\mathbf{M}_{c}$ are still
less relevant. Their definition applies only to a given cavity geometry,
and were included to make a complete description of $F$, so that
Eqs. (\ref{eq:wab-path}, \ref{eq:dw-path}) are guaranteed. The loss
of relevance for incoming terms in $\mathbf{M}_{c}$ relates with
the SU hypothesis.

Now, we take into account all these questions to analyze the thermodynamic
behavior of 2-HS systems in \textbf{Region 1}. The spheroidal cavity
will be excluded from the thermodynamic analysis because we do not
find a small set $\mathbf{M}$ that enable the unified study of this
and other geometries. We adopt the small set of measures $\mathbf{M}=\left\{ V,A,L_{e},R\right\} $
where the last parameter is the radius of curvature of the (curved)
surface. Measure $R$ is frequently used in the study of fluid systems
in contact with simple curved surfaces as such with cylindrical or
spherical symmetries \cite{Poniewierski_1997,Blokhuis_2007}. We also
select a rule to identify the dependence of $F$ on the adopted set
of measures $\mathbf{M}$. It is based on re-writing Eq. (\ref{eq:b2Vgen})
in the form\begin{equation}
Vb_{2}(pore)=Vb_{2}-A\, a(R)+L_{e}\,\ell(R)\;.\label{eq:Vb-Th1}\end{equation}
Again, the adopted $\mathbf{M}$ and the identification rule are non-unique.
In the Appendix \ref{sec:Appendix-A} a different $\mathbf{M}$ is
analyzed. For the adopted $\mathbf{M}=\left\{ V,A,L_{e},R\right\} $,
we can define the $F$ derivatives related with the volumetric-work,
surface-area-work, edges-length-work and radius-of-curvature-work
\begin{equation}
-P=\left.\frac{\partial F}{\partial V}\right|_{T,\mathbf{M}-V}\:,\label{eq:press00}\end{equation}
\begin{equation}
\gamma=\left.\frac{\partial F}{\partial A}\right|_{T,\mathbf{M}-A}\:,\label{eq:surften00}\end{equation}
\begin{equation}
\tau=\left.\frac{\partial F}{\partial Le}\right|_{T,\mathbf{M}-Le}\:,\label{eq:lineten00}\end{equation}
\begin{equation}
C_{R}=\left.\frac{\partial F}{\partial R}\right|_{T,\mathbf{M}-R}\:.\label{eq:surfcurvcorr00}\end{equation}
From Eq. (\ref{eq:Lapl00}) we relate the difference of pressures
$\Delta P_{\hat{\mathbf{X}}}$ for an infinitesimal deformation in
$\hat{\mathbf{X}}$ direction with $\gamma$, $\tau$, etc. by\begin{equation}
\Delta P_{\hat{\mathbf{X}}}=\gamma\,\frac{\nabla_{\mathbf{X}}A\cdot\hat{\mathbf{X}}}{\nabla_{\mathbf{X}}V\cdot\hat{\mathbf{X}}}+\tau\,\frac{\nabla_{\mathbf{X}}Le\cdot\hat{\mathbf{X}}}{\nabla_{\mathbf{X}}V\cdot\hat{\mathbf{X}}}+\ldots\:,\label{eq:Lapl000}\end{equation}
Now it is apparent that Eq. (\ref{eq:Lapl000}) is a generalization
of the Laplace equation obtained for a macroscopic fluid system in
contact with a spherical wall \cite{Henderson_1983,Henderson_1986_inCroxton86}.
An interesting fact is that the EOS given in Eqs. (\ref{eq:press00}-\ref{eq:surfcurvcorr00})
may be strongly dependent on the details of the fluid system. On the
other hand, the relation between $\Delta P_{\hat{\mathbf{X}}}$, $\gamma$,
$\tau$, etc. given by the Laplace-like equation (\ref{eq:Lapl000})
only depends on the geometry of the cavity and the adopted $\mathbf{M}$.
For example, given a cuboidal pore it remains unperturbed if we confine
2-HS, an N-Lennard-Jones, or any other fluid. Before analyzing each
confinement geometry we wish to state that, for all the studied cavities,
the thermodynamic pressure from Eq. (\ref{eq:press00}) is\begin{equation}
\beta P=Z_{2}^{-1}\,2\left(V-b_{2}\right)\:.\label{eq:Pr-all}\end{equation}
This is our first global or thermodynamic EOS for the 2-HS system.
The same expression was obtained in Eq. (\ref{eq:Pt-p}) for the local
pressure in the constant density region when we analyze cavities of
any shape. We find that both pressures are equal, which shows the
consistence of the present thermodynamic study. A similar result for
spherical confinement was previously obtained \cite{Urrutia_2010}.
Based on the universal behavior of Eqs. (\ref{eq:b2Vgen}, \ref{eq:Vb-Th1})
and the consistence between the local pressure in the constant density
region and thermodynamic pressure in all the studied cavity geometries,
we confirm that the adopted volume measure is correct in the spirit
of SUSI. Therefore, taking the volume measure $V=Z_{1}$ we argue
that the identity between both pressures should be true for a 2-HS
system in \textit{any cavity shape}. In the next paragraphs we perform
the thermodynamic analysis for each pore shape. We fix $\sigma=1$
to keep notation simple.

\subsection{The cuboidal pore\label{sub:Thermo-cuboid}}

The cuboidal pore does not involve $R$, then $\mathbf{M}=\left\{ V,\, A,\, Le\right\} $.
We obtain the thermodynamic pressure of Eq. (\ref{eq:Pr-all}) and
also,\begin{equation}
\beta\gamma=-Z_{2}^{-1}\,2a_{2}\:,\label{eq:SurfT-cub}\end{equation}
\begin{equation}
\beta\tau=Z_{2}^{-1}\,2\ell_{2}\:.\label{eq:LineT-cub}\end{equation}
The three EOS relate the pressure, surface tension, and line-tension
with the measures $\left\{ V,\, A,\, Le\right\} $ of the system.
They apply to any cuboidal pore, in particular these equations are
valid for the cubic confinement. Surface tension $\gamma$ of Eq.
(\ref{eq:SurfT-cub}) is in coincidence with Eq. (\ref{eq:SurfTP}),
it is negative for large enough cavities. A Simple inspection shows
that for large cuboids the EOS scales $\beta P\simeq\rho+b_{2}\rho^{2}/2$,
$\beta\gamma\simeq-a_{2}\rho^{2}/2$ and $\beta\tau\simeq\ell_{2}\rho^{2}/2$
. For $\bar{P}_{w,\,\hat{\mathbf{X}}}$ we may find in the literature
two frequently used deformations. Adopting the length parameters $\mathbf{X}=\left\{ L_{x},\, L_{y},\, L_{z}\right\} $,
the first one is like a piston expansion transformation and reads
$\hat{\mathbf{X}}=(1,0,0)$. From Eq. (\ref{eq:Gammaw-0}) \begin{equation}
\beta\bar{P}_{w,\,\hat{\mathbf{X}}}=Z_{2}^{-1}\left[2\left(V-b_{2}\right)+\frac{2a_{2}2(L_{y}+L_{z})-2\ell_{2}4}{L_{y}L_{z}}\right]\:,\label{eq:Pw-cub1}\end{equation}
with this choice of $\hat{\mathbf{X}}$ magnitudes $P$, $\gamma$,
$\tau$, and $\bar{P}_{w,\,\hat{\mathbf{X}}}$ are related each other
by\begin{equation}
\Delta P_{\hat{\mathbf{X}}}=\gamma\,\left(\frac{2}{L_{y}}+\frac{2}{L_{z}}\right)+\tau\,\frac{4}{L_{y}L_{z}}\:.\label{eq:Lapl-cub1}\end{equation}
Equivalent results may be obtained with $\hat{\mathbf{X}}=(0,1,0)$
or $\hat{\mathbf{X}}=(0,0,1)$. The second option is an isotropic
expansion with $\hat{\mathbf{X}}=(1,1,1)$, which produces\begin{equation}
\beta\bar{P}_{w,\,\hat{\mathbf{X}}}=-Z_{2}^{-1}\left[2\left(V-b\right)+\frac{2a_{2}4(L_{x}+L_{y}+L_{z})-2\ell_{2}12}{L_{y}L_{z}+L_{x}L_{z}+L_{x}L_{y}}\right]\:,\label{eq:Pw-cub2}\end{equation}
\begin{equation}
\Delta P_{\hat{\mathbf{X}}}=\gamma\,\frac{2Le}{A}+\tau\,\frac{24}{A}\:.\label{eq:Lapl-cub2}\end{equation}
We now analize $\hat{\mathbf{X}}=(1,0,0)$, $\hat{\mathbf{X}}=(1,1,1)$
starting with a cubical cavity $L=L_{x}=L_{y}=L_{z}$. In this case
Eqs. (\ref{eq:Pw-cub1}) and (\ref{eq:Pw-cub2}) converge to a single
expression. The same applies also to Eqs. (\ref{eq:Lapl-cub1}, \ref{eq:Lapl-cub2})
which can be simplified because $(A-A_{x})/V=4L^{-1}$, $Le_{x}/V=4L^{-2}$,
and $Le/A=2\, L^{-1}$. Therefore, for all the studied $\hat{\mathbf{X}}$
for a cubical cavity we obtain\begin{equation}
\beta\bar{P}_{w,\,\hat{\mathbf{X}}}=-Z_{2}^{-1}\left[2\left(V-b\right)+8\frac{a_{2}L-\ell_{2}}{L^{2}}\right]\:,\label{eq:Pw-cubic}\end{equation}
\begin{equation}
\Delta P_{\hat{\mathbf{X}}}=\gamma\,\frac{4}{L}+\tau\,\frac{4}{L^{2}}\:.\label{eq:Lapl-cubic}\end{equation}
The same expressions (\ref{eq:Pr-all}, \ref{eq:SurfT-cub}, \ref{eq:LineT-cub},
\ref{eq:Pw-cubic}, \ref{eq:Lapl-cubic}) are obtained if we start
from the beginning the analysis of a cubical cavity with $\mathbf{X}=\{L\}$
and $\hat{\mathbf{X}}=(1)$, which shows the robustness of the procedure.

\subsection{The cylindrical pore\label{sub:Thermo-cyl}}

From the same basic set of measures we recognize that the planar and
curved surfaces, with areas $A_{p}$ and $A_{c}$, respectively, are
geometrically and therefore thermodynamically different. Then we split
the total area in two, adopting $\mathbf{M}=\left\{ V,\, A_{p},A_{c},\, Le,\, R\right\} $.
For $P$ and $\gamma_{p}$ we obtain expressions identical to Eqs.
(\ref{eq:Pr-all}, \ref{eq:SurfT-cub}). Other EOS are\begin{equation}
\beta\gamma_{c}=-Z_{2}^{-1}2a_{2}\left(1-c_{I}R^{-2}\right)\:,\label{eq:SurfT-cyl}\end{equation}
\begin{equation}
\beta\tau=Z_{2}^{-1}2\ell_{2}\left(1-c_{I\! I}R^{-2}\right)\:,\label{eq:LineT-cyl}\end{equation}
\begin{equation}
\beta C_{R}=-Z_{2}^{-1}R^{-3}\left(c_{I\! I\! I}A_{c}+c_{I\! V}Le\right)\:,\label{eq:CurvT-cyl}\end{equation}
where $c_{I}=F(\mathsf{s})/48$, $c_{I\! I}=G(\mathsf{s})\,\pi/14$,
$c_{I\! I\! I}=\pi/96\,\left(F(\mathsf{s})+\mathsf{s}F'(\mathsf{s})/2\right)$,
and $c_{I\! V}=\pi/210\,\left(G(\mathsf{s})+\mathsf{s}G'(\mathsf{s})/2\right)$.
All of these coefficients are positive smooth functions with small
values, e.g. $c_{I}<0.03$, which shows that $\gamma_{p}\simeq\gamma_{c}$.
We can also extract the curvature dependence of $\gamma$ \begin{equation}
\frac{\gamma_{c}}{\gamma_{p}}-1=-\frac{1}{48}R^{-2}+O(R^{-4})\:.\label{eq:SurfTdiff-cyl}\end{equation}
Taking $\mathbf{X}=\left\{ L_{h}/2,\, R\right\} $ for this geometry,
we see three simple choices for $\hat{\mathbf{X}}$. The piston expansion
$\hat{\mathbf{X}}=(1,0)$ provides\begin{equation}
\Delta P_{\hat{\mathbf{X}}}=\gamma_{p}\,\frac{2}{R}\:.\label{eq:Lapl-cyl1}\end{equation}
For the lateral $\hat{\mathbf{X}}=(0,1)$ and isotropic $\hat{\mathbf{X}}=(1,1)$
expansions we obtain\begin{equation}
\Delta P_{\hat{\mathbf{X}}}=\gamma_{p}\frac{2}{L_{h}}+\gamma_{c}\frac{1}{R}+\tau\,\frac{2}{L_{h}R}+C_{R}\frac{1}{A_{c}}\:,\label{eq:Lapl-cyl2}\end{equation}
\begin{equation}
\Delta P_{\hat{\mathbf{X}}}=\gamma_{p}\frac{Le}{A}+\gamma_{c}\,\frac{Le+2\pi L_{h}}{A}+\tau\,\frac{4\pi}{A}+C_{R}\frac{1}{A}\:.\label{eq:Lapl-cyl3}\end{equation}
For the cylinder with maximum area at fixed volume $L_{h}/2=R$ and
Eq. (\ref{eq:Lapl-cyl3}) reduces to\begin{equation}
\Delta P_{\hat{\mathbf{X}}}=\left(\gamma_{p}+2\gamma_{c}\right)\,\frac{2}{3R}+\left(\tau+\frac{C_{R}}{4\pi}\right)\frac{2}{3R^{2}}\:,\label{eq:Lapl-cylsim}\end{equation}
which may be still obtained analyzing the cylindrical pore from the
beginning with $L_{h}/2=R$, $\mathbf{X}=\left\{ R\right\} $ and
$\hat{\mathbf{X}}=(1)$.

\subsection{The spherical and spherical+core pores\label{sub:Thermo-sph}}

The spherical pore has $Le=0$, which reduces the measures to $\mathbf{M}=\left\{ V,\, A,\, R\right\} $.
The expression for $P$ is identical to Eq. (\ref{eq:Pr-all}) to
which we add

\begin{equation}
\beta\gamma=-Z_{2}^{-1}2a_{2}\left(1-18^{-1}R^{-2}\right)\:,\label{eq:SurfT-sph}\end{equation}
\begin{equation}
\beta C_{R}=-Z_{2}^{-1}2a_{2}A\,9^{-1}R^{-3}\:.\label{eq:CurvTR-sph}\end{equation}
For $\mathbf{X}=\left\{ R\right\} $ and $\hat{\mathbf{X}}=(1)$ the
difference of pressures is\begin{equation}
\Delta P_{\hat{\mathbf{X}}}=\gamma\,\frac{2}{R}+\frac{C_{R}}{A}\:,\label{eq:Lapl-sph}\end{equation}
and the same result is obtained for any other choice of $\mathbf{X}$
and $\hat{\mathbf{X}}$. Note that using the Eqs. (\ref{eq:SurfT-sph},
\ref{eq:CurvTR-sph}) we may transform $\frac{C_{R}}{A}$ to obtain\begin{equation}
\Delta P_{\hat{\mathbf{X}}}=\gamma\,\left(\frac{2}{R}+\beta C_{R}\right)+\frac{\partial\gamma}{\partial R}\:.\label{eq:Lapl-sph2}\end{equation}
Eq. (\ref{eq:Lapl-sph2}) is very similar to the Laplace equation
for a macroscopic fluid in contact with a spherical wall \cite{Henderson_1983,Henderson_1986_inCroxton86,Blokhuis_2007}.
Now, we will analyze the 2-HS system in a spherical pore with a hard
core. The shape of this pore involves two surfaces with different
curvature. Therefore, as we did for the cylindrical cavity we consider
two separate surface area measures. We adopt $\mathbf{M}=\left\{ V,\, A_{e},\, A_{i},\, R_{e},-R_{i}\right\} $,
where $-R_{i}$ is the (negative) radius of curvature of the internal
surface, and naturally, the labels $e$ and $i$ design the properties
of the external and internal surfaces, respectively. The pressure
$P$ is given in Eq. (\ref{eq:Pr-all}), while other EOS are\begin{equation}
\beta\gamma_{e}=-Z_{2}^{-1}2a_{2}\left(1-18^{-1}R_{e}^{-2}\right)\:,\label{eq:SurfT-sph+c1}\end{equation}
\begin{equation}
\beta\gamma_{i}=-Z_{2}^{-1}2a_{2}\left(1-18^{-1}R_{i}^{-2}\right)\:,\label{eq:SurfT-sph+c2}\end{equation}
\begin{equation}
\beta C_{Re}=-Z_{2}^{-1}2a_{2}A_{e}9^{-1}R_{e}^{-3}\:,\label{eq:CurvTR-sph+c1}\end{equation}
\begin{equation}
\beta C_{-Ri}=Z_{2}^{-1}2a_{2}A_{i}9^{-1}R_{i}^{-3}\:.\label{eq:CurvTR-sph+c2}\end{equation}
In these Eqs. we recognize that the opposite sign in both radius of
curvature do not affect surface tension expressions, but invert the
sign in curvature term. Adopting the length parameters $\mathbf{X}=\left\{ R_{e},-R_{i}\right\} $,
we can analyze three simple transformations $\hat{\mathbf{X}}=(1,0)$,
$\hat{\mathbf{X}}=(0,1)$, and $\hat{\mathbf{X}}=(1,1)$. For the
first two cases we find\begin{equation}
\Delta P_{\hat{\mathbf{X}}}=\gamma_{e}\,\frac{2}{R_{e}}+\frac{C_{Re}}{A_{e}}\:,\label{eq:Lapl-sph+c1}\end{equation}
\begin{equation}
\Delta P_{\hat{\mathbf{X}}}=-\gamma_{i}\,\frac{2}{R_{i}}+\frac{C_{-Ri}}{A_{i}}\:.\label{eq:Lapl-sph+c2}\end{equation}
Here the effect of the negative curvature radius in the Laplace-like
equations is apparent. The last transformation gives\begin{equation}
\Delta P_{\hat{\mathbf{X}}}=\frac{\gamma_{e}+\gamma_{i}}{R_{e}+R_{i}}+\frac{\gamma_{e}-\gamma_{i}}{L_{h}}+\frac{C_{Re}+C_{-Ri}}{A_{e}-A_{i}}\:.\label{eq:Lapl-sph+c3}\end{equation}

For all the studied simple geometrical confinement we obtained several
relation between intensive-like magnitudes that resemble the Laplace
equation. We must stress that previous to implement the thermodynamic
study of the system we has needed to choose both, a set $\mathbf{M}$
and an identification rule (see Eq. (\ref{eq:Vb-Th1})). Both choices
affect its thermodynamic description. In the Appendix \ref{sec:Appendix-A}
a different choice for $\mathbf{M}$ is taken, which produce other
set of EOS.

\begin{figure}[th]
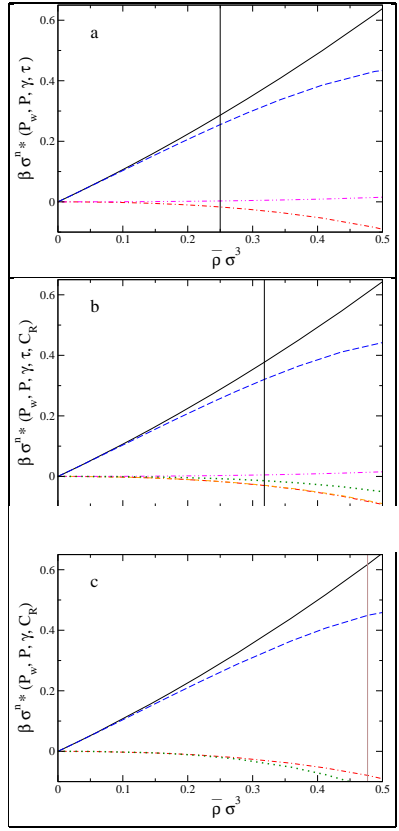

\centering{}\begin{tabular}{|c|}
\hline 
\includegraphics[clip,width=5cm]{P_shapes_cube}\tabularnewline
\hline 
\includegraphics[clip,width=5cm]{P_shapes_cyl}\tabularnewline
\hline 
\includegraphics[width=5cm]{P_shapes_sph}\tabularnewline
\hline
\end{tabular}\caption{(color online) Magnitudes related to the thermodynamic work: pressures,
wall-fluid surface tensions, line tensions and curvature-work. From
top to bottom a) is for a cube cavity, b) is for a symmetric cylindrical
cavity and c) is for a spherical cavity. The power $n$ is chosen
in each case to give a dimensionless magnitude.\label{fig:pressures}}

\end{figure}

In Figure \ref{fig:pressures} we plot pressure for work, pressure,
surface tension and other EOS for the cube, the $L_{h}=2R$ cylinder
and spherical cavities (subfigures a, b, and c, respectively) as
functions of the rough density. In continuous line we plot $\bar{P}_{w,\,\hat{\mathbf{X}}}$,
in dashed line $P$, in dot-dashed line $\gamma$, dot-dot-dashed
line shows $\tau$, and dotted line $C_{R}$. The vertical line shows
the end of the plateau of constant density $\rho_{0}$ and constant
pressure $P_{0}$. The most remarkable feature is that Figs. \ref{fig:pressures}a,
\ref{fig:pressures}b, and \ref{fig:pressures}c are very similar
in the density range $(0,\,0.5)$. A very small difference in $\bar{P}_{w,\,\hat{\mathbf{X}}}$
at $\bar{\rho}\sigma^{3}=0.5$ is due to the expected geometric dependence
of the ratio $V/A$. In Fig. \ref{fig:pressures}b both wall-fluid
surface tensions, $\gamma_{p}$ and $\gamma_{c}$, are plotted but
are indistinguishable. A clear difference between the three figures
appears in the curvature term, which is not present for the cube cavity
at Fig. \ref{fig:pressures}a. In Fig. \ref{fig:pressures}b $C_{R}$
mix two curvature contributions, one due to the curved surface and
other due to the curved edges. Even, in Fig. \ref{fig:pressures}c
$C_{R}$ is a purely surface curvature effect. A second difference
is the vertical line that shows the end of the plateau, which is a
purely local property of the 2-HS system. Note that $F(\mathbf{X})$
(and also $F(\mathbf{M})$) is an analytic function at this point
because their analytic domain extends to the end of Region 1. The
non-analytic point for Figs. \ref{fig:pressures}a, \ref{fig:pressures}b
and \ref{fig:pressures}c corresponds to maximum densities $\bar{\rho}\sigma^{3}=2,\,2.55$
and $3.82$, respectively. Beyond the vertical plateau-end-line the
identification of the thermodynamic pressure $P$ with the plateau's
pressure $P_{0}$ breaks down because the central plateau of constant
density disappears. If we wish to retain the identity beyond this
point we can regard about the analytic continuation of Eqs. (\ref{eq:rho0})
and (\ref{eq:Pt}). This approach may conduce to a non-monotonic behavior
of $P$ related in some cases (for the spherical cavity) with a negative
$\rho_{0}$, although, the total work $\bar{P}_{w,\,\hat{\mathbf{X}}}$
is not influenced by this question.

In consonance with Eq. (\ref{eq:SurfTdiff-cyl}) we visualize the
possibility of analyze a cavity that mix planar and curved spherical
surfaces. Such truncated-spherical cavity should have a $Z_{2}$
involving a complex dependence on some set of parameters $\mathbf{X}$.
By virtue of Eq. (\ref{eq:b2Vgen}) we infer that each surface makes
its own contribution to $Z_{2}$ allowing to obtain both the wall-fluid
surface tension related to the spherical surface $\gamma_{c}$ and
that corresponding to the planar one $\gamma_{p}$. Then $\gamma_{c}$
should be essentially given by Eq. (\ref{eq:SurfT-sph}) and $\gamma_{p}$
by Eq. (\ref{eq:SurfT-cub}) with a common unknown function $Z_{2}$,
therefore\begin{equation}
\frac{\gamma_{c}}{\gamma_{p}}-1=-\frac{1}{18}R^{-2}+O(R^{-4})\:.\label{eq:SurfTdiff-sph}\end{equation}
Other interesting cavity is the half-cylinder, it mix curved and linear
right anlgled edges. Even that we ignore $Z_{2}$ taking Eqs. (\ref{eq:LineT-cub})
and (\ref{eq:LineT-cyl}) we can obtain the curvature dependence of
$\tau$\begin{equation}
\frac{\tau_{c}}{\tau_{p}}-1=-\frac{\pi}{14}R^{-2}+O(R^{-4})\:.\label{eq:CurvTdiff-hcyl}\end{equation}
The idea of build mixed shape cavities allow us to explore several
confinement conditions involving complex geometrical shapes. As was
already stated at \cite{Urrutia_2010} and discussed in Sec. \ref{sec:Two-bodies}
some results of PW are easily mapped from the 2-HS system confined
in a bounded region to the 2-HS system confined to the conjugated
unbounded region. This is a consequence of the inside-outside symmetry.
Particularly, all the expressions which are independent of $Z_{1}$
and $Z_{2}$ are symmetric with respect to an inside-outside transformation.
Therefore, Eqs. (\ref{eq:SurfTdiff-cyl}, \ref{eq:SurfTdiff-sph})
and (\ref{eq:CurvTdiff-hcyl}) may also be applied to the conjugated
system where both particles are outside of $\Omega$. In general,
the thermodynamic description of the conjugated system is obtained
by mapping $Z_{1}$ and $Z_{2}$ (see Sec. \ref{sec:Two-bodies}),
and inverting the overall sign of $P$, $\bar{P}_{w,\,\hat{\mathbf{X}}}$
and $\Delta P_{\hat{\mathbf{X}}}$.

\subsection{Extrapolation to systems with many HS}

In \cite{Urrutia_2010} was recognized that some properties of the
2-HS systems can be mapped exactly to the many-HS systems in the low
density regime. We simply follow the arguments of that work. In large
inhomogeneous systems the thermodynamic limit is frequently considered,
and sometimes, becomes convenient to introduce a mathematical surface
where the surface tension is supposed to act. This is the so called
Gibbs dividing surface. In our previous analysis we have not introduced
a Gibbs dividing surface. Even so, if we are forced to define it we
must assume that our Gibbs dividing surface is placed in coincidence
with the surface of diverging external potential, e.g. for the spherical
cavity it is the surface of a sphere with radius $R$. The wall-fluid
surface tension of a HS fluid in contact with a curved wall and its
limiting zero curvature value at the same density relate by \begin{equation}
\frac{\gamma_{c}(R)}{\gamma_{c}(\infty)}-1=-c_{V}R^{-2}+O(R^{-4})+O(\rho)\:,\label{eq:SurfTdif}\end{equation}
where the geometric dependent coefficient is $c_{V}(sph)=1/18$ and
$c_{V}(cyl)=1/48+O(R^{-2})$. Both results apply to the HS fluid inside
the cavity, but also, for the fluid outside the cavity. This symmetry
is clear for the spherical surface, from the study of a fluid inside
the spherical cavity with a central core, see Eqs. (\ref{eq:SurfT-sph+c1},
\ref{eq:SurfT-sph+c2}). Even, it is a consequence of the more general
inside-outside symmetry. The central characteristics of Eq. (\ref{eq:SurfTdif})
is its zero order in density and second order in the radius of curvature
$R$. We are now able to extract an interesting property of the HS
fluid in contact with a curved hard wall. From Eq. (\ref{eq:SurfTdif})
the usual definition of the substrate-fluid Tolman length $\delta$
(a magnitude independent of the radius of curvature) is $\gamma_{c}/\gamma_{p}-1=-2\delta/R+O_{2}(R^{-1})$.
Therefore, we obtain\begin{equation}
\delta=0+O(\rho)\:,\label{eq:dTolman}\end{equation}
for both, spherical and cylindrical surfaces. It still applies for
HS fluid systems confined inside of the closed surface, and also,
for fluids outside it. Our\emph{ exact} result for $\delta$ is in
contradiction with the constant value $\delta=-\sigma/4$ obtained
in Eq. (35) of Ref. \cite{Bryk_2003} in the same limit. This difference
would be consequence of the unusual volume definition adopted which
does not refl{}ect the volume available to the liquid\textquoteright{}s
molecules (see Eq. (8) in \cite{Bryk_2003} and the comment below
Eq. (6) in Ref. \cite{Blokhuis_2007}). In Ref. \cite{Blokhuis_2007},
Blokhuis et. al. analyze the behavior of a liquid system of particles
interacting with a HS+attractive potential that mimics the London
dispersion forces in contact with a curved hard wall. Using density
functional calculations a limiting behavior of $\delta\simeq0$ independent
of the temperature is found (see Fig. 2 in that work) in good agreement
with Eq. (\ref{eq:dTolman}). In the same sense, our result for $\delta$
agrees with Fig. 9 of Ref. \cite{He_2008}. Other magnitudes can
also be evaluated. The line tension expressed to first non null order
in density is\begin{equation}
\beta\tau=\rho^{2}\frac{\ell_{2}}{2}\:.\label{eq:LineTrho}\end{equation}
In consonance with Eq. (\ref{eq:CurvTdiff-hcyl}), the first order
correction on the line tension due to the curvature of the edge with
a right dihedral angle is

\begin{equation}
\frac{\tau(R)}{\tau(\infty)}-1=-\frac{\pi}{14}R^{-2}+O(R^{-4})+O(\rho)\:,\label{eq:LineTdif}\end{equation}
which appears to be a novel result. The first non null curvature
dependence for the density at contact is \begin{equation}
\rho(r=0,R)-\rho(r=0,\infty)\simeq\rho^{2}\frac{a_{2}\sigma^{3}}{2}c_{VI}R^{-1}\simeq\eta^{2}\frac{9\, c_{VI}}{4\pi}\sigma R^{-1}+O(R^{-3})+O(\rho)\:,\label{eq:rhocdif}\end{equation}
with the packing fraction $\eta=(\pi/6)\,\sigma^{3}\rho$ and $c_{VI}=1,\,2$
for cylinder and spherical cavities, respectively. For a convex wall
we must invert the sign or simple change $R\rightarrow-R$. The Eq.
(\ref{eq:rhocdif}) is in concordance with first density order of
Eq. (36) in \cite{Bryk_2003} which analize a fluid in contact with
a convex hard cavity, but it is a new result for the HS fluid in a
spherical cavity and also for the fluid in contact with a convex or
concave cylindrical walls.

\section{Final Remarks\label{sec:Conclusions}}

The analytical evaluation of the canonical partition function for
the 2-HS confined system were performed for several cavities with
simple geometry. The cavities considered were the cuboidal, cylindrical
and ellipsoidal pores. The obtained expressions cover all density
range from infinite dilution to the jammed densest configuration.
The one body distribution function and pressure tensor were also analyzed.
As a byproduct, we have obtained expressions for the volume of intersection
between a sphere and a dihedron with right angle, between a sphere
and a right-angle vertex, and thus the expression for the intersecting
volume between a sphere and a box. To the best of our knowledge this
expressions were not previously published. The three studied cavities
were compared with the spherical and the spherical with a hard core,
cavities, hence the study of simple pore's geometry is completed.
The general behavior of all the available CI where analyzed by a graphical
representation, which shows how the $\mathbf{X}$-parameter space
breaks in several open analytic domains. Attention was also paid to
the CI solution for large cavities, to the characterization of the
non analytic domain and the dimensional crossovers.

Finally, we have focused on the thermodynamic properties of the 2-HS
confined system. Several questions about the free energy dependence
on geometrical parameters $\mathbf{X}$ and its thermodynamic meaningfulness
were discussed. We show the necessity of introduce a set of thermodynamic
measures $\mathbf{M}$ based in extensive-like magnitudes. These neat
defined measures constitute the basis of a consistent method developed
to make the thermodynamic study. We find that pressures, surface tension
and similar intensive-like magnitudes are then obtainable analytically.
A common feature was the arising of an exact expression resembling
the Laplace equation, which establishes the equilibrium between these
quantities. Finally, several connections to the many-HS system in
contact with curved hard walls were found. We have evaluated the first
curvature corrections to the surface tension, Tolman length and line
tension in the low density limit.

The solved integrals to obtain the CI of 2-HS system also shows the
complete dependence of $b_{2}(pore)$, the first non-trivial cluster
integral, for the many HS system in the cavity but also outside it.
For large enough cavity $b_{2}(pore)$ is analytic. Even that, for
smaller cavity's size $b_{2}(pore)$ is a non analytic function of
the $\mathbf{X}$ and $\mathbf{M}$-parameters. We are convinced that
any cluster integral $b_{j}(pore)$ behave the same behavior. Cluster
integrals are basic functions appearing in the virial expansion of
the so called \textit{real gases} EOS, thus, the study of the unanalicities
of $b_{j}(pore)$ could be of interest.

The performed study of free energy dependence of two-body simple systems
on the geometry of the container does not close the prospection. Indeed,
it shows that next steps should focus in the free energy contributions
of dihedral edges (straight and curved ones), non right vertex and
cone vertex. One of the conclusion of PW is that this future inspection
should be numerical.

\section*{Acknowledgment}

The author wishes to express his gratitude to Dr. Claudio Pastorino
and Dr. Gabriela Castelletti for valuable discussions about some of
the contents in PW. This work was supported in part by the Ministry
of Culture and Education of Argentina through Grants ANPCyT PICT Nos.
31980/05 and 2006-00492, and UBACyT No. x099.

{}

\appendix

\section{Thermodynamic properties with a different $\mathbf{M}$
\label{sec:Appendix-A}}

In Sec. \ref{sec:Thermodynamic-study} we have discussed why the thermodynamic
properties of the systems depends on the adopted set of measures.
Here, we investigate on this dependence adopting a different set of
measures to that used in Secs. \ref{sub:Thermo-cuboid}, \ref{sub:Thermo-cyl}
and \ref{sub:Thermo-sph}. Now, we utilize a measure set $\mathbf{M}$
currently used for the study of systems with spherical interfaces.
We adopt $\mathbf{M}=(V,\, A)$, and add to this the parameters used
for the description of the studied geometric confinement. As far as,
we will restrict ourselves to the full symmetric cavities, cubic,
cylindrical and spherical only one characteristic length parameter
is needed. The unique parameter is $\mathbf{X}=\{R\}$, where $L=2R$
for the cube and $L_{h}=2R$ for the cylinder. Following the same
ideas depicted in Sec. \ref{sec:Thermodynamic-study}, taking an area
common factor in Eq. (\ref{eq:b2Vgen}) it transforms to\begin{equation}
Vb_{2}(pore)=Vb_{2}-A\left(a_{2}-\ell_{2}Le\, A^{-1}+c_{2,1}A^{-1}-c_{2,2}Le\, R^{-2}A^{-1}\right)=Vb_{2}-A\, a(R)\;.\label{eqap:Vb}\end{equation}
Here, the expression between parenthesis only depends on $R$. We
then analyze with $\mathbf{M}=(V,\, A,\, R)$ the isotropic expansion
$\hat{\mathbf{X}}=(1)$. For all the cavities we obtain Eq. (\ref{eq:Pr-all})
for the pressure and\begin{equation}
\beta\gamma=-Z_{2}^{-1}\,2a(R)\:,\label{eqap:SurfT}\end{equation}
\begin{equation}
\beta C_{R}=-Z_{2}^{-1}\,2A\frac{da}{dR}\:.\label{eqap:LineR}\end{equation}
We may highlight that $a(R)$ is a known function and therefore $\gamma$
and $C_{R}$ are analytically known in the three analyzed cavities.
The Laplace-type equation is \begin{eqnarray}
\Delta P_{\hat{\mathbf{X}}} & = & \gamma\,\frac{2}{R}+C_{R}\frac{1}{A}\label{eqap:Lapl}\\
 & = & \gamma\,\left(\frac{2}{R}+\beta C_{R}\right)+\frac{\partial\gamma}{\partial R}\:.\nonumber \end{eqnarray}
Last expression without the $\beta C_{R}$ term was obtained in some
refined studies of spherical cavities in the bulk of fluid systems,
and also, in studies of spherical drops surrounded by its vapor. Notably,
in our systems, which does not need to be spherical $\beta C_{R}\sim R^{-6}$
a higher order term in the characteristic length of the cavity.

\bibliographystyle{plain}

\end{document}